\documentclass{WileyMSP-template}
\usepackage{amsmath,eqparbox}
\usepackage{braket}
\usepackage{amssymb}
\usepackage{mathtools}
\usepackage{xcolor}
\usepackage{bm}
\usepackage{appendix}
\usepackage{multirow}
\usepackage{tabularx}

\newcommand{\reffigs}[2]{Figs. (\ref{#1}) and (\ref{#2})}

\newcommand{\eqrefs}[2]{Eqs.~\eqref{#1} and \eqref{#2}}

\newcommand{\eref}[1]{Eq.~\eqref{#1}}

\newcommand{\lowerbossphantom}{\vphantom{\bar{\bar{x}}}}
\newcommand{\upperbossphantom}{\vphantom{\dagger}}
\newcommand{\tempop}[3][\textstyle]{\settowidth{\dimen1}{$#1\hat{#2}$}\makebox[\dimen1][l]{$#1\hat{#2\mspace{#3}}$}}
\newcommand{\xop}[1]{{\mathchoice{\tempop[\displaystyle]{#1}{3.5mu}}{\tempop{#1}{3.5mu}}{\tempop[\scriptstyle]{#1}{3.5mu}}{\tempop[\scriptscriptstyle]{#1}{3mu}}}}
\newcommand{\chat}[1]{\ensuremath{\xop{#1}}}
\newcommand{\aop}[2]{\ensuremath{\chat{c}_{#1#2\lowerbossphantom}^{\upperbossphantom}}}
\newcommand{\cop}[2]{\ensuremath{\chat{c}_{#1#2\lowerbossphantom}^{\dagger\upperbossphantom}}}
\newcommand{\cbar}[1]{\ensuremath{\xbar{#1}}}

\newcommand{\tempbar}[3][\textstyle]{\settowidth{\dimen1}{$#1\bar{#2}$}\makebox[\dimen1][l]{$#1\bar{#2\mspace{#3}}$}}
 
\newcommand{\xbar}[1]{{\mathchoice{\tempbar[\displaystyle]{#1}{3.5mu}}{\tempbar{#1}{3.5mu}}{\tempbar[\scriptstyle]{#1}{3.5mu}}{\tempbar[\scriptscriptstyle]{#1}{3mu}}}} 
\renewcommand{\i}{\mathrm{i}}
\renewcommand{\d}{\mathrm{d}}
\newcommand{\tn}[1]{\textnormal{#1}}

\begin{document}
\pagestyle{fancy}

\title{Accelerating Nonequilibrium Green Functions Simulations: \\The G1-G2 Scheme and Beyond
}
\maketitle   





\author{Michael Bonitz*},
\author{Jan-Philip Joost},
\author{Christopher Makait},
\author{Erik Schroedter},
\author{Tim Kalsberger}, and
\author{Karsten Balzer$^\dagger$}

\begin{affiliations}
Institut für Theoretische Physik und Astrophysik, Christian-Albrechts Universität Kiel, Germany\\
 Kiel Nano, Surface and Interface Science KiNSIS, Kiel University, Kiel, Germany\\
 $^\dagger$Rechenzentrum, Christian-Albrechts-Universität Kiel, D-24098 Kiel, Germany\\
*Email Address: bonitz@physik.uni-kiel.de

\end{affiliations}

\keywords{Nonequilibrium Green functions, Keldysh technique, G1--G2 scheme}

\abstract{The theory of Nonequilibrium Green functions (NEGF) has seen a rapid development over the recent three decades. Applications include diverse correlated many-body systems in and out of equilibrium. Very good agreement with experiments and available exact theoretical results could be demonstrated if the proper selfenergy approximations were used.
However, full two-time NEGF simulations are computationally costly, as they suffer from a cubic scaling of the computation time with the simulation duration.
Recently we have introduced the G1-G2 scheme that exactly reformulates the generalized Kadanoff-Baym ansatz with Hartree-Fock propagators (HF-GKBA) into time-local equations, which achieves time-linear scaling and allows for a dramatic speedup and extension of the simulations  [Schluenzen et al., Phys. Rev. Lett. \textbf{124}, 076601 (2020)]. Remarkably, this scaling is achieved quickly, and also for high-level selfenergies, including the nonequilibrium $GW$ and $T$-matrix approximations [Joost et al., Phys. Rev. B \textbf{101}, 245101 (2020)]. Even the dynamically screened ladder approximation is now feasible [Joost et al., Phys. Rev. B \textbf{105}, 165155 (2022)], and also applications to
electron-boson systems were demonstrated. Here we present an overview on recent results that were achieved with the G1--G2 scheme. We discuss problems and open questions and present further ideas of how to overcome the current limitations of the scheme and present. We illustrate the G1--G2 scheme by presenting applying it to the excitation dynamics of Hubbard clusters, to optical excitation of graphene, and to charge transfer during stopping of ions by correlated materials.
}

\section{Introduction}\label{s:intro}

Correlated many-particle systems are of rapidly growing interest in many fields including condensed matter physics \cite{ctmc_rubtsov_05,jensen_ultrafast_2013}, atoms in optical lattices \cite{xia_quantum_2015,schluenzen_prb16}, nuclear physics \cite{garny_2009} or dense plasmas and warm dense matter \cite{Graziani2014FrontiersAC,dornheim_physrep_18}.
The dynamics of correlated quantum many-particle systems following an external excitation can be described by a variety of approaches, including time-dependent density functional theory (TD-DFT), density matrix renormalization group approaches or nonequilibrium Green functions (NEGF). Here we focus on NEGF simulations that, over the last three decades, have been actively used to simulate correlated many-particle systems in many fields, for recent reviews, see Refs.~\cite{schluenzen_jpcm_19,hirsbrunner_2019}. The direct numerical solution was pioneered by P. Danielewicz \cite{DANIELEWICZ_84_ap2}, S. K\"ohler \cite{koehler_prc_95}, W. Sch\"afer \cite{schaefer_wegener}, N.H. Kwong \cite{koehler-kwong-code}, and M. Bonitz and co-workers \cite{bonitz-etal.96jpcm,semkat_99_pre}. A new surge of activity set in with the application of NEGF simulations to finite non-uniform systems such as small atoms, molecules or molecular transport by R. van Leeuwen and co-workers \cite{dahlen_prl_07, dahlen_solving_2007, myo_prb_09, stan_jcp_09} that led to many new applications, e.g. to quantum dots \cite{bonitz_prb_7,balzer_prb_9}, 
or to the interaction of atoms and molecules with laser pulses \cite{balzer_pra_10,balzer_pra_10_2}, as well as to improved numerical solution schemes, e.g. \cite{schluenzen_prb17_comment}.

However, NEGF simulations for the two-time single-particle Green functions are hampered by an unfavorable cubic scaling with the simulation duration (number of time steps $N_\tn{t}$), already for the simplest electron-electron scattering approximation (selfenergies beyond Hartree-Fock). This is in contrast to many other time-dependent methods in many-body physics, including the solution of the time-dependent Schrödinger equation (TDSE) or TD-DFT which scale linearly with $N_\tn{t}$.
The NEGF scaling can be reduced to quadratic scaling by applying the generalized Kadanoff-Baym ansatz (GKBA) \cite{lipavski_prb_86}, however, this is possible only for the simple second-order Born selfenergy (SOA). For more advanced scattering models such as $T$-matrix or $GW$ approximation the scaling remains cubic \cite{schluenzen_jpcm_19}.
Still, the GKBA has been picked up by many groups and has allowed for a significant extension of the scope of problems that can be tackled with NEGF simulations. They include physical processes in spatially uniform macroscopic systems such electron-phonon and electron-electron scattering in optically excited semiconductors, e.g. \cite{haug_pss_92,haug-ell_prb_92,haug_ssc_96,gartner_prb_99,hartmann_pss_92,bonitz-etal.96jpcm,kwong-etal.98pss,banyai_prl_98,manzke_pss_95,zimmermann_jl_98,binder_pqe_97,gartner_jpcs_06,gartner_prb_06}, nuclear matter \cite{koehler_prc_95,koehler_pre_96}, and dense plasmas \cite{kremp-etal.97ap,kremp_99_pre,haberland_01_pre,semkat_00_jmp,semkat_99_pre}. More recently, the GKBA was also applied to finite non-uniform problems including atoms, molecules, quantum dots and clusters \cite{ hermanns_psc_12,balzer_2013_nonequilibrium,hermanns_jpcs13,jahnke_prb_13,latini_charge_2014,perfetto_pra_15,covito_pra_18}, to charge transport 
\cite{yao_low-temperature_2015, spicka_relation_2021, ridley_many-body_2022}, to the interaction of electrons with phonons and photons \cite{melo_unified_2016}, to  C60  \cite{bostrom_charge_2018}, to topological states in graphene  \cite{schuler_how_2020}, and to excitonic states \cite{murakami_ultrafast_2020}. We also mention recent work on the general properties of the GKBA
\cite{reeves_unimportance_2023}, 
for more examples, see Sec.~\ref{ss:gkba-nonuniform}.

A further dramatic speedup was reported recently by Schl\"unzen et al. \cite{schluenzen_prl_20} who discovered that the GKBA can be exactly reformulated in a time-local form -- the G1-G2 scheme -- i.e. coupled equations for the single-particle NEGF and the correlated part of the two-particle NEGF, that leads to time-linear scaling. With this NEGF simulations achieved the same scaling as TDSE or TD-DFT simulations. Most remarkably, this scaling is not only achieved for SOA selfenergies but for more advanced selfenergies as well \cite{joost_prb_20,joost_prb_22}, see Sec.~\ref{s:g1-g2}.

This paper is structured as follows: In Sec.~\ref{s:theoretical_framework} we introduce the theory of nonequilibrium (Keldysh) Green functions and discuss the most important electron-electron correlation selfenergies. Section~\ref{s:GKBA} introduces the HF-GKBA and discusses applications to uniform and finite systems. Then, in Sec.~\ref{s:g1-g2}, we introduce the G1-G2 scheme, discuss its extension to important selfenergy approximations, as well as stability issues. The scheme is illustrated for the dynamics of small Hubbard clusters, for ion stopping and quasi-1D plasmas and for the excitation of graphene by a short laser pulse. 
The main bottleneck of the G1-G2 scheme -- the large memory consumption -- is addressed in Sec.~\ref{s:embedding}. There we discuss a time local embedding scheme which allows to simulate the ultrafast charge transfer during the impact of ions on correlated materials. A second approach to overcome the large memory consumption is introduced in Sec.~\ref{s:quantum-fluctuations} where we introduce our NEGF-based quantum fluctuations approach. The paper concludes with a discussion and outlook, in Sec.~\ref{s:discussion}.

\section{Theory} \label{s:theoretical_framework}
We consider a general many-particle system and describe it in standard second quantization.
The basic quantities are the one-particle and two-particle nonequilibrium Green functions. We start by recalling their definitions and the equations of motion of the single-particle NEGF.

\subsection{Nonequilibrium Green Functions} \label{ss:negf}
We consider a generic Hamiltonian of an $N$-particle system given by
\begin{equation}
    \hat{H}(t)= \sum_{ij}h^{(0)}_{ij}(t)\hat{c}^\dagger_i\hat{c}_j+\frac{1}{2}\sum_{ijkl}w_{ijkl}(t)\hat{c}^\dagger_i\hat{c}^\dagger_j\hat{c}_l\hat{c}_k\,,
\label{eq:h}
\end{equation}
where, $h^{(0)}$ is the single-particle contribution which may be time dependent due to the interaction of the particles with external electromagnetic fields, e.g.~\cite{kremp_99_pre,haberland_01_pre}, charged particle impact (stopping) \cite{balzer_prb16,balzer_prl_18,schluenzen_cpp_18,borkowski_pss_22}, or the rapid variation (quench) of system parameters such as the confinement potential  \cite{schneider_fermionic_2012,schluenzen_prb16,schluenzen_prb17}.
Further, $w$ denotes the pair interaction which carries a time dependence arising from interaction quenches \cite{gericke_jpa03,moeckel_prl08,schollwoeck_prb_19}
or from the numerical preparation of a correlated initial state. In many cases an efficient procedure is to start from an uncorrelated initial state and to build up correlations dynamically via ``adiabatic switching'', e.g.~\cite{schluenzen_jpcm_19}. This approach will be used throughout this paper.

\begin{figure}[h]
\centering
\includegraphics[width=0.5\columnwidth]{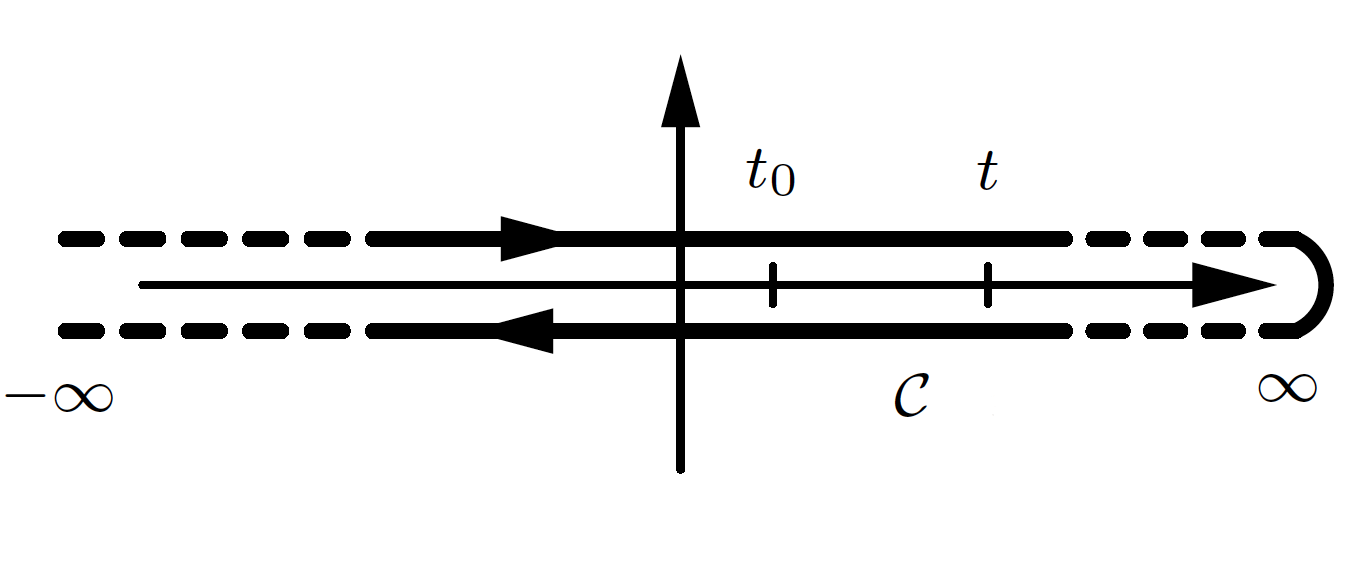}
\caption{Keldysh ``round-trip'' time contour that is used in NEGF theory to treat initial correlations via ``adiabatic switching'' of the pair interaction, starting from an uncorrelated state in the remote past, for more details, see Refs.~\cite{stefanucci_nonequilibrium_2013,balzer-book,bonitz_pss_19_keldysh,joost_prb_20}.
}
\label{fig:contour}
\end{figure}

The matrix indices and summations in the Hamiltonian \eqref{eq:h}
refer to an arbitrary complete orthonormal system of single-particle orbitals $|i\rangle$ for which we define
 creation ($\chat c^\dagger_i$) and annihilation ($\chat c_i$) operators that obey Bose or Fermi statistics. Using the standard Heisenberg procedure, these operators are made time dependent and are used to  define the one-body nonequilibrium Green function where all time arguments $z, z'$ are defined on the Keldysh contour $\mathcal{C}$~\cite{schluenzen_jpcm_19} (see Fig.~\ref{fig:contour}),
\begin{align}
    G_{ij}(z,z')=\frac{1}{\i\hbar}\left\langle \mathcal{T}_\mathcal{C} \left\{\chat{c}_i(z)\chat{c}^\dagger_j(z')\right\} \right\rangle\,,
\end{align}
where, $\mathcal{T}_\mathcal{C}$ is the time-ordering operator on the contour, and the averaging is performed with the correlated unperturbed $N$-particle density operator of the system.
The two-particle Green function, $G^{(2)}$, is defined analogously by
\begin{align}\label{eq:g2-def}
    G^{(2)}_{ijkl}(z_1,z_2,z_3,z_4)
    &=\frac{1}{\left(\i\hbar\right)^2}\left\langle \mathcal{T}_\mathcal{C} \left\{\chat{c}_i(z_1)\chat{c}_j(z_2)\chat{c}^\dagger_l(z_4)\chat{c}^\dagger_k(z_3)\right\} \right\rangle\\
    &=G^{(2),\tn{H}}_{ijkl}(z_1,z_2,z_3,z_4) \pm G^{(2),\tn{F}}_{ijkl}(z_1,z_2,z_3,z_4)
    + \mathcal{G}_{ijkl}(z_1,z_2,z_3,z_4)\,,
    \label{eq:g2-corr-def}
\end{align}
where in the second line the Hartree ($G^{(2),\mathrm{H}}$), Fock ($G^{(2),\mathrm{F}}$), and correlation part ($\mathcal{G}$) of the two-particle Green function are introduced, with
\begin{align}\label{eq:g2h-keldysh}
    G^{(2),\tn{H}}_{ijkl}(z_1,z_2,z_3,z_4) &= G_{ik}(z_1,z_3)G_{jl}(z_2,z_4)\,,\\\label{eq:g2f-keldysh}
    G^{(2),\tn{F}}_{ijkl}(z_1,z_2,z_3,z_4) &= G_{il}(z_1,z_4)G_{jk}(z_2,z_3)\,.
\end{align}
As an alternative to Eq.~(\ref{eq:g2-corr-def}), the two-particle Green function can be subdivided into a Hartree contribution and the rest -- the exchange-correlation function, $L$,
\begin{align}
    L_{ijkl}(z_1,z_2,z_3,z_4)\coloneqq&\, G^{(2)}_{ijkl}(z_1,z_2,z_3,z_4)
    -G_{ik}(z_1,z_3)G_{jl}(z_2,z_4)\,,\label{eq:definition_XC_function}
\end{align}
which comprises correlations and exchange (Fock) contributions, as is common e.g. in density functional theory. The function $L$ is also the central ingredient in the theory of fluctuations and in the quantum fluctuations approach that will be discussed in Sec.~\ref{s:quantum-fluctuations}.

\subsection{Keldysh--Kadanoff-Baym Equations} \label{ss:KBE}

The equations of motion for the NEGF are the Keldysh--Kadanoff--Baym  equations (KBE)\footnote{Throughout this work, ``$\pm$'' refers to bosons/fermions.} \cite{kadanoff-baym-book,keldysh64,bonitz_pss_19_keldysh}
\begin{align}
    \sum_k 
    \left[ \i\hbar \frac{\d}{\d z} \delta_{ik} - h^{(0)}_{ik}(z) \right] G_{kj}(z,z') - \delta_{ij} \delta_\mathcal{C}(z,z') 
    \label{eq:KBE1}
    &=\pm \i\hbar \sum_{klp} \int_\mathcal{C} \d \cbar{z}\, w_{iklp}(z,\cbar{z})G^{(2)}_{lpjk}(z,\cbar{z},z',\cbar{z}^+)\\
    & = \, \sum_k \int_\mathcal{C} \d \cbar{z}\, \Sigma_{ik}(z,\cbar{z})  G_{kj}(\cbar{z},z')\, , 
    \label{eq:kbe-sigma-form1} \\
   \sum_k 
   G_{ik}(z,z') \left[- \i\hbar \frac{\overset{\leftarrow}{\d}}{\d z'} \delta_{kj} - h^{(0)}_{kj}(z') \right] - \delta_{ij} \delta_\mathcal{C}(z,z') 
   \label{eq:KBE2} 
   &=\pm \i\hbar \sum_{klp} \int_\mathcal{C} \d \cbar{z}\, G^{(2)}_{iklp}(z,\cbar{z}^-,z',\cbar{z}) w_{lpjk}(\cbar{z},z') \\
    &= \sum_k \int_\mathcal{C} \d \cbar{z}\, G_{ik}(z,\cbar{z}) \Sigma_{kj}(\cbar{z},z') 
   \,,
   \label{eq:kbe-sigma-form2}
\end{align} 
where ${z^\pm \coloneqq z \pm \epsilon}$ with $\epsilon \to +0$, and
we introduced a two-time version of the interaction potential using the delta function on the Keldysh contour, $w_{ijkl}(z,z') = \delta_\mathcal{C}(z,z') w_{ijkl}(z)$, see, e.g. Refs.~\cite{schluenzen_jpcm_19,schluenzen_cpp16,stefanucci_nonequilibrium_2013}.

Note that we have presented two forms of the r.h.s. of the KBE, i.e. of the collision integral. The first lines, ~\eqrefs{eq:KBE1}{eq:KBE2} contain the two-particle NEGF, Eq.~\eqref{eq:g2-def}. The second form of the r.h.s., \eqrefs{eq:kbe-sigma-form1}{eq:kbe-sigma-form2}, contains the selfenergy $\Sigma$ which is introduced in most NEGF formulations with the goal to eliminate the two-particle Green function. Below, in Sec.~\ref{s:sigmas}, we will consider several approximations for $\Sigma$. Here we already note that the  dependence of the single-particle Green function on two time arguments, combined with the time integral on the r.h.s. of Eqs.~(\refeq{eq:kbe-sigma-form1}) and (\refeq{eq:kbe-sigma-form2}), gives rise to a cubic scaling, $N_t^3$, of the computing time with the number of time steps $N_t$. As we will show below, in Sec.~\ref{s:g1-g2}, with the introduction of the G1-G2 scheme this scaling can be reduced to $N^1_t$, regardless of the choice of the selfenergy \cite{schluenzen_prl_20,joost_prb_20}. 

\subsection{Selfenergy approximations}\label{s:sigmas}
\begin{figure}[t]
\centering
\includegraphics[width=0.6\columnwidth]{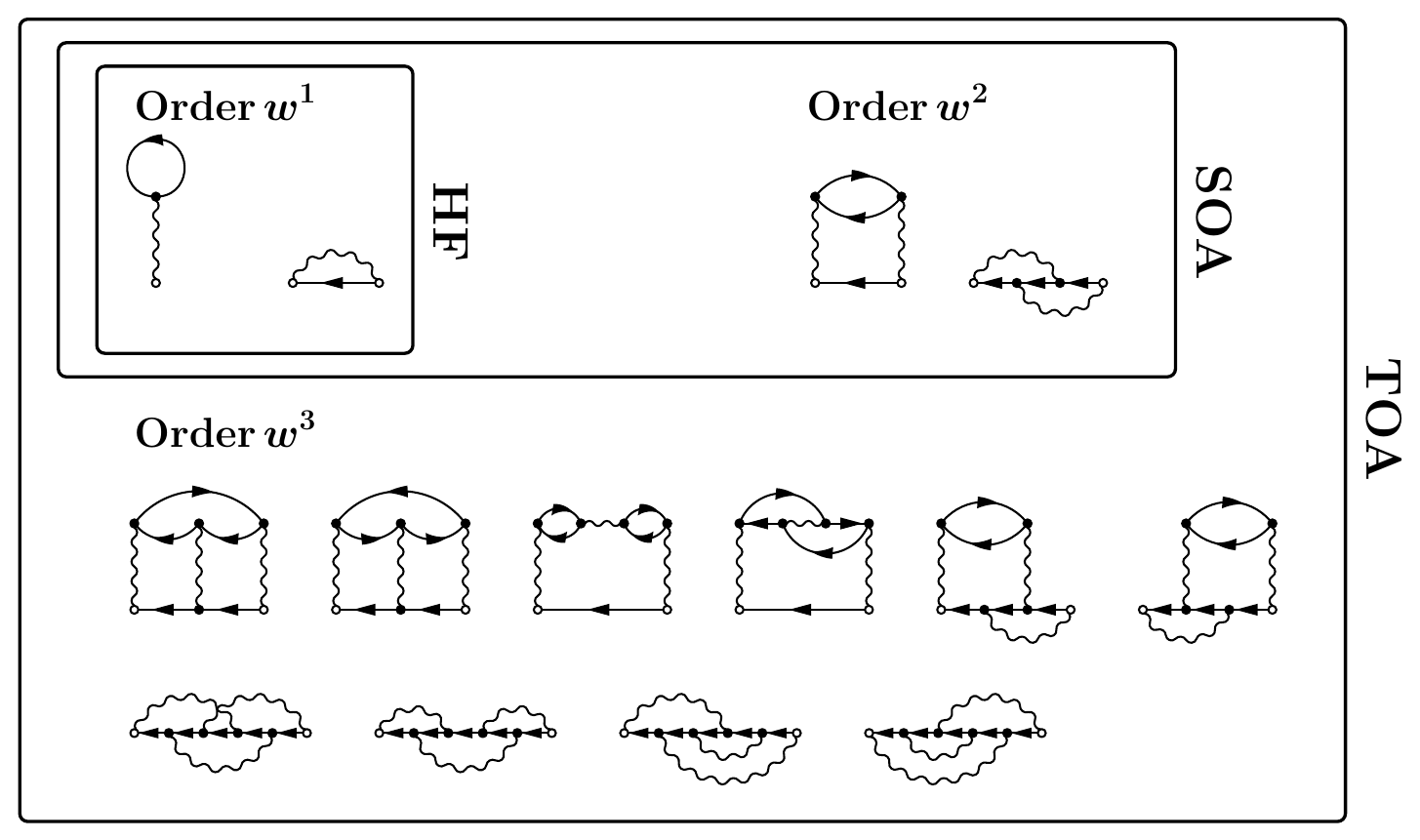}
\caption{Selfenergy diagrams for the perturbative (with respect to powers of the interaction) approach up to order three. Hartree--Fock (HF) contains contributions of first order in $w$, the second-order selfenergy (SOA) contains second-order diagrams (together with the first order), whereas the third-order approximation (TOA) contains all diagrams up to  order $w^3$. Note that the second diagrams of order $w^1$ and $w^2$ describe exchange processes, respectively. For order $w^3$, exchange processes are included in the latter six diagrams. Figure from Ref.~\cite{joost_prb_22}. $\copyright$ 2022 by The American Physical Society.}
\label{fig:diagrams_order}
\end{figure}

A graphical over\-view of the most important selfenergy approximations in terms of Feynman diagrams is presented in \reffigs{fig:diagrams_order}{fig:diagrams_resummation}. A main selection criterion is that each of the approximations is conserving, i.e. conserves particle number, momentum and total energy.
To shorten the presentation we only provide the results for the $>$ and $<$ component, $\Sigma^\gtrless(t,t')$ which follows from $\Sigma(z,z')$ by taking the time arguments on different branches of the contour in Fig.~\ref{fig:contour}, for details see Refs.~\cite{stefanucci_nonequilibrium_2013,schluenzen_jpcm_19}.
\subsubsection{Real time single-particle Green functions}\label{ss:g1}
Equations (\refeq{eq:KBE1}) and (\refeq{eq:KBE2}) for the one-particle NEGF are formulated on the Keldysh contour, cf. Fig.~\ref{fig:contour}. They are equivalent to equations for Keldysh Green function matrices of real-time arguments, where the matrix components differ by the location of the time arguments on the contour \cite{keldysh64,bonitz_pss_19_keldysh}. This gives rise to the correlation functions, $G^\gtrless(t,t')$, and the retarded and advanced functions, $G^{\rm R/A}(t,t')$,
\begin{align}
    G^<_{ij}(t,t') &=\pm \frac{1}{\i\hbar}\left\langle \chat c_j^\dagger(t') \chat c_i(t)\right\rangle\,,
    \\
    G^>_{ij}(t,t') &= \frac{1}{\i\hbar}\left\langle \chat c_i(t) \chat c_j^\dagger(t')\right\rangle  \,,
    \\
    G^{\rm R/A}_{ij}(t,t') &=
    \pm \Theta[\pm(t-t')]\left\{G^>_{ij}(t,t')-G^<_{ij}(t,t')\right\}\,,
    \label{eq:gra-def}
\end{align}
for details see the text books \cite{stefanucci_nonequilibrium_2013,balzer-book}.

Let us note the following important property of the correlation functions on the time diagonal:
\begin{align}
G_{ij}^<(t,t) \equiv  G_{ij}^<(t) &= G_{ij}^>(t) - \frac{1}{\i\hbar}\delta_{ij} = \pm \frac{1}{\i\hbar} n_{ij}(t)\,,
\label{eq:g-glsymm}
\end{align}
where $n_{ij}$ is the single-particle density matrix. Thus, $G^\gtrless$ have a clear physical meaning and are directly related to observables. In the same way the two-particle Green function defined in Eq.~(\ref{eq:g2-def}) can be expressed for real times. On the time diagonal it is given by
\begin{align}
   G_{ijkl}^{(2)}(t) &\coloneqq G_{ijkl}^{(2)} (t,t,t,t) = \frac{1}{\left(\mathrm{i}\hbar\right)^2} \left\langle \hat{c}^\dagger_k(t) \hat{c}^\dagger_l(t) \hat{c}_j(t) \hat{c}_i(t)\right\rangle \,, \label{G2less}
\end{align}
and can be decomposed into Hartree, Fock, and correlation part
\begin{align}\label{eq:G2_rtc}
    G_{ijkl}^{(2)}(t) &= G^<_{ij}(t,t)G^<_{kl}(t,t) \pm G^<_{il}(t,t)G^<_{jk}(t,t) + \mathcal{G}_{ijkl}(t)\,.
\end{align}


In the following, we provide the most relevant selfenergy approximations in terms of these functions.

\subsubsection{Hartree--Fock selfenergy}\label{ss:hf}
The selfenergy terms that are of first-order in the pair potential, $w$, describe particle interaction on the mean-field level. They are combined in the so-called Hartree--Fock (HF) selfenergy, which for a time-diagonal interaction tensor only has a single-time-dependent (delta) component for the real time $t$,
\begin{align}
     \Sigma^{\tn{HF},\delta}_{ij}(t) =& \pm \i \hbar \sum_{kl} w^\pm_{ikjl}(t) G^<_{lk}(t,t) \, . \label{eq:sigma_hf}
    \end{align}
For this reason, the first-order terms can be directly included into an effective single-particle Hamiltonian of the form
\begin{align}
 h^\mathrm{HF}_{ij}(t) = h^{(0)}_{ij}(t) \pm \mathrm{i}\hbar \sum_{kl} w^\pm_{ikjl}(t) G^<_{lk}(t,t)\,, \label{eq:h_HF}
\end{align}
where we introduced the (anti-)symmetrized matrix element of the pair potential,
\begin{align}
 w^\pm_{ijkl}(t) &\coloneqq w_{ijkl}(t) \pm w_{ijlk}(t)
\label{eq:wpm-def}
=w_{ijkl}(t) \pm w_{jikl}(t)\,.
\end{align}
This potential has the symmetries
\begin{align}
    w^\pm_{ijkl}(t) &= \pm w^\pm_{ijlk}(t) = \pm w^\pm_{jikl}(t)\,, \label{eq:wpm-symm}
\end{align}
and, for fermions, moreover, $w^\pm_{ijkk}(t)=  w^\pm_{iikl}(t)= 0$. 

\subsubsection{Second-order Born selfenergy (SOA)}\label{ss:soa}
The simplest selfenergy that includes correlations and, thus, allows to describe dissipation and relaxation effects (selfenergy beyond Hartree--Fock) is given by the second-order Born approximation~\cite{schluenzen_cpp16}, 
\begin{align}
 \Sigma^\gtrless_{ij}\left(t,t'\right) 
 &= \pm\left(\i\hbar\right)^2 \sum_{klpqrs} \, w_{iklp}\left(t\right) w^\pm_{qrjs}\left(t'\right) 
 \label{eq:sigma-soa} 
 G^\gtrless_{lq}\left(t,t'\right) G^\gtrless_{pr}\left(t,t'\right) G^\lessgtr_{sk}\left(t',t\right)\, .
\end{align}
We will use the notation ``SOA'' for the selfenergy that includes all terms up to second order (including HF).
SOA provides the starting point for all following approximations.
Note that the potential $w^\pm$, Eq.~\eqref{eq:wpm-def}, gives rise to two contributions---the direct SOA and the associated exchange diagram which are shown in Fig.~\ref{fig:diagrams_order}.

\subsubsection{Third-order approximation (TOA)}\label{ss:toa}
\begin{figure}[t]
\centering
\includegraphics[width=0.5\columnwidth]{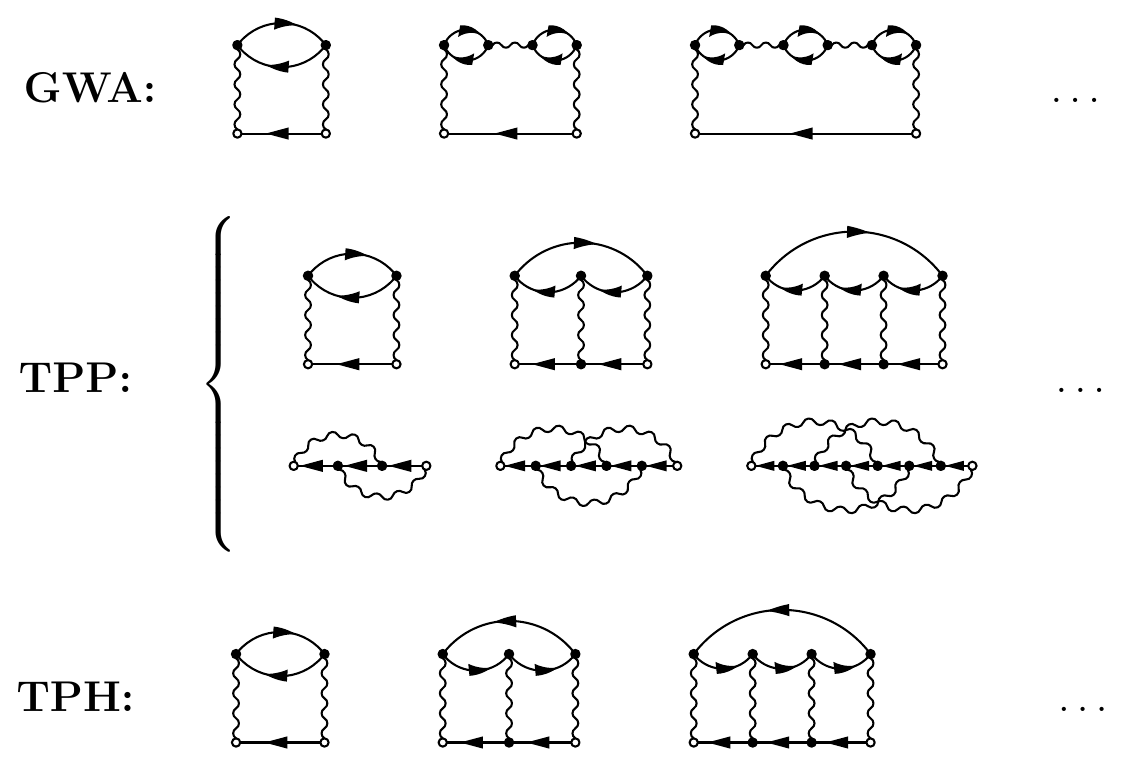}
\caption{Selfenergy diagrams for the three resummation approaches, starting from the second-order contributions. Dots indicate continuation of the sums to infinite order. Note that for TPP it is possible to also include the corresponding exchange diagrams (second line). Figure from Ref.~\cite{joost_prb_22}. $\copyright$ 2022 by The American Physical Society.}
\label{fig:diagrams_resummation}
\end{figure}

The third-order approximation for the selfenergy allows to significantly improve the accuracy of simulations, compared to SOA. It contains all diagrams that include up to three interaction lines, cf. Fig.~\ref{fig:diagrams_order}. There exist 10 skeletonic diagrams that are of order $w^3$ and which are also part of the $GW$, TPP and TPH approximations, cf. Fig.~\ref{fig:diagrams_resummation}. Thus, TOA contains the starting terms of the ladder and bubble sums.
TOA was first introduced and tested in Ref.~\cite{schluenzen_prb17} and was found to be very accurate for weak and moderate coupling, independently of the filling (density) \cite{schluenzen_jpcm_19}.

\subsubsection{Infinite series summations. Dynamical screening and strong coupling}\label{ss:negf-gf-gh}
Aside from approximations for the selfenergy that arise from perturbation theory with respect to the pair potential, there exists another class of approximations that result from summation of an infinite series of diagrams. The first example is the polarization approximation (summation of bubble diagrams, $GW$ approximation) that allows to include dynamical screening effects which is important, in particular, for charged particles in plasmas, condensed matter or in macromolecules. $GW$ is a weak coupling approximation, but includes a selfconsistently screened pair potential $W$. To account for strong coupling effects, the second example is the particle--particle $T$ matrix that results from summing up the entire Born series. Finally, we will consider the second flavor of the $T$ matrix---the particle--hole $T$-matrix approximation. These approximation use a static potential as an input. Since these are standard approximations, a derivation is not necessary, e.g. \cite{schluenzen_cpp16,schluenzen_jpcm_19}. Instead, below we list the compact final result, together with the associated diagrams in Fig.~\ref{fig:diagrams_resummation}.
To shorten the notation, we introduce the following products of single-particle Green functions
\begin{align}
    {G}^{\tn{H},\gtrless}_{ijkl}(t,t') &\coloneqq G^\gtrless_{ik}(t,t') G^\gtrless_{jl}(t,t')\,,
    \label{eq:g2h-def}\\
    {G}^{\tn{F},\gtrless}_{ijkl}(t,t') &\coloneqq G^\gtrless_{il}(t,t') G^\lessgtr_{jk}(t',t)\,,
    \label{eq:g2f-def}
\end{align}
where ${G}^{\tn{H},\gtrless}_{ijkl}$ is just the Hartree part of the two-particle Green function, whereas ${G}^{\tn{F},\gtrless}_{ijkl}$ is the Fock part. These are special cases of the Keldysh contour versions of the Hartree and Fock Green functions, cf. Eqs.~(\ref{eq:g2h-keldysh}) and (\ref{eq:g2f-keldysh}).

\subsubsection{Dynamical screening. $GW$ selfenergy}\label{ss:negf-gw}
The selfenergy in $GW$ approximation (GWA) is defined in terms of the dynamically screened potential $W$ 
\begin{align}
 \Sigma_{ij}^{\rm{GWA},\gtrless}(t,t') = \i \hbar \sum_{kl} W^\gtrless_{ilkj}(t,t') G^\gtrless_{kl}(t,t')\, ,
\label{eq:sigma-gw}
\end{align}
that obeys the following integral equation (Dyson equation)
\begin{align}
\label{eq:W-def}
    & W^\gtrless_{ijkl}(t,t') = 
    \pm \i \hbar \sum_{pqrs} w_{ipkq}(t) w_{rjsl}(t') G^{\tn{F},\gtrless}_{qspr}(t, t') \pm \i\hbar \sum_{pqrs} w_{ipkq}(t)\times\\ &\qquad\Bigg\{ \int_{t_0}^t \d\cbar{t} \, \Big[ G^{\tn{F},>}_{qspr}(t, \cbar t) - G^{\tn{F},<}_{qspr}(t, \cbar t) \Big] W^\gtrless_{rjsl}(\cbar{t},t') \nonumber 
    +\int_{t_0}^{t'} \d \cbar{t} \, G^{\tn{F},\gtrless}_{qspr}(t, \cbar t) \left[W^<_{rjsl}(\cbar{t},t') - W^>_{rjsl}(\cbar{t},t') \right]\Bigg\}\,. \nonumber
\end{align}
Here, the first term on the right coincides with the second order direct Born diagram (SOA) whereas the integral term gives rise to an infinite sum of additional diagrams that follow iteratively, starting by inserting the second-order terms for $W$, under the integral. The first diagrams are sketched in Fig.~\ref{fig:diagrams_resummation}. Note that here we did not use the (anti-)symmetrized potential $w^\pm$. The $GW$ approximation is a key to describing screening effects in systems with long-range Coulomb interaction such as metals, semiconductors or plasmas. In fact, the integral term in Eq.~(\ref{eq:W-def}) describes the induced potential created by a charged particle in the medium, i.e. polarization effects. Therefore, the $GW$ selfenergy is directly related to other many-body approaches to screening and plasmons, including Hedin's theory \cite{hedin_pr_65} or classical kinetic equations of the Balescu-Lenard type \cite{balescu_60,lenard_60}. An alternative approach to screening and plasmons is linear response theory of density fluctuations, e.g. \cite{dornheim_prl_18,hamann_cpp_20} which will be discussed in Sec.~\ref{s:quantum-fluctuations}.
The NEGF approach allows for a nonequilibrium generalization of screening effects and incorporates such effects as the build up of the screening cloud in an optically excited semiconductor, e.g. \cite{banyai_prl_98}, and plasma instabilities, e.g. \cite{bonitz-etal.93prl,bonitz-etal.94pre,bonitz94pp,bonitz-etal.94prb}. In the NEGF approach, the 
plasmon spectrum is contained in the screened potential, $W^<(T,\tau)$, where we introduced center of mass ($T$) and relative ($\tau$) times. Then, the nonequilibrium plasmon spectrum at time $T$ follows by Fourier transforming $\tau \to \omega$. Also, the inverse dielectric function can be directly computed from the screened potential and the bare potential in Fourier space. For example, for a spatially uniform system, $\epsilon^{-1}(\omega,q)=W(\omega,q)/w(q)$.

Note that the $GW$ approximation is a weak coupling approximation which only approximately accounts for correlation effects on the plasmon spectrum \cite{hamann_cpp_20}. To go beyond the weak coupling limit requires to include ladder-type diagrams that are part of the $T$-matrix selfenergy. Combination of dynamical screening and strong coupling is achieved with the DSL approximation, cf. Secs.~\ref{ss:negf-dsl} and \ref{ss:dsl}.

Finally, the dependence of $W$ on two times and the time integral on the r.h.s. of Eq.~(\ref{eq:W-def}) imply that the computational effort for evaluating the $GW$ selfenergy scales cubically with the simulation duration. 

\subsubsection{Strong coupling. Particle--particle $T$-matrix selfenergy (TPP)}\label{ss:negf-tpp}
The definition of the TPP selfenergy has a similar structure as $GW$, but the screened potential is replaced by the particle--particle $T$ matrix,
\begin{align}
 \Sigma_{ij}^{{\rm TPP},\gtrless}(t,t') = \i \hbar \sum_{kl} T^{\tn{pp},\gtrless}_{ikjl}(t,t') G^\lessgtr_{lk}(t',t)\, ,
 \label{eq:sigma-tpp}
\end{align}
which obeys a slightly different integral equation \cite{kraeft-green-book,schluenzen_cpp16},
\begin{align}\label{eq:t-matrix-eq}
  & T^{\tn{pp},\gtrless}_{ijkl}(t,t') = 
  \i \hbar \sum_{pqrs} w_{ijpq}(t) G^{\tn{H},\gtrless}_{pqrs}(t, t') w^\pm_{rskl}(t') + \i\hbar \sum_{pqrs} w_{ijpq}(t) \times 
  \\
  &\qquad
  \Bigg\{ \int_{t_0}^t \d \cbar t\, \Big[ G^{\tn{H},>}_{pqrs}(t, \cbar t) - G^{\tn{H},<}_{pqrs}(t, \cbar t) \Big] T^{\tn{pp},\gtrless}_{rskl}(\cbar t, t') 
  +\int_{t_0}^{t'} \d \cbar t\, G^{\tn{H},\gtrless}_{pqrs}(t, \cbar t) \Big[ T^{\tn{pp},<}_{rskl}(\cbar t, t') - T^{\tn{pp},>}_{rskl}(\cbar t, t')\Big] \Bigg\}\, .\nonumber
\end{align}
The main difference to the Dyson equation for the screened potential $W$ is the replacement of the Fock Green function (\ref{eq:g2f-def}) by the Hartree Green function (\ref{eq:g2h-def}), ${G}^{\tn{F},\gtrless}\to {G}^{\tn{H},\gtrless}$. 
The corresponding equation for the $T$-matrix is closely related to the Lippmann-Schwinger equation of scattering theory, e.g. \cite{bonitz_qkt}.
Furthermore, the first term here contains the potential $w^\pm$, thus it includes the exchange diagram. As a consequence, each diagram of the iteration series is complemented by an exchange diagram, cf. Fig.~\ref{fig:diagrams_resummation}, second and third line, respectively. Note that the first term in Eq.~(\ref{eq:t-matrix-eq}) alone would recover the SOA diagram (which is second order in the interaction) which is valid only for weak interparticle coupling, corresponding to small-angle scattering. On the other hand, inserting this result for $T^{\tn{pp}}$, in the other terms, iteratively gives rise to higher order contributions which describe large-angle scattering and strong coupling phenomena. Moreover, if the $T$-matrix selfenergy is applied to two-component systems of oppositely charged particles, the solution of the KBE does include bound states such as atoms and excitons which can be identified in the spectral function, e.g.~\cite{schmielau_pngf2}. Moreover, the $T$-matrix approximation takes into account the influence of the surrounding medium, via Coulomb correlations and Pauli blocking, on the bound states which may lead to lowering of the ionization energy and Mott effect (pressure ionization). Finally, in a charged particle system, the potential appearing in Eq.~(\ref{eq:t-matrix-eq}) has to be statically screened to remove Coulomb divergencies by using a Yukawa or Debye-type short-range potential. The incorporation of dynamical screening effects that were discussed in Sec.~\ref{ss:negf-gw} is achieved within the DSL approximation, cf. Secs.~\ref{ss:negf-dsl} and \ref{ss:dsl}. 

Finally, the dependence of the T-matrix  on two times and the time integral on the r.h.s. of Eq.~(\ref{eq:t-matrix-eq}) imply that the computational effort for evaluating the $T$-matrix selfenergy scales cubically with the simulation duration, as in the case of the GW-approximation.

\subsubsection{Particle--hole $T$-matrix selfenergy}\label{ss:negf-teh}
The par\-ticle--hole $T$ matrix is defined analogously to the particle--particle $T$ matrix,
\begin{align}
 \Sigma_{ij}^{\rm{TPH},\gtrless}(t,t') = \i \hbar \sum_{kl} T^{\tn{ph},\gtrless}_{ikjl}(t,t') G^\gtrless_{lk}(t,t')\, ,
 \label{eq:sigma-tph}
\end{align}
with the main difference given by the appearance of $\mathcal{G}^{\tn{F},\gtrless}$ in the Lippmann--Schwinger equation, 
 \begin{align}
     &T_{ijkl}^{\tn{ph},\gtrless}(t,t') = 
     \pm \i \hbar \sum_{pqrs} w_{iqpl}(t) G^{\tn{F},\gtrless}_{psqr}(t,t') w_{rjks}(t') + \i\hbar \sum_{pqrs} w_{iqpl}(t)\times\\ &\qquad\bigg\{ \int_{t_0}^t \d \cbar t\, \Big[ G^{\tn{F},>}_{psqr} (t,\cbar t) - G^{\tn{F},<}_{psqr} (t,\cbar t) \Big] T^{\tn{ph},\gtrless}_{rjks}(\cbar t, t') 
     +\int_{t_0}^{t'} \d \cbar t \, G^{\tn{F},\gtrless}_{psqr} (t,\cbar t) \Big[ T^{\tn{ph},<}_{rjks}(\cbar t, t') - T^{\tn{ph},>}_{rjks}(\cbar t, t')\Big]\bigg\} \, .
     \nonumber
 \end{align}
 Note that, in contrast to the particle--particle $T$ matrix, in this definition no exchange contributions are included. 
 While it is possible to sum up an additional diagram series by using the (anti-)symmetrized potential, $w^\pm$, instead of $w$, this would lead to a violation of physical conservation laws~\cite{schluenzen_phd_21}.

\subsubsection{Combining strong coupling and dynamical screening}\label{ss:negf-dsl}
An important task of many-body theory, in particular for systems with long-range Coulomb interaction,  is to combine strong coupling and dynamical screening effects. As we discussed before in Secs.~\ref{ss:negf-gw} and \ref{ss:negf-tpp} this would allow to incorporate strong coupling effects into the theory of dynamical screening. On the other hand, it would allow to account for dynamical screening effects in the $T$-matrix selfenergy. This is, of course, a formidable problem for systems in nonequilibrium.
There exist several approximate solutions. One is the third-order approximation (TOA) that was discussed above in Sec.~\ref{ss:toa} which contains diagrams from both approximations, up to the third order. Another approximate solution is provided by the FLEX (fluctuating exchange) scheme, e.g. \cite{schluenzen_jpcm_19,stahl_prb_21}. A combination of strong coupling and dynamical screening (the dynamically screened ladder approximation, DSL) for the case of excitons in thermal equilibrium has been formulated in terms of a Bethe--Salpeter equation  \cite{zimmermann_pss_78,haug_78}. 
In Ref.~\cite{joost_prb_22} it was demonstrated that a G1--G2 scheme on the level of the  nonequilibrium DSL approximation is straightforwardly derived which also allows to incorporate exchange diagrams into the particle--hole $T$-matrix and $GW$ approximation in a conserving manner, see Sec.~\ref{ss:dsl}. Within the NEGF approach a corresponding DSL selfenergy  was later derived from the Bethe-Salpeter equation in Ref.~\cite{joost_phd_2022}.

\section{Generalized Kadanoff--Baym Ansatz (GKBA)} \label{s:GKBA}
\subsection{Idea}
The GKBA was introduced by Lipavsky, Spicka and Velicky in 1986 \cite{lipavski_prb_86} and has been widely used since then.
Within the GKBA the time-off-diagonal lesser and greater single-particle NEGFs are reconstructed from their time-diagonal values (i.e. the single-particle density matrix) using the retarded and advanced Green functions, i.e., 
\begin{equation}
    G^\gtrless_{ij}(t,t') =\mathrm{i}\hbar \Big\{ G^\mathrm{R}_{ik}(t,t')G^\gtrless_{kj}(t')-G^\gtrless_{ik}(t)G^\mathrm{A}_{kj}(t,t')\Big\}\,,
\label{eq:gkba}
\end{equation}
where additional terms containing double time integrals over the selfenergy are being omitted, for the complete reconstruction equations, see Refs.~\cite{lipavski_prb_86,schluenzen_jpcm_19}. Note that this ansatz is independent of the choice of the selfenergy approximation.
 The ansatz (\ref{eq:gkba}) allows one to eliminate the time-off-diagonal values of $G^\gtrless$ and express them in terms of their values on the time diagonal.
However, this reconstruction requires knowledge of the retarded and the advanced NEGFs that still depend on two times. One, therefore, needs approximations that allow for the calculation of these components without having to solve the general two-time equations. This is usually done within the \textit{free GKBA} or the \textit{Hartree--Fock GKBA} (HF-GKBA) by treating the real-time components of the single-particle NEGF on the time off-diagonal, respectively, on the non-interacting or Hartree--Fock level. In the following, we will concentrate mostly on the HF-GKBA where the greater and lesser Green functions obey simple Hartree--Fock-type equations:
\begin{align}
    \mathrm{i}\hbar\partial_t G^\gtrless_{ij}(t,t')&= h^\mathrm{HF}_{ik}(t)G^\gtrless_{kj}(t,t')\,, \label{eq:HF_KBE_1}\\
    -\mathrm{i}\hbar\partial_{t'}G^\gtrless_{ij}(t,t')&=G^\gtrless_{ik}(t,t')h^\mathrm{HF}_{kj}(t')\,.\label{eq:HF_KBE_2}
\end{align}
The accuracy of the GKBA, Eq.~(\ref{eq:gkba}), and of the choice of Hartree-Fock propagators can be tested against full two-time simulations or other benchmark results. An example of an optically excited semiconductor is presented in Fig.~\ref{fig:gkba-test-eh} and discussed in Sec.~\ref{sss:eh-plasma}. Detailed tests for finite Hubbard clusters were performed in Ref.~\cite{schluenzen_prb17}.

\subsection{Application of the GKBA to spatially uniform systems}\label{ss:gkba-uniform}
\subsubsection{Dense plasmas and electron-hole plasmas}\label{sss:eh-plasma}
\begin{figure}
    \centering
    \includegraphics[width=0.7\textwidth]{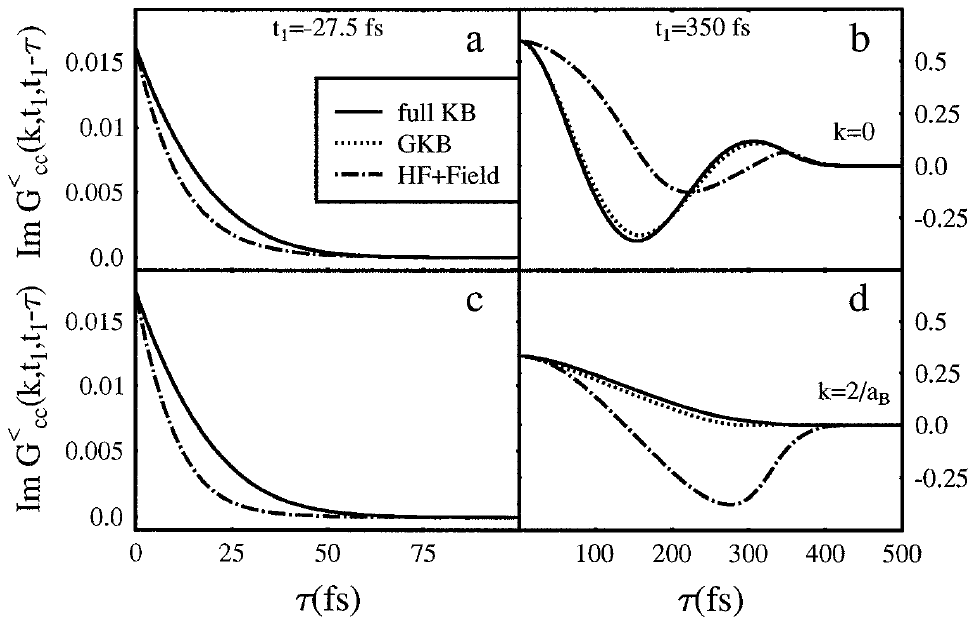}
    \caption{Test of the free GKBA for a laser excited electron-hole plasma in a GaAs quantum well with carrier density $n=2\cdot 10^{11}$cm$^{-2}$. The pulse is peaked at $t=0$ with a duration of $50$fs and a photon energy of $25$meV above the gap ($E_g=1.5$eV).  We plot the imaginary part of $G^<(k,t_1,t_2=t_1-\tau)$, starting at the time diagonal ($\tau=0$) and going back in time $t_2$. The curves correspond to the full two-time KBE result, the GKBA with exact $G^R$ (``GKB'') and HF-GKBA. a), c) correspond to $t_1 =  -27.5 $fs and b), d) to $t_1 = 350$ fs and to the momenta $k=0$ (parts a, b) and $k =2/a_B$ (parts c, d), respectively. From Ref.~\cite{kwong-etal.98pss}.}
    \label{fig:gkba-test-eh}
\end{figure}

After the GKBA was introduced in Ref.~\cite{lipavski_prb_86}, it was first applied to spatially uniform electron-hole plasmas in optically excited semiconductors. First numerical tests of the free GKBA (GKBA with free propagators) against two-time NEGF simulations where performed in Ref.
\cite{bonitz-etal.96jpcm}. 
Applications to electron-hole plasmas including excitation by a laser pulse were presented by Kwong \textit{et al.} in Ref.~\cite{kwong-etal.98pss}, whereas the time-dependent momentum orientation relaxation of the electron and hole distributions due to Coulomb scattering was studied by Binder \textit{et al.} \cite{binder-etal.97prb}.
A typical example of a laser excited electron-hole plasma in a single quantum well \cite{kwong-etal.98pss} is shown in Fig.~\ref{fig:gkba-test-eh}. The system is excited with a $50~$fs pulse, and the imaginary part of $G^<$ in the conduction band is shown shortly before the pulse maximum, during carrier generation (left), and long after the pulse (right). The top panel depicts results for the band minimum ($k=0$) and the bottom panel for momenta high in the band, $k=2/a_B$, where $a_B$ is the exciton Bohr radius. This figure allows one to test the quality of the GKBA against full two-time interband KBE simulations (full lines). One version of the GKBA corresponds to the use of Hartree-Fock propagators that include the laser field (dash-dotted lines) which exhibits systematic deviations from the KBE result that are increasing during the excitation. The second set of curves shown with the dotted line corresponds to the GKBA with the most accurate propagators: here we used (the interband generalization of) the  ansatz (\ref{eq:gkba}) where the propagators are taken from the full two-time simulation. Thus, this result directly allows one to test relevance of the integral terms that are omitted in the GKBA, Eq.~(\ref{eq:gkba}). As one can see from the figure, at early times, the curves ``full KB'' and ``GKB'' are indistinguishable and, even for later times, the differences are very small, confirming the excellent accuracy of the ansatz, for the present system parameters. More details are given in Ref.~\cite{kwong-etal.98pss}.

\subsubsection{Quantum plasmas in strong electromagnetic fields. Gauge-invariant GKBA}
Another early application of the \textit{free GKBA} was to dense plasmas quantum plasmas subject to a strong laser field by Kremp \textit{et al.} \cite{kremp_99_pre}. In contrast to the above mentioned semiconductor applications where the laser field drives optical interband transitions, here the field accelerates electrons and ions (intraband processes). The main interest in the late 1990s was the modification of Coulomb collisions by the field and the collisional absorption of field energy and particle acceleration and heating. To avoid ambiguities in treating field effects arising from the gauge freedom, the authors of Ref.~\cite{kremp_99_pre} developed a gauge-invariant modification of the GKBA, see also Refs.~\cite{haberland_01_pre,bonitz_99_cpp}.
For a many-particle system in a spatially uniform field described by the vector potential $\textbf{A}(t)$ 
the retarded and advanced free propagators become, using momentum representation \cite{kremp_99_pre},
\begin{align}
    G_a^{R/A}(\textbf{k},\tau,T) &= \mp \frac{i}{\hbar}\Theta(\pm \tau)\exp\left\{ -\frac{i}{\hbar}\left( \frac{k^2}{2m_a}\tau + S_a(\textbf{A};\tau,T)\right) \right\}\,,
    \label{eq:gra-gauge-inv}
    \\
    S_a(\textbf{A};\tau,T) &= \frac{e_a^2}{2m_a c^2}\left[
    \int_{T-\frac{\tau}{2}}^{T+\frac{\tau}{2}} A^2(t')dt' + \frac{1}{\tau}\left(\int_{T-\frac{\tau}{2}}^{T+\frac{\tau}{2}} \textbf{A}(t') dt'\right)^2\right]\,,
\end{align}
where ``a'' denotes the particle species (electrons or ions), $k$ the momentum, $T=(t+t')/2$ the center of mass time and $\tau=t-t'$ the relative time. The first term in the exponent of Eq.~(\ref{eq:gra-gauge-inv}) is just the ideal kinetic energy whereas $S_a$ denotes the additional energy gained by the particles in the electromagnetic field which may contain energy shifts and sidebands in the case of a periodic field. For example, for a monochromatic field $\textbf{E}(t)=\textbf{E}_0\cos\Omega t$, corresponding to $\textbf{A}(t)=-\frac{c\textbf{E}_0}{\Omega}\sin\Omega t$, the field contribution to the propagators becomes \cite{kremp_99_pre}
\begin{align}
    S_a(\textbf{A};\tau,T) &= U_a^{\tn{pond}}\tau\left\{1- \frac{\sin{\Omega\tau} \cos{2\Omega T}}{\Omega\tau} + \frac{8 \sin^2{\Omega T}\sin^2{\Omega\tau/2}}{(\Omega\tau)^2}\right\}\,,\qquad U_a^{\tn{pond}} = \frac{e_a^2 E_0^2}{4m_a\Omega^2}\,,
\end{align}
where $U_a^{\tn{pond}}$ is the well-known ponderomotive energy, i.e. the mean kinetic energy (per period) a charged particle gains in a monochromatic field. In this case the results are expected to agree with the results of Floquet theory which has been broadly applied to the interaction of atoms with light, e.g. \cite{Burke_1991} and more recently also to materials~\cite{Giovannini_2020} and band engineering using optimized light fields~\cite{Castro_2023}.

The strength of the NEGF approach is that it allows to straightforwardly extend these investigations to correlation effects in many-electron systems. The time-local result in the presence of an arbitrary field is obtained with the gauge-invariant GKBA \cite{kremp_99_pre},
\begin{align}
    G_a^<(\textbf{k},t,t') = - G_a^R(\textbf{k},t,t') f_a\left[
    \textbf{k}-\frac{e_a}{c}\textbf{A}(t')+ \frac{e_a}{c}\int_{t'}^t d\bar t \frac{\textbf{A}(\bar t)}{t-t'}
    \right]\,,
    \label{eq:gkba-gauge-inv}
\end{align}
which applies to $t<t'$, where the momentum argument of the distribution function is shifted by the electric field, and similarly, the retarded function is affected by the field, cf. Eq.~(\ref{eq:gra-gauge-inv}). With this the time-local kinetic equation for the distribution function becomes
\begin{align}
    \frac{\partial f_a}{\partial T}(\textbf{k},T) + e_a \textbf{E}(T) \nabla_k f_a(\textbf{k},T) &= 2 \tn{Re}\int_{t_0}^T d\bar t\left\{ 
    G_a^>\left(\textbf{k}+ \Delta \textbf{k}(\textbf{A};T,\bar t); T, \bar t \right) \Sigma_a^<\left(\textbf{k}+\Delta \textbf{k}(\textbf{A};\bar t,T); \bar t, T \right)
    - \left( > \leftrightarrow < \right) \right\} \equiv \sum_b I_{ab}(\textbf{k},T)\,,\\
\mbox{where} \quad \Delta \textbf{k}(\textbf{A};t_1,t_2) &=    \frac{e_a}{c}\textbf{A}(t_2)-\frac{e_a}{c}\int_{t_1}^{t_2} dt'\frac{\textbf{A}(t')}{t_2- t_1}\,,
\label{eq:gi-kin-eq}
\end{align}
which holds for an arbitrary selfenergy. The gauge-invariant GKBA result for the collision integrals $I_{ab}$ is obtained by inserting Eq.~(\ref{eq:gkba-gauge-inv}) for all appearances of $G^\gtrless$ under the integral. Thus, we are able to capture the effect of the field on any two-particle collision process.

We provide, as an example, the result for collisions described in direct second order Born approximation (SOA):
\begin{align}
   I_{ab}(\textbf{k}_a,T) &= \frac{2}{\hbar^2} \int \frac{d^3k_b d^3 k'_a d^3 k'_b}{(2\pi\hbar)^6}|V_{ab}(\textbf{q})|^2 \delta(k_a+k_b- k'_a - k'_b)\int_{t_0}^T dt' \cos\left\{ \frac{E_{ab}-E'_{ab}}{\hbar}(t-t')-\textbf{q}\textbf{R}_{ab}(T,t')\right\}\Phi_{ab}(t')\,,
   \\
   \textbf{R}_{ab}(T,t') &= \left( \frac{e_a}{m_a}-\frac{e_b}{m_b}\right)\int_{t'}^T d\bar t\int_{\bar t}^T dt'' \textbf{E}(t'') \longrightarrow \left( \frac{e_a}{m_a}-\frac{e_b}{m_b}\right)\textbf{E}_0 \left[ (T-t')\frac{\sin{\Omega T}}{\Omega} + \frac{\cos{\Omega T}-\cos{\Omega t'}}{\Omega^2}
   \right]
   \,,\label{eq:rab-definition}
\end{align}
where $\textbf{q} = \textbf{k}_a-\textbf{k}'_a$, $V_{ab}$ is the statically screened Coulomb potential, and $\Phi_{ab}(t)= \{f_a(k_a')f_b(k_b')[1-f_a(k_a)]
[1-f_b(k_a)] - f_a(k_a)f_b(k_b[1-f_a(k'_a)]
[1-f_b(k'_a)]\}|_t$. The collision integral contains the standard non-Markovian time integral with the cosine of the energy difference of two colliding particles. The first contribution to the cosine contains just the standard kinetic energy difference, with $E_{ab}=\frac{k_a^2}{2m_a}+\frac{k_b^2}{2m_b}$ which, in the Markov limit, gives rise to an energy delta function (Fermi's golden rule). Interestingly,  the second term is the modification to the energy balance produced by the field: $\textbf{R}_{ab}$ is field induced change of the distance of the colliding particles ``a'' and ``b'' during the time interval $[t',T]$.
The rightmost expression in Eq.~(\ref{eq:rab-definition}) corresponds to a harmonic field. This field influence during the finite duration of the collision process is often called ``intra-collisional field effect'' and is of high interest in laser plasmas but also in semiconductors exposed to strong infrared fields \cite{bertoncini_prb_91,rossi_sst_04} and is also relevant for impact ionization \cite{madureira_jap_01}. While these complex effects are usually difficult to describe theoretically, within the gauge-invariant GKBA they appear naturally. Other effects caused by the field are ``collisional broadening'' of the energy delta function and also multi-photon processes. To see the appearance of multi-photon effects in the two-particle collisions it is sufficient to recall the standard identity
\begin{align}
    e^{\pm i z \cos{\Omega t}} = \sum_{n=-\infty}^\infty (\pm i)^n J_n(z)\, e^{\mp i n \Omega t} \,,
\nonumber
\end{align}
where $J_n$ denotes the Bessel function of $n$-th order. This expression directly appears in the case of a harmonic field [rightmost expression in Eq.~(\ref{eq:rab-definition})] where the argument $z$ is directly proportional to the field amplitude. Thus, not only the single-particle spectral function develops photon side-bands. Also, during the collision process, additional scattering channels open,  corresponding to Coulomb collisions during which a number of $n$ photons is absorbed or emitted. An extension of the gauge-invariant GKBA to GW selfenergies was presented in Ref.~\cite{bonitz_99_cpp} and numerical solutions of the kinetic equation (\ref{eq:gi-kin-eq}) in Ref.~\cite{haberland_01_pre}.

\subsection{Application of the GKBA to finite non-uniform systems}
\label{ss:gkba-nonuniform}
Let us now turn to finite non-uniform systems. The first applications of the HF-GKBA to the latter were presented in
Refs.~\cite{hermanns_psc_12,balzer-book}.

applications to quantum wells and quantum dots \cite{hermanns_psc_12,balzer_epl_12,hermanns_prb14}, electronic double excitations and nonlinear response of small Hubbard clusters \cite{bonitz_cpp13,balzer_jpcs13}, 
transient photoabsorption of atoms and charge migration in molecules \cite{perfetto_pra_15,perfetto_jctc_19}.
On the other hand, there appeared more general theoretical investigations concerning the comparison of the GKBA
to the independent theory of density operators (BBGKY hierarchy), time reversal invariance \cite{bonitz_cpp18},  energy conservation and initial correlations \cite{hermanns_jpcs13,hermanns_prb14,karlsson_prb_18,hopjan_initial_2019,tuovinen_pss_19}.
A particularly interesting observation was that the problem of unphysical damping of two-time solutions of the KBE for small Hubbard clusters \cite{von_friesen_successes_2009,von-friesen_prb10} did not occur when the GKBA was applied \cite{hermanns_jpcs13,hermanns_prb14,schluenzen_prb17}. Detailed comparisons of the HF-GKBA to full two-time KBE simulations were performed in Refs.~\cite{schluenzen_prb17,tuovinen_prb_20}.

\section{The G1--G2 scheme}\label{s:g1-g2}

The G1--G2 scheme was introduced by Schl\"unzen, Joost and Bonitz in Ref.~\cite{schluenzen_prl_20}. A detailed analysis for various selfenergies was given in Ref.~\cite{joost_prb_20}, and an extension to the dynamically screened ladder approximation was presented in Ref.~\cite{joost_prb_22}. Here we present a brief introduction, including important selfenergy approximations. At the end of this section, we discuss limitations, open problems and possible solutions to them.

\subsection{Idea}\label{ss:g1g2-idea}

The G1--G2 scheme was developed with the goal to find a representation of the HF-GKBA that scales only linearly with the propagation time. Within the two-time NEGF approach this is prevented by the occurrence of two-time quantities and memory integrals, cf. Eqs.~(\ref{eq:KBE1}) to (\ref{eq:kbe-sigma-form2}). This is true even when considering only the KBE on the time diagonal, 
\begin{align}
 \i\hbar \frac{\d}{\d t}G_{ij}^<(t) - \left[ h^\mathrm{HF},G^< \right]_{ij}(t) = \big[I + I^\dagger\big]_{ij}(t)\;,\;\;\;\mbox{with}\;\;
 \left[ h^\mathrm{HF},G^< \right]_{ij}(t) =\sum_k \left\{ h^\mathrm{HF}_{ik}(t) G^<_{kj}(t) - G^<_{ik}(t)h^\mathrm{HF}_{kj}(t)
 \right\}\,,
 \label{eq:eom_gone}
\end{align}
due to the collision integral appearing on the right-hand side, which contains an integral over the history of the system,
\begin{align}
 \label{eq:definition_I}
 I_{ij}(t) = \sum_{k} \int_{t_0}^t \mathrm{d}\cbar{t} \left[ \Sigma_{ik}^>(t,\cbar{t}) G_{kj}^<(\cbar{t},t) - \Sigma_{ik}^<(t,\cbar{t}) G_{kj}^>(\cbar{t},t) \right] = \pm\i\hbar \sum_{klp} w_{iklp}(t) \mathcal{G}_{lpjk}(t)\,.
\end{align}
Note that the quantity $\Sigma$ appearing in Eqs.~(\ref{eq:KBE1})--(\ref{eq:kbe-sigma-form2}) describes the total selfenergy. In contrast, the lesser and greater components of the selfenergy $\Sigma$ entering the collision integral in Eq.~(\ref{eq:definition_I}) contain only the correlation contribution. The mean-field part is contained in the Hartree--Fock Hamiltonian $h^\mathrm{HF}$ in the commutator of Eq.~(\ref{eq:eom_gone}).
The central idea of the G1--G2 scheme is to express the collision integral in terms of the correlation part of the two-particle Green function,
\begin{align}
    \mathcal{G}_{ijkl}(t) = G^{(2)}_{ijkl}(t) - G^<_{ik}(t)G^<_{jl}(t) \mp G^<_{il}(t)G^<_{jk}(t)\,,
\end{align}
thereby eliminating the selfenergy. When applying the HF-GKBA, this formulation results in a cost-efficient linear scaling with respect to the propagation time. In the following, this concept is illustrated for the simplest case of the second-order Born selfenergy, cf. Eq.~(\ref{eq:sigma-soa}), where the collision integral is given by
\begin{align}
 I^{\rm SOA}_{ij}(t) =
 & \pm\left(\i\hbar\right)^2 \sum_{klpqrsu} \, w_{iklp}\left(t\right) \int_{t_0}^t \d\cbar t\, w^\pm_{qrsu}\left(\cbar t\right) \Big[ G^>_{lq}\left(t,\cbar t\right) G^>_{pr}\left(t,\cbar t\right) G^<_{uk}\left(\cbar t,t\right) G^<_{sj}\left(\cbar t,t\right) - G^<_{lq}\left(t,\cbar t\right) G^<_{pr}\left(t,\cbar t\right) G^>_{uk}\left(\cbar t,t\right) G^>_{sj}\left(\cbar t,t\right) \Big] \, .
\end{align}
By comparing this expression to Eq.~(\ref{eq:definition_I}) one can identify the correlated part of the two-particle Green function within the second-order Born approximation \cite{bonitz_jpcs_13},
\begin{align}\label{eq:g2-integral}
 \mathcal{G}^{\rm SOA}_{ijkl}(t) =& \i\hbar\sum_{pqrs} \int_{t_0}^t \d\cbar t\, w^\pm_{pqrs}\left(\cbar t\right) 
\Big[ G^>_{ip}(t,\cbar t) G^>_{jq}(t,\cbar t) G^<_{rk}(\cbar t,t) G^<_{sl}(\cbar t,t) - G^<_{ip}(t,\cbar t) G^<_{jq}(t,\cbar t) G^>_{rk}(\cbar t,t) G^>_{sl}(\cbar t,t) \Big] \, .
\end{align}
As this expression still contains two-time quantities and a memory integral, at first glance it seems that it has no advantage compared to the selfenergy formulation. However, it turns out that when applying the HF-GKBA, cf. Eq.~(\ref{eq:gkba}), for the occurring single-particle Green functions, the differential equation of Eq.~(\ref{eq:g2-integral}) does not contain a time integral.
When differentiating $\mathcal{G}^{\rm SOA}$ with respect to the time $t$, two contributions occur, the first from the single-particle Green functions, and the second from the integral boundary. A straightforward calculation reveals that the resulting ordinary differential equation is of the form \cite{bonitz_qkt,schluenzen_prl_20,joost_prb_20}, 
\begin{align}
 \i\hbar  \frac{\d}{\d t} \mathcal{G}^{\rm SOA}_{ijkl}(t) &- \Big[ h^{(2),\tn{HF}},\mathcal{G}^{\rm SOA}\Big]_{ijkl}(t) = \left(\i\hbar\right)^2\sum_{pqrs} w^\pm_{pqrs}(t) \Big[ G^>_{ip}(t) G^>_{jq}(t) G^<_{rk}(t) G^<_{sl}(t) - G^<_{ip}(t) G^<_{jq}(t) G^>_{rk}(t) G^>_{sl}(t) \Big]  \label{eq:G1-G2_soa}\,,
\end{align}
with the effective two-particle Hamiltonian
\begin{align}
 h^{(2),\tn{HF}}_{ijkl}(t) &= h^\tn{HF}_{ik}(t)\delta_{jl} + h^\tn{HF}_{jl}(t)\delta_{ik}\,.
 \label{eq:h2-hf}
\end{align}
Indeed, the effort of solving Eq.~(\ref{eq:G1-G2_soa}) scales linearly with the number of time steps as it contains only single-time quantities. Moreover, it is important to note that no further approximations beyond the HF-GKBA were used to derive Eq.~(\ref{eq:G1-G2_soa}), which makes the system of equations (\ref{eq:eom_gone}) and (\ref{eq:G1-G2_soa}) an equivalent reformulation of the two-time HF-GKBA with second-order Born selfenergy.\\
From a numerical point of view it is advantageous to replace all occurrences of $G^>$ in Eq.~(\ref{eq:G1-G2_soa}) by $G^<$ using Eq.~(\ref{eq:g-glsymm}), which reduces the number of sums to evaluate,
\begin{align}
 \i\hbar  \frac{\d}{\d t} \mathcal{G}^{\rm SOA}_{ijkl}(t) - \Big[ h^{(2),\tn{HF}},\mathcal{G}^{\rm SOA} \Big]_{ijkl}(t) = \hat\Psi^\pm_{ijkl}(t) := & \sum_{pq} \Big[w^\pm_{ijpq}(t) G^<_{pk}(t) G^<_{ql}(t) - G^<_{ip}(t) G^<_{jq}(t) w^\pm_{pqkl}(t)\Big]  \nonumber\\\nonumber
 &+\i\hbar\sum_{pqr} \Big[ w^\pm_{ipqr}(t) G^<_{jp}(t) G^<_{qk}(t) G^<_{rl}(t) + w^\pm_{pjqr}(t) G^<_{ip}(t) G^<_{qk}(t) G^<_{rl}(t)\Big]  \\
 &-i\hbar\sum_{pqrs} \Big[ G^<_{ip}(t) G^<_{jq}(t) G^<_{rl}(t) w^\pm_{pqkr}(t) + G^<_{ip}(t) G^<_{jq}(t) G^<_{rk}(t) w^\pm_{pqrl}(t)\Big]  \label{eq:G1-G2_soa_delta}\,.
\end{align}
Here we introduced the inhomogeneity, $\hat\Psi^\pm$, which contains terms that do not involve $\mathcal{G}$.

\begin{figure}
    \centering
    \includegraphics[width=0.6\textwidth]{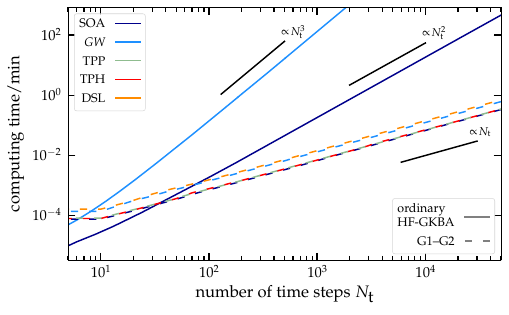}
    \caption{Comparison of the CPU time scaling between the ordinary implementation (involving the memory integral) of the HF-GKBA, cf.~Sec.~\ref{s:GKBA}, and the G1--G2 scheme with respect to the number of time steps $N_\tn{t}$. Calculations were performed for a 10-site Hubbard chain. The linear time scaling of G1--G2 is confirmed across all approximations, while the standard HF-GKBA, cf. Sec.~\ref{s:GKBA}, scales as $N_\tn{t}^2$, for 2B, and $N^3_\tn{t}$, for $GW$. Two-time KBE simulations, cf. Sec.~\ref{ss:KBE}, scale cubically, even for the simplest case of SOA. Adapted from Ref.~\cite{joost_prb_20}, licensed under CC BY 4.0.}
    \label{fig:scaling}
\end{figure}

\textbf{Verification of time-linear scaling of the CPU time.}
As we have seen, the G1--G2 scheme allows one to express the second order Born selfenergy approximation on the HF-GKBA level in the form of two coupled ordinary differential equations, Eqs. (\ref{eq:eom_gone}) and (\ref{eq:G1-G2_soa_delta}). These equations are time local, so the numerical effort to solve them should scale linearly with $N_\tn{t}$. However, it cannot be ruled out that this scaling is achieved only asymptotically, for large $N_\tn{t}$. Fortunately, this is not the case. As was shown in Ref.~\cite{schluenzen_prl_20} time linear scaling is achieved already after a few tens of time steps. This is demonstrated for SOA selfenergies, for a small Hubbard chain in Fig.~\ref{fig:scaling}, cf. the dark blue lines. While the standard HF-GKBA (involving a non-Markovian time integral) quickly approaches the expected quadratic scaling, the G1--G2 simulation scales, indeed, linearly with $N_\tn{t}$ and is faster than the former already after $30$ time steps. As we will see below, the same behavior is recovered for selfenergy approximations beyond SOA.

\subsection{G1--G2 scheme with selfenergies beyond SOA}\label{s:g1g2-approx}
In order to generalize the G1--G2 scheme to more advanced selfenergies it is advantageous to rewrite the differential equation of $\mathcal{G}$ as 
\begin{align}\label{eq:g1g2_eom2}
    \mathrm{i}\hbar\frac{\mathrm{d}}{\mathrm{d}t}\mathcal{G}_{ijkl}(t)&= \left[h^{\tn{(2),HF}}(t),\mathcal{G}(t)\right]_{ijkl} + I^{(2)}_{ijkl}(t) + I^{(2)}_{jilk}(t) - \left[I^{(2)}_{klij}(t)\right]^* - \left[I^{(2)}_{lkji}(t)\right]^*\,,
\end{align}
where, in analogy to Eq.~(\ref{eq:eom_gone}), a new term, $I^{(2)}$, was introduced which contains contributions from the three-particle density operator. When ignoring all three-particle correlations, this collision term can be expressed as
\begin{align}\label{eq:g1g2_I2}
    I^{(2)}_{ijkl}(t) &= \Psi^\pm_{ijkl}(t) + L^\tn{pp}_{ijkl}(t) + P^\mathrm{ph}_{ijkl}(t) + P^{GW}_{ijkl}(t)\,.
\end{align}
The first term can be identified as the second-order Born contribution, 
\begin{align}\label{eq:g1g2_2B}
    \Psi^\pm_{ijkl}(t) &\coloneqq \frac{1}{2}\sum_{pq} w^\pm_{ijpq}(t)  G^<_{pk}(t)G^<_{ql}(t) + \mathrm{i}\hbar \sum_{pqr} w^\pm_{ipqr} G^<_{jp}(t) G^<_{qk}(t) G^<_{rl}(t)\,,
\end{align}
which, when considered on its own, leads to Eq.~(\ref{eq:G1-G2_soa_delta}). This term is related to $\hat\Psi^\pm$, introduced in Eq.~(\ref{eq:G1-G2_soa}), by $\hat\Psi^\pm_{ijkl}=\Psi^\pm_{ijkl}+\Psi^\pm_{jilk}-\Psi^{\pm*}_{klij} - \Psi^{\pm*}_{lkji}$.

The subsequent terms are the ``ladder'' terms, related to the particle--particle $T$-matrix,
\begin{align}\label{eq:g1g2_TPP}
    L^\tn{pp}_{ijkl}(t) \coloneqq \frac{1}{2}\sum_{pq} w_{ijpq}(t) \mathcal{G}_{pqkl}(t) + \mathrm{i}\hbar \sum_{pqr} w_{ipqr}(t) G^<_{jp}(t) \mathcal{G}_{qrkl}(t)\,,
\end{align}
the particle--hole $T$-matrix,
\begin{align}\label{eq:g1g2_TPH}
    P^\mathrm{ph}_{ijkl}(t) \coloneqq \mathrm{i}\hbar \sum_{pqr} w^\pm_{ipqr}(t) G^<_{rl}(t) \mathcal{G}_{qjkp}(t)\,,
\end{align}
and the $GW$ channel (polarization terms, ``bubble'' diagrams),
\begin{align}\label{eq:g1g2_GW}
    P^{GW}_{ijkl}(t) \coloneqq \pm \mathrm{i}\hbar \sum_{pqr} w^\pm_{ipqr}(t) G^<_{qk}(t) \mathcal{G}_{rjpl}(t)\,.
\end{align}
Using Eqs. (\ref{eq:g1g2_eom2})--(\ref{eq:g1g2_GW}), all selfenergy approximations that were presented in Sec.~\ref{s:sigmas} can be expressed in the frame of the G1--G2 scheme. The explicit form of the collision term $I^{(2)}$ corresponding to each selfenergy approximation is given in Tab.~\ref{tab:g1g2approx}.\\
Note that the $GW$ and TPH approximation are defined without exchange contribution in Sec.~\ref{s:sigmas}, resulting in the following expressions for $I^{(2)}$ in the G1--G2 scheme,
\begin{align}\label{eq:g1g2_GW_exp}
    I^{(2),GW}_{ijkl}(t) &\coloneqq \frac{1}{2}\sum_{pq} w_{ijpq}(t)  G^<_{pk}(t)G^<_{ql}(t) + \mathrm{i}\hbar \sum_{pqr} w_{ipqr} G^<_{jp}(t) G^<_{qk}(t) G^<_{rl}(t) \pm \mathrm{i}\hbar \sum_{pqr} w_{ipqr}(t) G^<_{qk}(t) \mathcal{G}_{rjpl}(t)\,,\\\label{eq:g1g2_TPH_exp}
    I^{(2),\mathrm{TPH}}_{ijkl}(t) &\coloneqq \frac{1}{2}\sum_{pq} w_{ijpq}(t)  G^<_{pk}(t)G^<_{ql}(t) + \mathrm{i}\hbar \sum_{pqr} w_{ipqr} G^<_{jp}(t) G^<_{qk}(t) G^<_{rl}(t) + \mathrm{i}\hbar \sum_{pqr} w_{ipqr}(t) G^<_{rl}(t) \mathcal{G}_{qjkp}(t)\,.
\end{align}
Defining those two approximations including exchange would result in a violation of energy conservation as shown in Ref.~\cite{schluenzen_phd_21}. However, an energy conserving approximation can be constructed by combining the $GW$ and TPH contributions including their respective exchange contributions, which is here denoted as the fully antisymmetrized polarization (ASP) approximation. \\
Adding all contributions of Eqs.~(\ref{eq:g1g2_2B})--(\ref{eq:g1g2_GW}) to the two-particle collision integral results in the dynamically screened ladder (DSL) approximation, which combines the dynamical screening of $GW$ with scattering effects of the particle–particle and particle–hole ladders. The TOA and FLEX approximations can be expressed as special variants of DSL where  $\mathcal{G}$ does not enter Eqs.~(\ref{eq:g1g2_2B})--(\ref{eq:g1g2_GW}) self-consistently, but instead, additional independent calculations have to be performed on the level of the approximations given in Tab.~\ref{tab:g1g2approx}, for details see the entries in the table.\\

\textbf{Verification of time-linear scaling of the CPU time.} As we did for SOA selfenergies in Sec.~\ref{ss:g1g2-idea}, we again test the scaling of the G1--G2 simulations with $N_\tn{t}$, but now for the more advanced selfenergy approximations studied in this section.
The tremendous numerical improvement of the G1--G2 scheme, as compared to the non-Markovian version of the HF-GKBA, is again confirmed in Fig.~\ref{fig:scaling} for the example of a 10-site Hubbard chain. The results confirm that, in the case of the non-Markovian HF-GKBA, the computing time of the $GW$ approximation increases cubically with the number of time steps (see the bright blue curve), and the same is observed for the $T$-matrix selfenergy (not shown). In contrast, within the G1--G2 scheme all approximations -- SOA, $GW$ and particle-particle as well as particle-hole $T$-matrix -- exhibit a linear scaling with the propagation time. As a result, already at less than 50 time steps the time-linear scheme is faster than the two-time approach in all considered cases. Remarkably, this linear scaling is achieved also for the more involved dynamically screened ladder approximation (DSL) that is studied in Sec.~\ref{ss:dsl}. DSL simulations are not possible with the two-time KBE and within the non-Markovian HF-GKBA.

\begin{table}[t]
\def\arraystretch{1.5}
\centering
\begin{tabular}{c||l}
  \textbf{Approximation} & \textbf{Collision term} \\  \hline 
 SOA & $I^{(2)} = \Psi^\pm$\\\hline
 TPP & $I^{(2)} = \Psi^\pm + L^\tn{pp}$\\\hline
 ASP & $I^{(2)} = \Psi^\pm + P^\tn{ph} + P^{GW}$\\\hline
 TPH & $I^{(2)} = \Psi + P^\tn{ph}\phantom{P^{GW}} \qquad\qquad\qquad\quad (w^\pm\rightarrow w)$\\\hline
 $GW$ & $I^{(2)} = \Psi + P^{GW}\phantom{P^\tn{ph}} \qquad\qquad\qquad\quad (w^\pm\rightarrow w)$\\\hline
 TOA & $I^{(2)} = \Psi^\pm + L^\tn{pp}[\mathcal{G}^\tn{SOA}] + P^\tn{ph}[\mathcal{G}^\tn{SOA}] + P^{GW}[\mathcal{G}^\tn{SOA}]$\\\hline
 FLEX & $I^{(2)} = \Psi^\pm + L^\tn{pp}[\mathcal{G}^\tn{TPP}] + P^\tn{ph}[\mathcal{G}^\tn{TPH}] + P^{GW}[\mathcal{G}^{GW}]$\\\hline
 DSL & $I^{(2)} = \Psi^\pm + L^\tn{pp} + P^\tn{ph} + P^{GW}$
\end{tabular}
\caption{\label{tab:g1g2approx} Explicit form of the collision term $I^{(2)}$ entering Eq.~(\ref{eq:g1g2_eom2}) for different approximations of the G1--G2 scheme. The definition of the individual contributions is given in Eqs. (\ref{eq:g1g2_2B})--(\ref{eq:g1g2_GW}) Note that for $GW$ and TPH the standard pair potential $w$ has to be used. In contrast, the combination of both in the ASP approximation uses $w^\pm$. Adapted from Ref.~\cite{joost_phd_2022}, licensed under CC BY 4.0.
}
\end{table}

\subsection{Dynamically screened ladder approximation}\label{ss:dsl}
The dynamically screened ladder is the most sophisticated of the approximations presented in the previous section, as it contains all three-particle terms that can be expressed as products of single- and two-particle Green functions. Only three-particle correlation effects are neglected within the DSL approximation. Its relation to selfenergy approximations within many-body perturbation theory was established in Ref.~\cite{joost_phd_2022}. It corresponds to solving the Martin--Schwinger hierarchy employing the HF-GKBA and using a variant of the parquet approximation to the selfenergy. As such it contains not only the diagrams of the three infinite series presented in section \ref{ss:negf-gf-gh} to \ref{ss:negf-teh}, namely $GW$ and the particle--particle and particle--hole $T$ matrix, but also cross-coupling contributions. The latter distinguish DSL from the traditional FLEX selfenergy of NEGF theory, which ignores all coupling between the three infinite diagrammatic series~\cite{bickers_conserving_1989,bickers_conserving_1989-1}.

Note that the DSL approximation was not originally derived for the G1--G2 scheme. Instead it has a long history as a method for truncating the BBGKY hierarchy within density matrix theory~\cite{schmitt_truncation_1990,cassing_self-consistent_1992,tohyama_applications_2020}. In this role it was first introduced by Wang and Cassing in 1985 as NQCD (Nuclear Quantum Correlation Dynamics)~\cite{wang-cassing-85,cassing_dynamical_1988}. Subsequently, it found widespread use under the names TDDM~\cite{de_blasio_nonperturbative_1992,lacroix_nuclear_2004} and WC~\cite{akbari_challenges_2012,akbari_development_2012} approximation. It was later independently derived for equilibrium  theory~\cite{colmenero_approximating_1993,colmenero_approximating_1993-1} and is since known as the Valdemoro approximation in the context of the (anti-Hermitian) contracted Schrödinger equation~\cite{valdemoro_contracted_2007} and in the field of time-dependent two-particle reduced density matrix (TD-2RDM) theory for the description of atoms and molecules~\cite{lacknerphd,lackner_pra_17,lackner_propagating_2015,schafer-bung_correlated_2008}. It should be emphasized that all these different names describe the same approximation which is, in fact, fully identical to the G1--G2 scheme when using the DSL approximation defined in section~\ref{s:g1g2-approx}.



\subsection{Stability issues. Bottlenecks and open questions}

\begin{figure}
    \centering
    \includegraphics[width=0.55\textwidth]{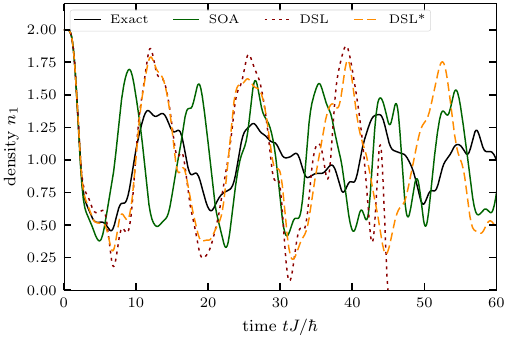}
    \caption{Density dynamics on the first site of a six-site Hubbard chain for $U=J$. In the initial state the first three sites are doubly occupied while the remaining sites are empty. The exact CI solution is compared to the SOA and DSL results, cf. Tab.~\ref{tab:g1g2approx}. The approximation DSL* is improved by enforcing contraction consistency and applying purification.}
    \label{fig:instability}
\end{figure}

\subsubsection{Stability issues of the G1--G2 scheme. $N$-representability}
Within NEGF theory, certain approximations are known to lead to unstable solutions. This is the case when the respective selfenergy is not positive semi-definite (PSD)~\cite{stefanucci_diagrammatic_2014,uimonen_diagrammatic_2015}.
For approximations in the G1--G2 scheme similar instabilities can be observed, especially for large interaction strengths. An example is shown in Fig.~\ref{fig:instability} for a six-site Hubbard model,
\begin{align}
 \chat{H}(t) = -J\sum_{\left<i,j\right>} \sum_\alpha \cop{i}{\alpha}\aop{j}{\alpha} + U(t) \sum_i \chat{n}_{i}^{\uparrow} \chat{n}_{i}^{\downarrow}\,,
\label{eq:h-hubbard}
\end{align}
which includes hopping processes between nearest-neighbor sites $\left<i,j\right>$ with amplitude $J$ and interactions of particles on the same site with strength $U$. Here, $U$ is set to $1J$ and in the initial state the first three sites are doubly occupied while the remaining sites are empty. While the SOA solutions does not capture the exact dynamics very well, the calculation remains stable for the complete duration of $60 tJ/\hbar$. In contrast, the DSL approximation reproduces the frequency of the exact density oscillations much better. However, the calculation becomes unstable after $45 tJ/\hbar$. As discussed in Ref.~\cite{joost_phd_2022} this can be related to the violation of $N$-representability during the propagation of the two-particle Green function.\\
For the $N$-representability condition to be met, the density matrix has to be derivable from an $N$-particle wave function. In practice, $N$-representability can be improved by two procedures. First, by ensuring contraction consistency between the two Green functions of adjacent particle numbers, i.e.
\begin{align}\label{eq:trace1}
    N &= \pm \mathrm{i}\hbar \sum_{p} G^<_{pp}(t)\,,\\   (N-1) G^<_{ij}(t) &= \pm \mathrm{i}\hbar \sum_{p} G^{(2)}_{ipjp}(t)\,,\label{eq:trace2}\\    (N-2) G^{(2)}_{ijkl}(t) &= \pm \mathrm{i}\hbar \sum_p G^{(3)}_{ijpklp}(t)\,,\label{eq:trace3}
\end{align}
where $N$ is the total particle number of the system.
In order to ensuring contraction consistency the collision integral in Eq.~(\ref{eq:g1g2_eom2}) is extended by a correction term
\begin{align}\label{eq:I_CC}
 I^{\tn{(2),CC}}_{ijkl}(t) \coloneqq \pm \mathrm{i}\hbar \sum_{pqr} w_{ipqr}(t) G^{(3),\tn{CC}}_{qrjkpl}(t)\,,
\end{align}
where the quantity $G^{(3),\tn{CC}}$ is constructed in such a way that it ensures Eqs.~(\ref{eq:trace1})--(\ref{eq:trace3}). More details on this procedure can be found in Ref.~\cite{joost_prb_22}. The second measure to improve $N$-representability is to enforce the two-positivity conditions
 \begin{alignat}{5}\label{eq:2p_positivity1}
      &\left(\tn{i}\hbar\right)^2 G^{(2)}_{ijkl}(t) &&=&&\eqmakebox[G2pphhph][r]{$\displaystyle \left(\tn{i}\hbar\right)^2 G^{\tn{pp}}_{ijkl}(t)$} &&\coloneqq&& \bra{\Psi} \hat{c}^\dagger_k(t) \hat{c}^\dagger_l(t) \hat{c}_j(t) \hat{c}_i(t) \ket{\Psi} \succeq 0\,,\\
      &\left(\tn{i}\hbar\right)^2 G_{ijkl}^{(2),>}(t) &&=&& \eqmakebox[G2pphhph][r]{$\displaystyle\left(\tn{i}\hbar\right)^2 G^{\tn{hh}}_{ijkl}(t)$} &&\coloneqq&& \bra{\Psi} \hat{c}_i(t) \hat{c}_j(t) \hat{c}^\dagger_l(t) \hat{c}^\dagger_k(t) \ket{\Psi} \succeq 0\,,\\
      & && && \eqmakebox[G2pphhph][r]{$\displaystyle\pm\left(\tn{i}\hbar\right)^2 G^{\tn{ph}}_{ijkl}(t)$} &&\coloneqq&& \bra{\Psi} \hat{c}^\dagger_k(t) \hat{c}_j(t) \hat{c}^\dagger_l(t)  \hat{c}_i(t) \ket{\Psi} \succeq 0\,.\label{eq:2p_positivity3}
\end{alignat}
In practice this is done by performing a so-called purification of the two-particle Green function, i.e., by removing all unphysical positive eigenvalues,
\begin{align}
    G^{(2),\mathrm{pur}}_{ijkl} = G^{(2)}_{ijkl} - G^{(2),\mathrm{pos}}_{ijkl}\,,
\end{align}
where $G^{(2),\mathrm{pos}}$ contains only the contribution of the positive eigenvalues of $G^{(2)}$. Again, a more detailed description of the procedure can be found in Ref.~\cite{joost_phd_2022}. Applying both measures, enforcing contraction consistency and performing purification, leads to the improved approximation DSL*. In Fig.~(\ref{fig:instability}) this results in a stable density dynamics up to the final time of $60 tJ/\hbar$, while the accuracy remains on the level of the original DSL approximation.



\subsubsection{Bottlenecks of the G1--G2 scheme and possible solutions}\label{ss:g1-g2-problems}
The dramatic speed up of NEGF simulations within the G1--G2 scheme to time linear scaling comes at a cost: a large computer memory is required to store the two-particle Green function. 
In fact, the propagation of the single-time pair correlation function $\mathcal{G}(t)$ requires to store a rank-4 tensor, $\mathcal{G}_{ijkl}$ which has the dimension $N_\mathrm{b}^4$. This is presently the main bottleneck of the G1--G2 scheme.
%
%
This problem can, of course, be tackled computationally, by using massively parallel computer hardware over which the tensor $\mathcal{G}$ is distributed. This, however, gives rise to a significant overhead for communication.

There exist two additional possible solutions of the memory problem that are independent of the hardware: the first is an embedding selfenergy approach \cite{balzer_prb_23}, which will be discussed in Sec.~\ref{s:embedding}. The second possible strategy is a novel quantum fluctuations approach \cite{schroedter_cmp_22} that will be summarized in Sec.~\ref{s:quantum-fluctuations}. 
But, before discussing these concepts, we illustrate additional applications of the scheme, in Sec.~\ref{sss:g1g2-uniform}.

\subsection{G1--G2 simulations of spatially uniform systems}\label{sss:g1g2-uniform}
An example where large basis sets occur are spatially uniform systems. Here it is natural to resort to a  momentum basis, see also Sec.~\ref{ss:gkba-uniform}. We consider two cases. The first are systems with continuous momentum states, such as plasmas, nuclear matter, the electron gas (jellium), or electron-hole plasmas.  
Here, as we will see below, presently only one-dimensional model systems can be treated \cite{makait_cpp_23}. 
This will be discussed in Sec.~\ref{sss:uniform-1d}. The second example are lattice systems, such as atoms in optical lattices or 2D quantum materials. This will be discussed in Sec.~\ref{sss:uniform-lattice}.

In uniform systems, the single-particle and two-particle Green functions take on the form

\begin{align}
 G^\gtrless_{\mathbf{p}\alpha,\mathbf{p}'\alpha'}(t)&=\vcentcolon G^\gtrless_{\mathbf{p}\alpha}(t)\delta_{\mathbf{p},\mathbf{p}'}\delta_{\alpha,\alpha'}\,,\\
 \mathcal{G}_{\mathbf{k}\alpha,\mathbf{p}\beta,\mathbf{k}'\alpha',\mathbf{p}'\beta'}(t)&=\vcentcolon\mathcal{G}^{\alpha\beta}_{\mathbf{k}',\mathbf{p}',\mathbf{k}'-\mathbf{k}}(t)\delta_{\alpha\alpha'}\delta_{\beta\beta'}\delta_{\mathbf{k}+\mathbf{p},\mathbf{k}'+\mathbf{p}'}\label{eq:G2StructureMomRep}\,.
\end{align}
Here, bold-faced latin letters like $\mathbf{k},\mathbf{p},\mathbf{q}$ denote momenta, and greek letters like $\alpha,\beta$ denote the spin, and eventually particle species.

Conventional NEGF calculations take advantage of the system symmetries: $G_{\mathbf{p}\sigma}$ is a function of spin and one momentum (momentum distribution function or Wigner distribution). In particular, if the system is assumed to fulfill spatial symmetries, e.g. isotropy or cylinder symmetry, $G$ can be fully described by a reduced set of $\mathbf{p}$-vectors that span a representative portion of momentum space. For example, in the isotropic 3D system, $G$ can be characterized by the radial $p_r$-momentum, i.e. by a 1D-array of variables in a simulation program. Unfortunately, this advantage quickly diminishes in the G1--G2 scheme since we also have to store the two-particle Grenn function: $\mathcal{G}_{\mathbf{kpq}}^{\alpha\beta}$ is a function of three momenta. In a 3D isotropic system, we can find a representative of every triplet $(\mathbf{k},\mathbf{p},\mathbf{q})$ in a 6-dimensional subspace of the full 9-dimensional space of the momentum triplets. Usually, one aims to treat several hundreds of discrete momentum values in a program, which yields the necessity to store and compute many $10^{12}$'s of values of $\mathcal{G}_{\mathbf{kpq}}$. For current computers, this is not feasible. This problem is less severe in lower dimensions, especially in 1D, cf. Table \ref{tab:SOAScalings} for SOA, $GW$, and DSL scalings.

In the following subsection, where we investigate a quasi-1D quantum plasma, we will make use of this particular structure. For the application to uniform graphene, Sec. \ref{sss:uniform-lattice}, or other multi-band systems we need to generalize the equations: The $G^\gtrless$ and $\mathcal{G}$ obtain additional band indices, and $w$ cannot be written in the form $w_\mathbf{q}$ but has full dependence on all three momenta, $w_\mathbf{kpq}.$

\begin{table}[t]
    \centering
\begin{tabular}{|cc|ccc|cc|cc|c|}
\hline
    & &\multicolumn{5}{c|}{CPU time} & \multicolumn{3}{c|}{RAM} \\
    \multicolumn{2}{|c|}{Framework} &  \multicolumn{3}{c}{GKBA} & \multicolumn{2}{c|}{G1--G2} & \multicolumn{2}{c}{GKBA} & G1--G2 \\
\hline
    \multicolumn{2}{|c|}{Approximation} & SOA(d) & SOA & $GW$ & SOA/$GW^{(*)}$ & DSL & SOA & $GW$ & All\\
\hline
    1D & & $N_t^2N_x\ln{N_x}$ & $N_t^2 N_x^3$ & $N_t^3N_x\ln{N_x}$ & $N_tN_x^3$ & $N_tN_x^4$ & $N_tN_x$ & $N_t^2N_x$ & $N_x^3$\\
\hline
    \multirow{2}{*}{2D} & isotropic   & $N_t^2N_x\ln N_x$ & $N_t^2 N_x^5$ & $N_t^3N_x\ln N_x$ & $N_tN_x^5$ & $N_tN_x^7$ & $N_tN_x$ & $N_t^2N_x$ & $N_x^5$ \\
                        & anisotropic & $N_t^2N_x^2\ln N_x$ & $N_t^2N_x^6$ & $N_t^3N_x^2\ln N_x$ & $N_tN_x^6$ & $N_tN_x^8$ & $N_tN_x^2$ & $N_t^2N_x^2$ & $N_x^6$\\
\hline
    \multirow{3}{*}{3D} & isotropic   & $N_t^2N_x\ln{N_x}$ & $N_t^2N_x^6$ & $N_t^3N_x\ln{N_x}$ & $N_tN_x^6$ & $N_tN_x^9$ & $N_tN_x$ & $N_t^2N_x$ & $N_x^6$ \\
                        & cylindric   & $N_t^2N_x^2\ln N_x$ & $N_t^2N_x^8$ & $N_t^2N_x^2\ln N_x$ & $N_tN_x^8$ & $N_tN_x^{11}$ & $N_tN_x^2$ & $N_t^2N_x^2$ & $N_x^8$ \\
                        & anisotropic & $N_t^2N_x^3\ln{N_x}$ & $N_t^2N_x^9$ & $N_t^3N_x^3\ln{N_x}$ & $N_tN_x^9$ & $N_tN_x^{12}$ & $N_tN_x^3$ & $N_t^2N_x^3$ & $N_x^9$ \\
\hline
\end{tabular}
\caption{Orders ($\mathcal{O}$) of the numerical scalings of the CPU time and RAM consumption simulations of uniform systems with $w_\mathbf{kpq} \equiv w_\mathbf{q}$ in different dimensions and symmetries. `GKBA' denotes the HF-GKBA within the standard non-Markovian formalism. `SOA(d)' denotes scalings of the direct SOA diagram, while `SOA' also includes the exchange diagram. G1--G2 scalings for both types are identical. 'GKBA(d)': CPU scalings are based on Fast Fourier Transform techniques used for the efficient computation of convolutions. Scalings are based on a Cartesian grid with $N_x$ grid points per axis. The number of time steps is denoted $N_t$. $(*)$ For more general interaction matrices, $w_\mathbf{kpq}$, cf. Sec.~\ref{sss:uniform-lattice}, the DSL columns remain unchanged, whereas $GW$ in the G1--G2 scheme exhibits the same scalings as DSL. 
}
    \label{tab:SOAScalings}
\end{table}

\subsubsection{Spatially uniform continuous systems}\label{sss:uniform-1d}
The equations of motion, in the G1--G2 scheme, for uniform systems,  are of the form
\begin{align}
    \mathrm{i}\hbar\frac{\text{d}}{\mathrm{d}t}G^\gtrless_{\mathbf{p}\alpha}(t)=[I+I^\dagger]_{\mathbf{p}\alpha}(t),\qquad I_{\mathbf{p}\alpha}(t)=\pm\mathrm{i}\hbar  Z_\alpha\sum\limits_{\mathbf{kq},\beta}Z_\beta w_{\mathbf{q}}(t)\,\mathcal{G}^{\beta\alpha}_{\mathbf{kpq}}(t),
\end{align}
for the single-particle Green function, whereas the $\mathcal{G}$ equation within the $GW$ approximation is given by\footnote{The usual $GW$ approximation uses the non-antisymmetrized source term. We give here the antisymmetrized variant since it appears in the SOA and TMA.},
\begin{align}
    \mathrm{i}\hbar \frac{\text{d}}{\text{d}t}\mathcal{G}^{\alpha\beta}_{\mathbf{kpq}}(t)-\left[h^{\text{HF},(2)},\mathcal{G}\right]_{\mathbf{kpq}}^{\alpha\beta}(t)&=\hat\Psi^{\pm,\alpha\beta}_{\mathbf{kpq}}(t)+\Pi^{\alpha\beta}_{\mathbf{kpq}}(t)\,,
\end{align}
with the definitions 
\begin{align}
    \left[h^{\text{HF},(2)},\mathcal{G}\right]_\mathbf{kpq}^{\alpha\beta}(t)&=\mathcal{G}^{\alpha\beta}_{\mathbf{kpq}}(t)\left[h^\text{HF}_{\mathbf{k}-\mathbf{q},\alpha}(t)+h^\text{HF}_{\mathbf{p}+\mathbf{q},\beta}(t)-h^\text{HF}_{\mathbf{k},\alpha}(t)-h^\text{HF}_{\mathbf{p},\beta}(t)\right]=:\mathcal{G}_{\mathbf{kpq}}^{\alpha\beta}(t)\hbar\omega_{\mathbf{kpq}}^{\alpha\beta}(t)\,,
    \label{eq:2pCommutatorMomRep}
\end{align}
and
\begin{align}\nonumber
    \hat\Psi^{\pm,\alpha\beta}_{\mathbf{kpq}}(t)&=(\mathrm{i}\hbar)^2\left[w_{|\mathbf{q}|}^{\alpha\beta}(t)\pm \delta_{\alpha\beta}w^{\alpha\alpha}_{|\mathbf{k}-\mathbf{p}-\mathbf{q}|}(t)\right]\cdot\left[G^>_{\mathbf{k}-\mathbf{q},\alpha}(t)\,G^>_{\mathbf{p}+\mathbf{q},\beta}(t)\,G^<_{\mathbf{k},\alpha}(t)\,G^<_{\mathbf{p},\beta}(t)-G^<_{\mathbf{k}-\mathbf{q},\alpha}(t)\,G^<_{\mathbf{p}+\mathbf{q},\beta}(t)\,G^>_{\mathbf{k},\alpha}(t)\,G^>_{\mathbf{p},\beta}(t)\right]\,,\\
    \Pi^{\alpha\beta}_{\mathbf{kpq}}(t)&=\pi_{\mathbf{kpq}}^{\alpha\beta}(t)-\left[\pi_{\mathbf{p}+\mathbf{q},\mathbf{k}-\mathbf{q},\mathbf{q}}^{\beta\alpha}(t)\right]^*,\quad\text{where}\quad \pi_{\mathbf{kpq}}^{\alpha\beta}=(\pm)_\beta(\mathrm{i}\hbar)^2\left[G^>_{\mathbf{p}+\mathbf{q},\beta}(t)\,G^<_{\mathbf{p},\beta}(t)-G^<_{\mathbf{p}+\mathbf{q},\beta}(t)\,G^>_{\mathbf{p},\beta}(t)\right]\sum\limits_{\mathbf{p}'\gamma}w_{|\mathbf{q}|}^{\alpha\gamma}(t)\,\mathcal{G}^{\alpha\gamma}_{\mathbf{k}\mathbf{p}'\mathbf{q}}(t)\,,
    \nonumber
\end{align}
where $\hat\Psi^{\pm,\alpha\beta}_{\mathbf{kpq}}$ is the momentum representation of $\hat\Psi^{\pm}_{ijkl}$, cf. Eq.~(\ref{eq:G1-G2_soa_delta}).
It should be noted that the two $T$-matrix approximations can also be evaluated but, for the momentum representation, they are more expensive than $GW$ and SOA by one order in $N_\mathrm{b}$.
\begin{figure}[t]
    \centering
    \includegraphics[width=0.8\textwidth]{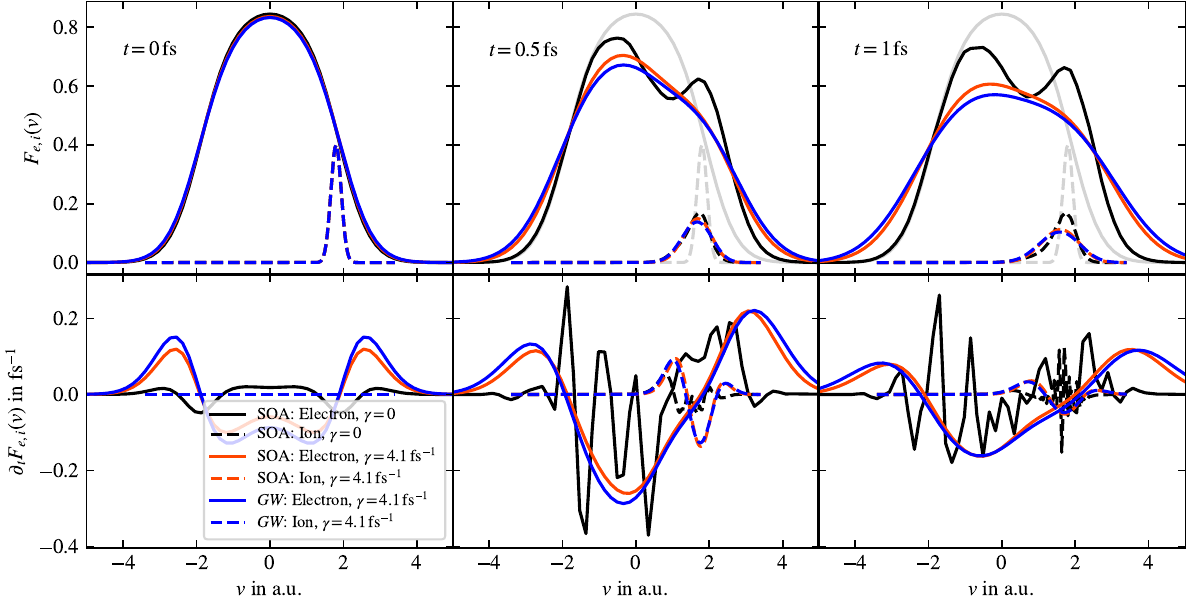}
    \caption{Snapshots of the dynamics of a quasi-1D two-component spin-symmetric electron-ion plasma corresponding to energy dissipation of a nearly monoenergetic ion beam. Shown is the time evolution of the velocity distribution functions, $F_{e,i}(v)=\mathrm{Im}G_{m_{e,i}\cdot v}$ for two cases: the statically screened Second Order Born approximation (SOA), and the dynamically screened $GW$ Approximation.
    Top row: distribution functions. Bottom row: time derivatives.
    Up to $t=0$ we performed an adiabatic switching of e-e and i-i, correlations. At $t=0$, we switched on the e-i interaction. $\gamma$ is an additional damping parameter in the HF-GKBA, the usual HF-GKBA corresponds to $\gamma=0$. The oscillations in the derivative at $t=0.5\,\mathrm{fs}$ and $t=1\,\mathrm{fs}$ are due to aliasing and rapidly change  in amplitude and sign over time. As a consequence, thermalization is artificially halted in a situation where the electronic distribution has two peaks. In contrast, in the damped system, with $\gamma=4.1\,\mathrm{fs}^{-1}$, thermalization does continue. The light gray lines mark the initial distribution. Momentum grid spacing is given by $\frac{1}{6}\hbar a_B^{-1}$, and $N=201$ basis vectors ($\sim 14\,\mathrm{GB}$ in RK4) were used for the discretized 1D-momentum space.}
    \label{fig:1DSim}
\end{figure}

The major contributions to the dynamics come from the SOA terms. There, one can show that
\begin{align}
    \mathcal{G}_\mathbf{kpq}^{\alpha\beta}(t)=\mathcal{G}_\mathbf{kpq}^{\mathrm{EQ},\alpha\beta}[G^\gtrless(t)]+\Delta \mathcal{G}_\mathbf{kpq}^{\alpha\beta}(t)\exp\left(-i\omega_\mathbf{kpq}^{\alpha\beta}t\right)\,,
\end{align}
where $\mathcal{G}^{\mathrm{EQ}}[G^\gtrless(t)]$ describes equilibrium correlations and $\mathcal{G}^\mathrm{EQ}$ and $\Delta \mathcal{G}(t)$ are functions that evolve on the same time scale as the distribution function $G^\gtrless$. For large $t$, a small variation of $\mathbf{k},\mathbf{p},\mathbf{q}$ leads to a significant variation of the phase factor, $\exp (-i\omega_{\mathbf{kpq}}^{\alpha\beta}t)$. At long simulation times, these oscillations become more and more dense. At some point, they cannot be properly resolved anymore and so-called aliasing appears, consequently the simulations become unreliable. We will discuss this problem and how to deal with it in Fig.~\ref{fig:1DSim}.

In many physical situations, the Markov limit gives a good approximate description of the behavior. The underlying assumptions are \cite{bonitz_qkt,bonitz_cpp18} that  $\mathcal{G}$ evolves at least an order of magnitude faster than $G$, i.e. the correlation time is much shorter than the relaxation time \cite{bonitz96pla,bonitz-etal.96pla}, and that a sufficiently long propagation time has passed. Then one finds that only `on-shell' collisions effectively contribute, i.e. such collisions where single-particle energies of the two-colliding particles are conserved. This is expressed by an energy-$\delta$-function in the Markovian collision integral (SOA in the following, and neglecting HF-contributions in the GKBA),
\begin{align}
    I_{\mathbf{p}\alpha}(t)&=\frac{2}{\hbar}\sum\limits_{\beta} 
    \int\frac{\text{d}\mathbf{p}_2}{(2\pi\hbar)^d}\int\frac{\text{d}\mathbf{q}}{(2\pi\hbar)^d}w_{\mathbf{q}}\left[w_\mathbf{q}\pm \delta_{\alpha\beta}w_{\mathbf{p}-\mathbf{p}_2-\mathbf{q}}\right]\delta\left[\mathbf{q}\cdot\left(\mathbf{v}_2-\mathbf{v}\right)+\frac{m_\alpha+m_\beta}{m_\alpha m_\beta}q^2\right] \times\nonumber\\
    &\quad \left[G^<_{\mathbf{p}+\mathbf{q},\alpha}(t)\,G^<_{\mathbf{p}_2-\mathbf{q},\beta}(t)\,G^>_{\mathbf{p},\alpha}(t)\,G^>_{\mathbf{p}_2,\beta}(t)-G^>_{\mathbf{p}+\mathbf{q},\alpha}(t)\,G^>_{\mathbf{p}_2-\mathbf{q},\beta}(t)\,G^<_{\mathbf{p},\alpha}(t)\,G^<_{\mathbf{p}_2,\beta}(t)\right]\,.\label{eq:MarkovCollInt}
\end{align}
First of all, we are mostly interested in the small-$\mathbf{q}$ region, since $w_\mathbf{q}$ is peaked around $\mathbf{q}=0$. In this limit, the condition in the $\delta$-function can be approximated as $\mathbf{q}\cdot (\mathbf{v}_2-\mathbf{v})=0$, where $\textbf{v}=\textbf{p}/m$, is the velocity. While in higher dimensions, many combinations of $\mathbf{q},\mathbf{v}_2,\mathbf{v}$ can fulfill this condition and, therefore, lead to an effective scattering, in 1D it can only be fulfilled at $\mathbf{v}_2=\mathbf{v}.$ This rather strict condition significantly slows down the thermalization of 1D gases. 
In a full nonj-Markovian simulation,  delta-function-like behavior appears only in the long-time limit. Before that, it is a cardinal sine, i.e. collisions slightly violating single-particle energy conservation (``off-shell collisions'') are not prohibited. As long as the $\delta$-function is sufficiently broadened, the numerical treatment on a momentum grid is well-conditioned. On the other hand, on long time scales, the numerical treatment will fail  to resolve the $\delta$-function adequately, yielding propagation errors. In practice, these errors slow down the dynamics in an un-physical way. This is, of course, just another formulation of the aliasing problem.

We tested the G1--G2 scheme for a uniform two-component quasi 1D-system, where the particles are trapped in a radial confinement potential and can only freely move in one dimension. The interaction matrix elements $w_q=\langle k-q,p+q|\hat{w}|k,p\rangle$ sensitively depend on the geometry of the confinement. As an example, we considered a high-density quantum plasma with parameters $r_s=\bar r/A_B = 0.5$ and $\Theta=k_BT/E_F=1.62$, that lie on compression path inertial confinement fusion plasmas \cite{gomez_experimental_2014}. To realize a quasi-1D simulation with realistic parameters we chose, a strong harmonic confinement with a radius of $1\, a_B$ in order to guarantee that only the radial ground state of the trap is occupied. As initial condition we chose an equilibrium electron distribution that is disturbed by a nearly monoenergetic ion beam leading to beam stopping, energy dissipation and heating of the electrons which is demonstrated in Fig.~\ref{fig:1DSim}.
We implemented, both, the Second-Order Approximation and the $GW$ Approximation.
For the SOA, we used a statically screened interaction potential,  where the screening has been fitted to the 1D-Lindhard formula, the static and long-wavelength limit of the $GW$ screening, cf. Ref. \cite{makait_cpp_23} for more details about the fit. Figure \ref{fig:1DSim} shows that the time scale of the dynamics of such a statically screened SOA reproduces that of the fully dynamical $GW$A very well.
%
In the initial stage, many collisions between electrons and ions occur, generating a two-peak distribution function of the electrons. This two-peak distribution then does not thermalize further, as a result of aliasing.
In the same figure we demonstrate that aliasing can be removed and thermalization  continues if a small damping constant $\gamma$ is  added to the single-particle energies in the HF-GKBA,
\begin{align}
    h^\mathrm{HF}\longrightarrow h^\mathrm{HF} - i\hbar\gamma\,.
\end{align}
Such a damping constant effectively prevents the $\delta$-function in Eq.~\eqref{eq:MarkovCollInt} from becoming sharp and is, therefore, effective in reducing aliasing. However, such a damping constant and, in particular, its magnitude must be carefully derived from physical considerations, especially since adding such an imaginary energy correction violates total energy conservation \cite{bonitz_qkt,bonitz-etal.99epjb}. More systematic approaches to overcome aliasing will be discussed elsewhere. Despite the model character, this example shows, that the G1--G2 scheme allows to systematically study the short time dynamics in inertial confinement fusion plasmas, in particular, the effect of ignition by high energy ions.

\subsubsection{2D lattice systems
}\label{sss:uniform-lattice}
As a second example of a uniform system, we now consider two-dimensional lattice systems that are of relevance e.g. for atoms in optical lattices or monolayers of graphene or transition metal dichalcogenides (TMDCs).
Optical excitation of graphene, including carrier multiplication and excitonic effects, have been studied by various authors using density matrix methods or NEGF, e.g. \cite{WinzerCarrierMult2010,winzer_phd_13,perfetto_prl_22} and references therein.
The purpose of our analyis is to test the capabilities of the G1--G2 scheme for simulating optical excitation of 2D materials taking into account details of the band structure. We discuss the relation to earlier investigations at the end of the section.

The uniformity of macroscopic lattice systems means that all quantities are invariant with respect to simultaneous translation of all coordinates by the same lattice vector $\textbf{R}$: $f(\mathbf{r}_1+\mathbf{R},\mathbf{r}_2+\mathbf{R},...,\mathbf{r}_N+\mathbf{R})=f(\mathbf{r}_1,\mathbf{r}_2,...,\mathbf{r}_N).$ The properties of the system depend on the lattice geometry. Here we sketch the fundamentals regarding a 2D-hexagonal lattice as found in macroscopic graphene,
\begin{figure}[t]
    \centering
    \includegraphics[width=0.4\textwidth]{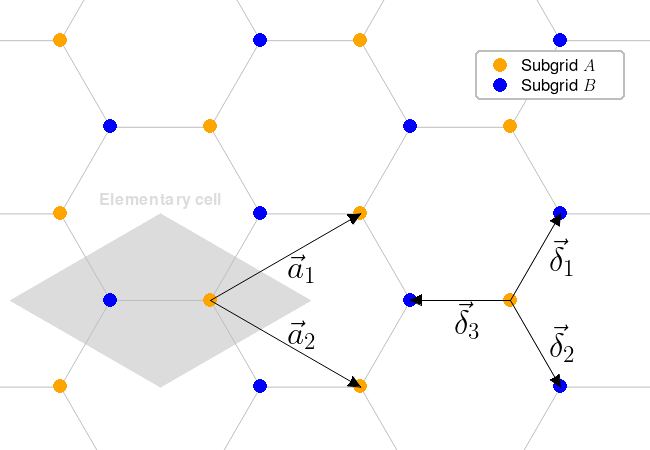}
    \includegraphics[width=0.5\textwidth, trim={0cm 0cm 0cm 0cm}, clip]{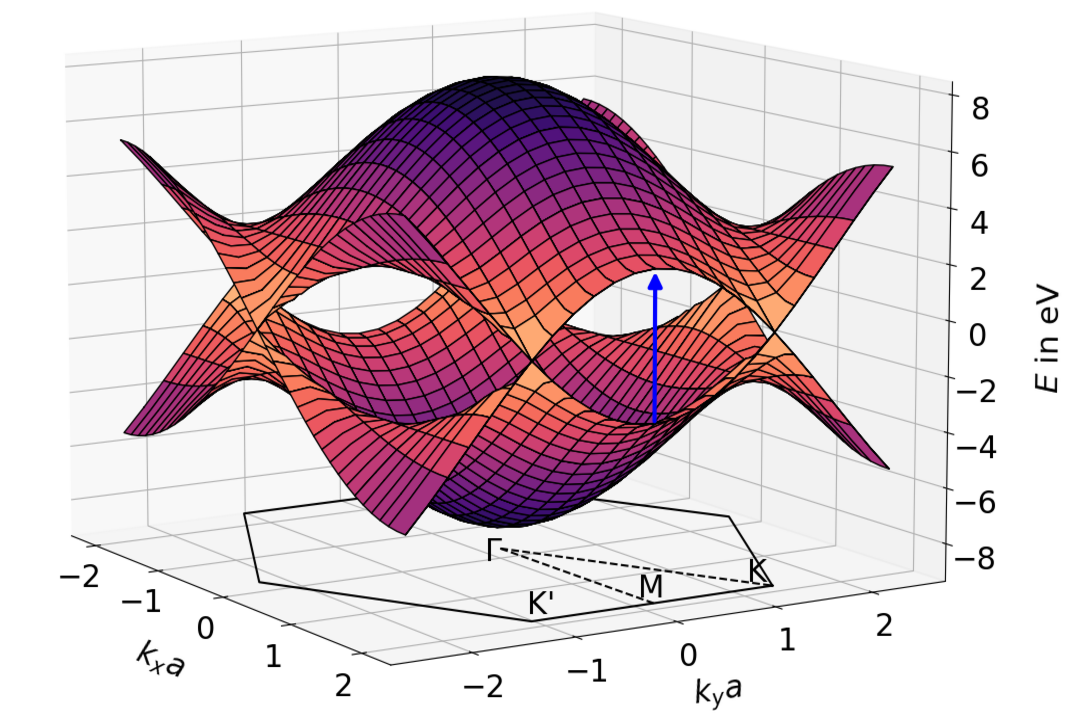}
    \caption{Left: Geometry and definitions of the vectors describing a graphene lattice. The two sublattices are marked by different colors. Right: Tight-Binding dispersion of graphene. We simulated the optical excitation of the M-point region and successive thermalization afterwards.}
    \label{fig:GrapheneLattice}
\end{figure}
where there are two atoms per unit cell. The tight-binding analysis of this lattice with only nearest neighbor hoppings included is typically done by separating the lattice into two sublattices, `A' and `B', cf. Fig.  \ref{fig:GrapheneLattice}, where `A' sites only couple to `B' sites and vice versa. Diagonalizing the Hamiltonian, one finds the plane wave states,
\begin{align}
    |\mathbf{k}\sigma\lambda\rangle = \frac{\Delta_\mathbf{k}}{\sqrt{2}|\Delta_{\mathbf{k}}|}|\,\mathbf{k}\sigma,A\,\rangle-\sigma_\lambda \frac{1}{\sqrt{2}}|\,\mathbf{k}\sigma,B\,\rangle\,,\label{eq:HoneycombStates}
\end{align}
where $|\mathbf{k}\sigma,A\rangle$ and $|\mathbf{k}\sigma,B\rangle$ are plane waves on the corresponding sublattice,  $\sigma_\lambda\in\{-1,+1\}$, and the $\mathbf{k}$-dependent phase shift between the two waves is given by $\Delta_\mathbf{k}\equiv\sum\limits_i e^{i\mathbf{k}\cdot\delta_i}$, cf. Fig. \ref{fig:GrapheneLattice} for the definition of the $\delta_i$'s. The corresponding tight-binding eigenenergies are given by
\begin{align}
    \epsilon^\lambda_\mathbf{k}=\sigma_\lambda t\sqrt{1+4\cos\left(\frac{3}{2}k_xa\right)\cos\left(\frac{\sqrt{3}}{2}k_ya\right)+4\cos^2\left(\frac{\sqrt{3}}{2}k_ya\right)}\,,
\end{align}
and are shown in Fig.~\ref{fig:GrapheneLattice}. Since in graphene in the ground state, only the states with negative energy are occupied, we will refer to the states with $\sigma_\lambda=-1$ as valence band ($v$) and to the states with $\sigma_\lambda=+1$ as conduction band ($c$). 

In the following, our goal is study the response of this system to a short laser pulse, by means of the G1--G2 scheme. To this end we need to specify the single-particle Hamiltonian and the field-matter interaction matrix in this basis.
By construction, the unperturbed single-particle Hamiltonian is diagonal in the momentum and spin indices. Additionally, we consider optical excitations via a time-dependent uniform electric field which we describe via a vector potential,  $\mathbf{A}(t)$\cite{Scully_Zubairy_1997,winzer_phd_13}, thus, the single-particle matrix element becomes\footnote{These matrix elements are valid for low to moderate field strength.}
\begin{align}
    \hat{h}^{(1)}_{\mathbf{k}\sigma\lambda\lambda'}=\epsilon_{\mathbf{k}}^\lambda\delta_{\lambda\lambda'}+
    \hbar \Omega_\mathbf{k}^{\lambda\lambda'}(t)
    \,,\qquad \Omega_\mathbf{k}^{\lambda\lambda'}(t)=i\frac{e}{m}\bm{M}_{\bm{k}}^{\lambda,\lambda'}\cdot\bm{A}(t)\,,\qquad 3\mathbf{E}=-\frac{\partial \mathbf{A}}{\partial t}\,.
\end{align}
Here $\Omega$ is the Rabi frequency which contains the intraband and interband momentum matrix elements
\begin{align}\label{eq:mvc}
    \bm{M}_{\bm{k}}^{v,v}&=-i\frac{M}{|\Delta_{\bm{k}}|}\Im\left[\Delta_{\bm{k}}^*\sum_{i=1}^3 e^{i\bm{k}\cdot\bm{\delta}_i}\frac{\bm{\delta}_i}{|\bm{\delta}_i|}\right]=-\bm{M}_{\bm{k}}^{c,c}\,,\\
    \bm{M}_{\bm{k}}^{v,c}&=-\frac{M}{|\Delta_{\bm{k}}|}\Re\left[\Delta_{\bm{k}}^*\sum_{i=1}^3 e^{i\bm{k}\cdot\bm{\delta}_i}\frac{\bm{\delta}_i}{|\bm{\delta}_i|}\right]=-\bm{M}_{\bm{k}}^{c,v}\,,
\end{align}
where the remaining parameter is chosen to be $M\approx 3.7\,\mathrm{nm}^{-1}$, cf. Ref. \cite{winzer_phd_13}. Note that we omitted the $A^2(t)$ term in the Hamiltonian since it vanishes in the G1-G2 equations of a uniform system. Obviously, light-matter coupling is only effective if the scalar product, $\mathbf{M}^{v,c}_\mathbf{k}\cdot\mathbf{A}$, is large. The interband matrix element, $\mathbf{M}^{v,c}$, is shown in Fig. \ref{fig:GrapheneLattice} by the green arrows, from which one can easily deduce which polarization of light is necessary to excite   electrons in a given momentum state. 

We now go beyond the tight-binding model and include interaction effects on the level of the Hubbard model, as introduced in Eq.~\eqref{eq:h-hubbard}. The matrix elements are easily evaluated in momentum basis, due to the diagonality of the interaction in the lattice basis, which yields
\begin{align}
    w_{\mathbf{kpq},\alpha\bar\alpha}^s=\frac{1}{N_xN_y}\frac{U}{4}\left(\frac{\Delta^*_{\mathbf{k}-\mathbf{q}}\Delta^*_{\mathbf{p}+\mathbf{q}}\Delta_{\mathbf{k}}\Delta_{\mathbf{p}}}{|\Delta^*_{\mathbf{k}-\mathbf{q}}\Delta^*_{\mathbf{p}+\mathbf{q}}\Delta_{\mathbf{k}}\Delta_{\mathbf{p}}|}+s\right),
\end{align}
and $s$ is the product of the signs of the four states involved. For the simulation of graphene, we use the parameter $U=1.6~J$, cf. Ref.~\cite{PhysRevLett.111.036601}. In particular, the interaction matrix elements differ strongly depending on whether an Auger-type ($s=-1$) or intraband ($s=+1$) collision is described. The spin indices $\alpha,\bar\alpha$ indicate that the interaction matrix is finite only for two electrons with opposite spin projection.


With these input parameters, it is straightforward to derive the G1--G2 Bloch equations. The procedure is the same as in the case of the semiconductor Bloch equations, cf. Sec.~\ref{sss:eh-plasma}, with the main difference being the special lattice geometry of graphene.  To this end, the G1--G2 equations \eqref{eq:eom_gone} and \eqref{eq:G1-G2_soa}  with electron-electron interactions in Second order Born approximation (SOA) are expanded in a $2\times 2$ band basis
\begin{align}
    i\hbar\frac{\text{d}}{\text{d}t}G^\gtrless_{\bm{k}, cc}(t)&=2\hbar\Im\left[(\Omega_{\bm{k}}^{v,c})^*G^\gtrless_{\bm{k}, vc}(t)\right]+\left(\bm{I}(t)+\bm{I}^\dagger(t)\right)_{\bm{k}, cc}\,,\\\nonumber
    i\hbar\frac{\text{d}}{\text{d}t}G^\gtrless_{\bm{k}, vv}(t)&=-2\hbar\Im\left[(\Omega_{\bm{k}}^{v,c})^*G^\gtrless_{\bm{k}, vc}(t)\right]+\left(\bm{I}(t)+\bm{I}^\dagger(t)\right)_{\bm{k}, vv}\,,\\
    i\hbar\frac{\text{d}}{\text{d}t}G^\gtrless_{\bm{k}, vc}(t)&=\left[\epsilon_{\bm{k}}^v-\epsilon_{\bm{k}}^c + 2\hbar\Omega_{\bm{k}}^{v,v} \right]G^\gtrless_{\bm{k}, vc}(t)+\hbar\Omega_{\bm{k}}^{v,c}(G^\gtrless_{\bm{k}, cc}(t)-G^\gtrless_{\bm{k}, vv}(t))+\left(\bm{I}(t)+\bm{I}^\dagger(t)\right)_{\bm{k}, vc}\,,\\
    i\hbar\frac{\text{d}}{\text{d}t}\mathcal{G}_{\bm{k}\bm{p}\bm{q}}^{\lambda_1\lambda_2\lambda_1'\lambda_2'}(t)&=\left\{\epsilon_{\bm{k}-\bm{q}}^{\lambda_1}+\epsilon_{\bm{p}+\bm{q}}^{\lambda_2}-\epsilon_{\bm{k}}^{\lambda_1'}-\epsilon_{\bm{p}}^{\lambda_2'}\right\}\mathcal{G}_{\bm{k}\bm{p}\bm{q}}^{\lambda_1\lambda_2\lambda_1'\lambda_2'}+\Psi_{\bm{k}\bm{p}\bm{q}}^{\lambda_1\lambda_2\lambda_1'\lambda_2'}
    \,,\nonumber\\
    &\hspace{1cm}+\hbar \sum_{\lambda}\left\{\Omega_{\bm{k}-\bm{q}}^{\lambda_1,\lambda}\mathcal{G}_{\bm{k}\bm{p}\bm{q}}^{\lambda,\lambda_2,\lambda'_1,\lambda_2'} + \Omega_{\bm{p}+\bm{q}}^{\lambda_2,\lambda}\mathcal{G}_{\bm{k}\bm{p}\bm{q}}^{\lambda_1,\lambda,\lambda_1',\lambda_2'} - \mathcal{G}_{\bm{k}\bm{p}\bm{q}}^{\lambda_1,\lambda_2,\lambda,\lambda_2'}\Omega_{\bm{k}}^{\lambda,\lambda_1'} - \mathcal{G}_{\bm{k}\bm{p}\bm{q}}^{\lambda_1,\lambda_2,\lambda_1',\lambda}\Omega_{\bm{p}}^{\lambda,\lambda_2'}\right\}\,. \label{eq:blochg2}
\end{align}
Since we assume spin symmetry in our simulation, we left out the spin indices. Also note that the function $\mathcal{G}$ in Eq.~\eqref{eq:blochg2} corresponds to the $\uparrow\downarrow\uparrow\downarrow$ spin component of the $\mathcal{G}$ matrix, which is the only non-vanishing component. 

In contrast to our G1--G2 analysis of the uniform gas in Sec.~\ref{sss:uniform-1d}, the collision integral $\mathbf{I}$ and the two-particle source term, $\Psi$, now also contain the full matrix product structure with regard to the band indices: 
\begin{align}
    I_{\mathbf{k},\lambda\lambda'}&=\pm\i\hbar \sum\limits_{{\mathbf{p}},\mathbf{q}\in 1\text{st BZ}\atop \nu,\nu',\bar{\nu}\in\{v,c\}}{\left[w^{\sigma_{\nu'}\cdot \sigma_{\bar{\nu}}\cdot \sigma_\lambda\cdot \sigma_\nu}_{\mathbf{kpq}}\right]}^*\mathcal{G}^{{\nu'}{\bar{\nu}}\lambda'\nu}_{\mathbf{kpq}}(t)\,,\\
    \Psi_{\mathbf{kpq}}^{\lambda_1\lambda_2,\lambda_1'\lambda_2'}&=(\i\hbar)^2\sum\limits_{\nu_1,\nu_2,\nu_1',\nu_2'\in\{v,c\}}G^>_{\mathbf{k}-\mathbf{q},\lambda_1\nu_1}(t)\,G^>_{\mathbf{p}+\mathbf{q},\lambda_2\nu_2}(t)w^{\sigma_{\nu_1}\cdot\sigma_{\nu_2}\cdot\sigma_{\nu_1'}\cdot\sigma_{\nu_2'}}_{\mathbf{kpq}}G^<_{\mathbf{k},\nu_1'\lambda_1'}(t)\,G^<_{\mathbf{p},\nu_2'\lambda_2'}(t)-(>\leftrightarrow <)\,.
\end{align}
\begin{figure}[t]
    \centering
    \includegraphics[width=0.45\textwidth]{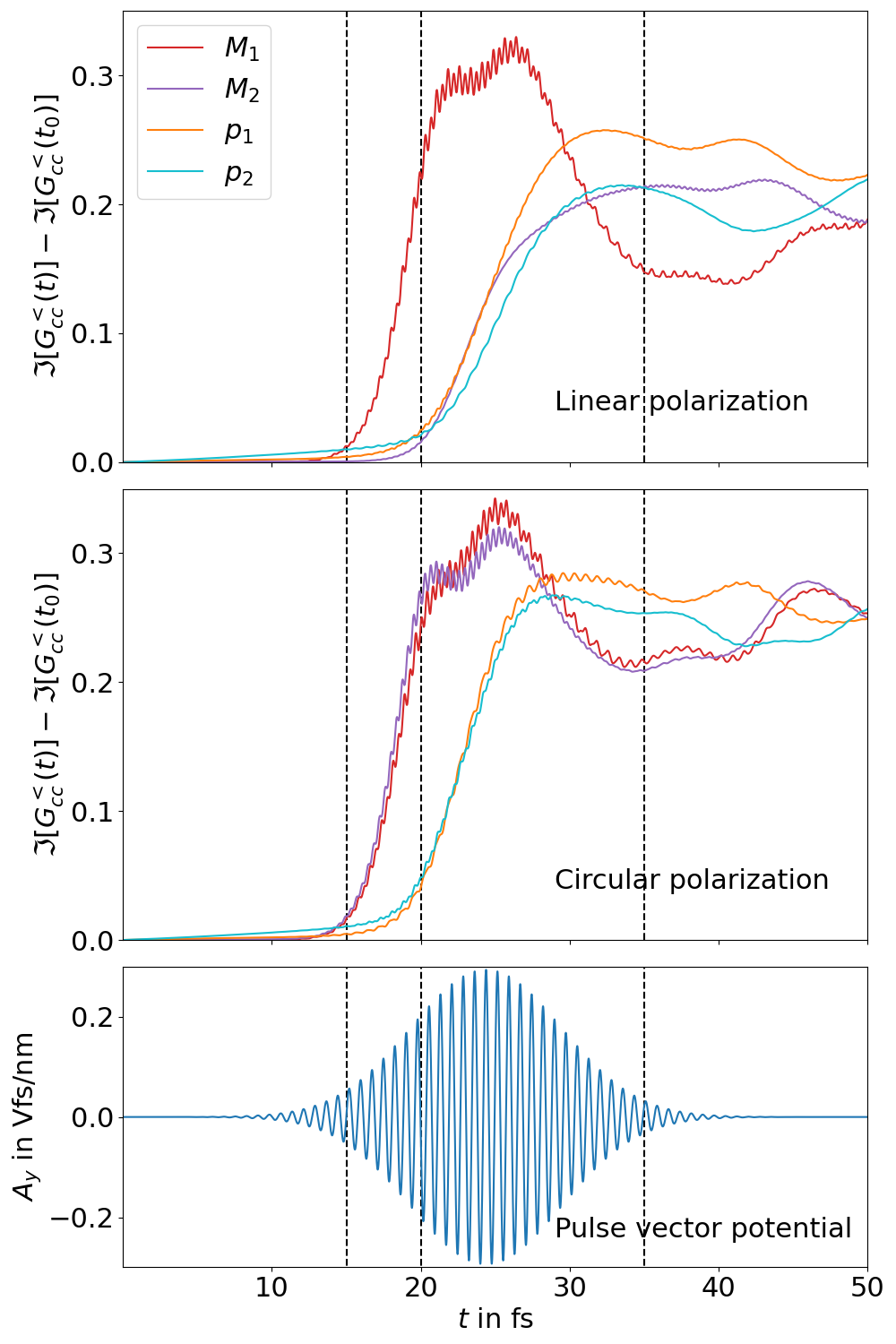}
    \includegraphics[width=0.35\textwidth, trim={0cm -3.5cm 2cm 0cm}, clip]{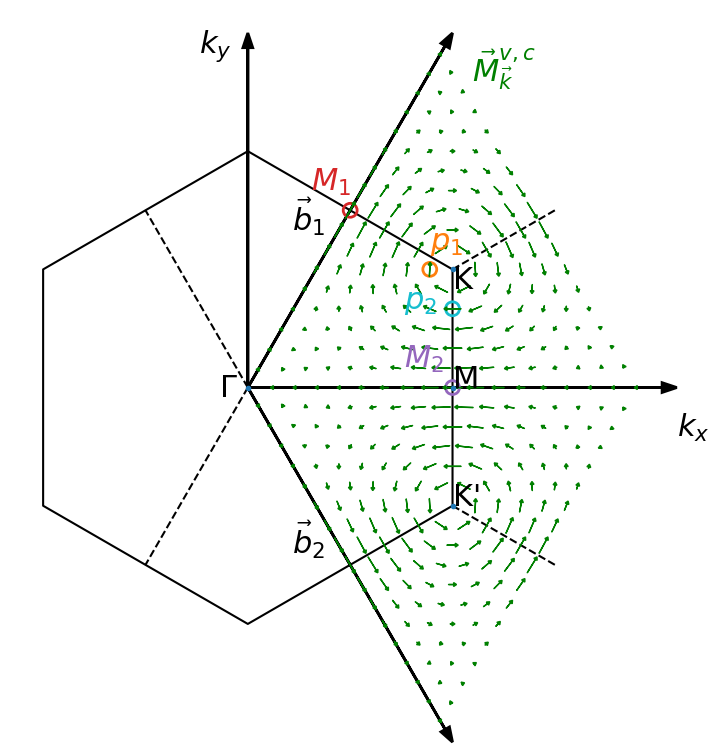}
    \caption{Left: time-dependent carrier density in the conduction band during and after the laser pulse
    , for four momenta that are shown in the right figure by the colored circles. Top figure: linearly polarized laser pulse. Middle figure: Circularly polarized light. Bottom: amplitude of the laser pulse where the three vertical lines are the time points shown in Fig.~\ref{fig:GrapheneLaserHeatmaps}. In both cases, the M-point is excited first, and the other $\mathbf{k}$-points are excited subsequently, via Auger collisions and carrier multiplication. Right: k-point mesh (green) used in the simulation: the rhomboidal variant of the first Brillouin zone is discretized by dividing up the reciprocal lattice vectors $\vec{b}_1$ and $\vec{b}_2$ into $N_x=N_y=18$ equidistant points. Green arrows: vectors of $\mathbf{M}^{v,c}_\mathbf{k}$, Eq.~\eqref{eq:mvc}. Light-matter interaction in a given $\mathbf{k}$-point is proportional to the scalar product of the electric field and $\mathbf{M}_\mathbf{k}^{v,c}$. 
}   \label{fig:grapheneSpecificKPoints}
\end{figure}
We note that the G1--G2 approach has been recently used with GW selfenergies to simulate carrier multiplication processes \cite{perfetto_prl_22}. However, the authors considered excitation of electrons in the vicinity of the Dirac points. Here, we focus on G1--G2 simulations of the whole Brillouin zone, fully including the nonlinear energy dispersion far away from the Dirac points. A similar analysis was performed in Ref.~\cite{PhysRevB.94.195438}, however, in the frame of a simple rate equation model.
On the technical side, we used the rhomboidal form of the first Brillouin zone and discretized it in a straightforward manner: we divided the reciprocal lattice vectors into $N_x$ equidistant points. The RAM scalings from Table \ref{tab:SOAScalings} in the 2D anisotropic row also apply here, showing that the required RAM scales as $\sim N_x^6$. In order to fit the program into a GPU, we had to restrict ourselves to $N_x=18$, which corresponds to a $\mathbf{k}$-point mesh of $18\times 18=324$ $\mathbf{k}$-points, for the first Brillouin zone, and $36\,\mathrm{GB}$ of memory on a GPU. 

\begin{figure}[t]
    \centering
    \includegraphics[width=0.9\textwidth]{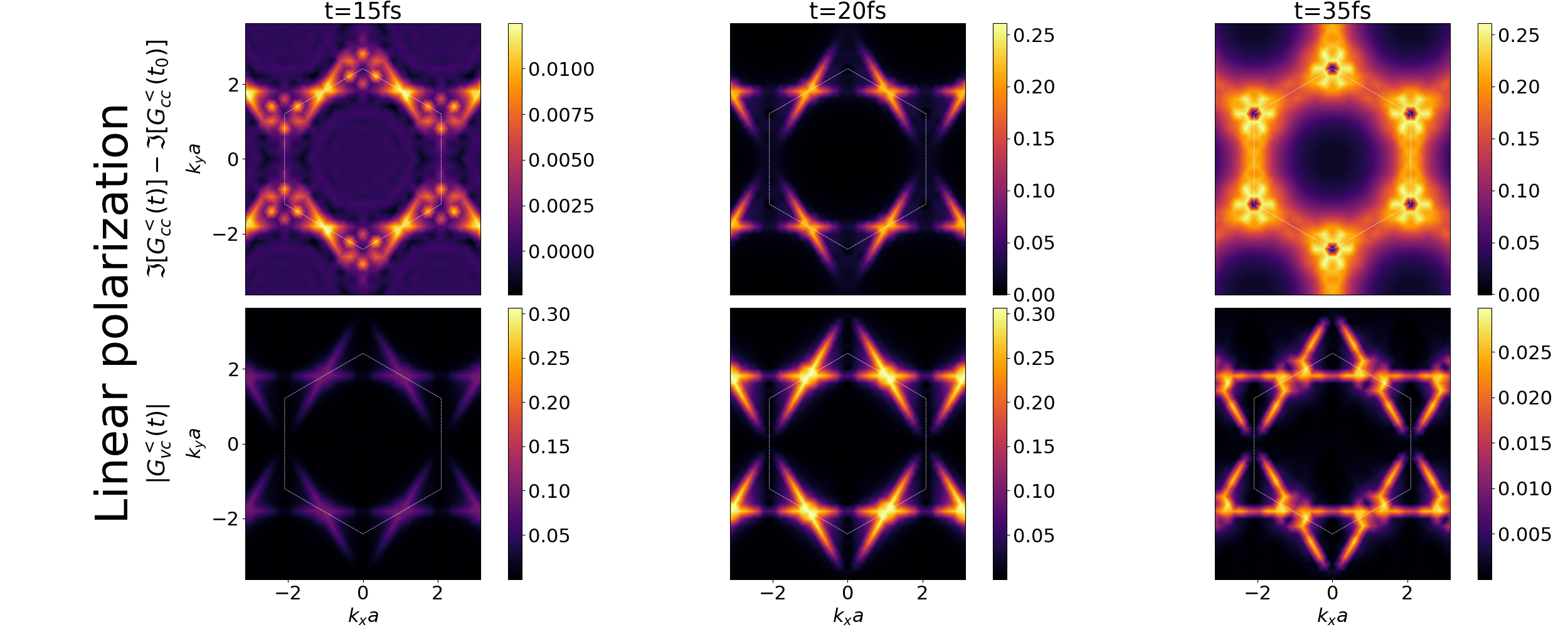}\\[2ex]
    \includegraphics[width=0.9\textwidth]{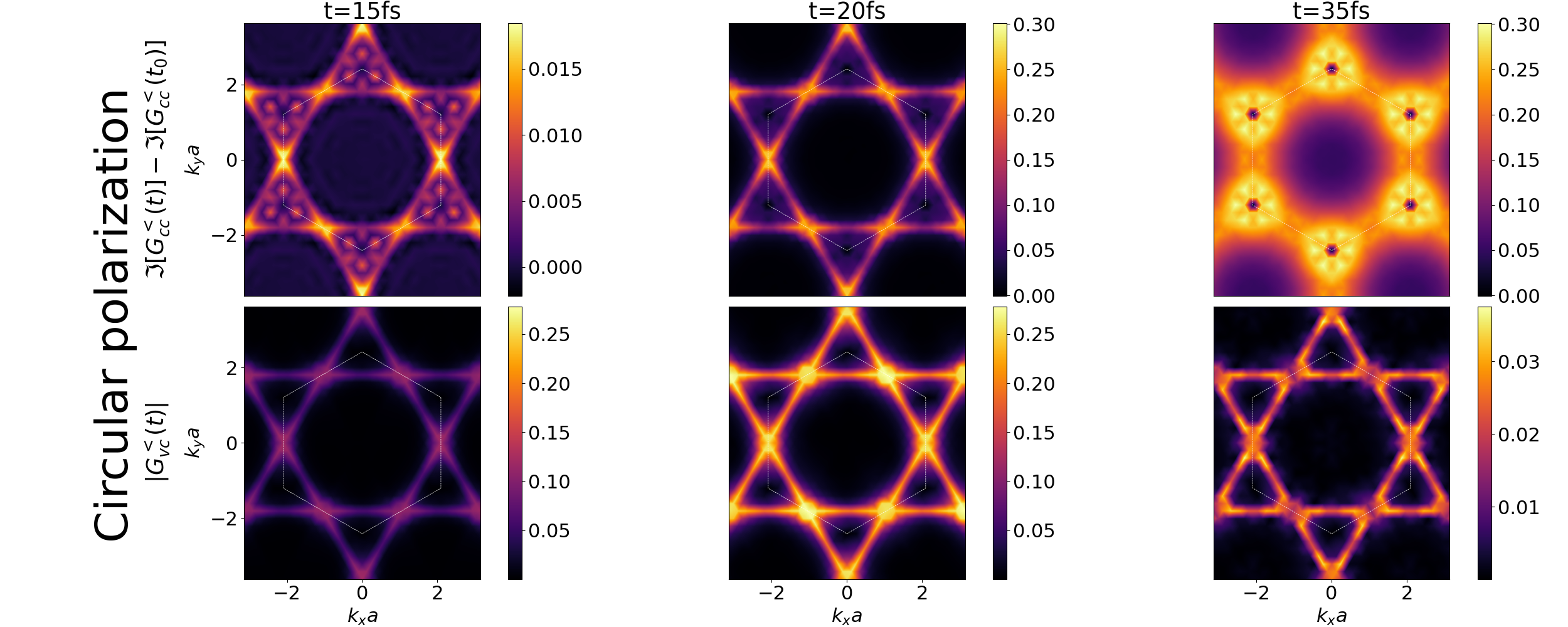}
    \caption{Optical excitation resonant at the M-points ($\lambda=230\,\mathrm{nm}$) with pulse duration $\Delta_t=10\,\text{fs}$, pulse maximum at $24.38\,\text{fs}$ and vector potential amplitude $A_0=0.3\frac{\text{Vfs}}{\text{nm}}$, cf. Fig.~\ref{fig:grapheneSpecificKPoints}. Top two rows:
    linear polarization in $y$-direction. Bottom two rows: circular polarization. Initially only areas with non-vanishing $\mathbf{M}_\mathbf{k}^{v,c}\cdot \mathbf{A}$ are excited, cf. Fig. \ref{fig:GrapheneLattice}. As a consequence, one of the M-points cannot be excited by the y-polarized laser, whereas the circularly polarized laser excites all M-points. }
    \label{fig:GrapheneLaserHeatmaps}
\end{figure}

A major advantage of the G1-G2 scheme is that long simulations are possible, because only the current time step has to be saved. In the examples studied below, we used $30\,000$ time steps which allows us to simulate the dynamics over $50~$fs, which includes  adiabatic switching (generation of the correlated initial state), the entire laser pulse duration, and a relaxation phase after the pulse, cf. Fig.~\ref{fig:grapheneSpecificKPoints}. 
After generation of a correlated ground state via adiabatic switching, we apply a laser pulse with a Gaussian envelope of FWHM $\Delta_t$ and an amplitude $A_0$. In Fig. \ref{fig:GrapheneLaserHeatmaps} we show the evolution of the carrier distributions and the inter-band polarization, for linearly and circularly polarized light, for three time points. A wavelength of $\lambda = 230\,\mathrm{nm}$ was chosen since this photon energy directly excites M-point electrons. Comparison of the excited areas of the Brillouin zone (two of the three M points) for the linearly polarized laser to $\mathbf{M}^{c,v}_\mathbf{k},$ cf. Fig. \ref{fig:grapheneSpecificKPoints}, shows that one can directly excite specific $\mathbf{k}$-points (here the M points) by properly aligning the polarization in such a way that $\mathbf{M}^{c,v}_\mathbf{k}\cdot \mathbf{A}$ is only large at desired $\mathbf{k}$-points. In that regard, the circularly polarized laser can excite all electrons that have the correct excitation energy. Our findings are also in agreement with Ref.~\cite{PhysRevB.94.195438}.

On a longer time scale we observe prethermalization: the high-energy electrons at the M points relax towards the Dirac points. Also, with the decay of the laser amplitude, the interband polarization quickly decays. At this point, an increase of the number of electrons in the conduction band is only due to electron-electron scattering, leading to carrier multiplication, cf. Refs.~\cite{WinzerCarrierMult2010,perfetto_prl_22}. This is clearly visible in Fig.~\ref{fig:grapheneSpecificKPoints}, where first only the high-energy M-point region is excited whereas, after the pulse, many low-energy $\mathbf{k}$-points follow while the population of the excited electrons at the M-point decreases.

To summarize, in this section we demonstrated that the G1--G2 scheme can  be efficiently applied to simulate optical excitation of 2D quantum  materials, thereby fully resolving the momentum geometry and following the dynamics over long periods of time. Using a momentum grid that covers not only the region around the Dirac points but the entire Brillouin zone allows for a realistic investigation of the response of 2D quantum materials to laser excitation of varying amplitude, frequency and polarization. In the future it will be interesting to use improved selfenergies, including the GW approximation \cite{perfetto_prl_22} and  DSL which are of comparable computational complexity, within the G1--G2 scheme, cf. Tab.~\ref{tab:SOAScalings}. Another question of current interest is the selective excitation of electrons in a certain valley which can be achieved by suitable laser pulse shapes, e.g.~Refs.~\cite{rost_fd_22,rost_prr_22,mrudul_jpb_21, mrudul_optica_21}. Further, the present G1--G2 analysis of 2D lattice models can be straightforwardly extended to transition metal dichalcogenides where additional effects such as chirality of valley excitons and phonons emerge, e.g.~
\cite{Caruso_ChiralValleyExcitons_2022,Caruso_NanoLett2023}.

%



\subsection{Other applications of the G1--G2 scheme}\label{ss:g1-g2-other}
The tremendous speed-up of NEGF calculations that was caused by the introduction of the G1--G2 scheme quickly sparked new theoretical developments. On the one hand the Nakatsuji--Yasuda~\cite{donsa_prr_23} and Faddeev~\cite{pavlyukh_prb_21} approximations were presented to partially include three-particle electronic correlation effects beyond the DSL approximation. On the other hand the G1--G2 scheme was extended to the description of bosons~\cite{pavlyukh_prb_22,pavlyukh_prbl_22}. Both the electron as well es the boson variant were implemented in the CHEERS code~\cite{pavlyukh2023cheers}. Subsequently, the G1--G2 scheme has been successfully applied to a variety of problems, such as the simulation of ion stopping in monolayers of graphene and MoS$_2$~\cite{borkowski_pss_22,niggas_prl_22}, the photoionization of organic molecules~\cite{pavlyukh_prb_21}, ultrafast electron-boson dynamics~\cite{karlsson_prl21,pavlyukh_time-linear_2022}, ultrafast carrier and optical excitation and exciton dynamics in 2D materials~\cite{perfetto_prl_22,perfetto_nanol_23}, and quantum transport \cite{tuovinen_prl_23}.

\section[Embedding approach]{Embedding approach for open or heterogeneous systems}\label{s:embedding}

For large systems and long simulation times, the amount of main memory needed to store the Green functions quickly becomes a limiting factor for NEGF calculations. For the G1--G2 scheme, the main bottleneck is the system size. In order to overcome this bottleneck, one can apply different embedding concepts which may drastically reduce the memory consumption. Before explaining the basic ideas we point out that embedding approaches are successfully being applied also in another context. 
For the description of uniform (translationally invariant lattice) systems, a very successful method is the formalism of nonequilibrium 
dynamical mean-field theory (DMFT)~\cite{freericks_2006,aoki_2014}, which takes into account only local correlations via a spatially local selfenergy ($\Sigma_{ij}\propto\delta_{ij}$) and has been employed mainly with focus on strongly correlated materials~\cite{vollhardt_2019}.
As a result of the local selfenergy approximation, which becomes exact in the limit of infinite coordination number~\cite{georges_1992}, the lattice KBE can be mapped onto a
local impurity problem in which the impurity site is embedded into an environment described by a two-time hybridization function. Though the obtained impurity problem is in general hard to solve, see e.g. Refs.~\cite{diag_mc_werner_09,eckstein_2010,tsuji_2013,gramsch_2013_hamiltonian} for methods and DMFT applications, the method requires to compute and propagate only local quantities or spatially very restricted NEGF components, when the method is extended to cluster schemes~\cite{tsuji_2014,eckstein_2016}.

On the other hand, for heterogeneous and open systems, which are in focus of the present section, the main idea is different: here it is advantageous to subdivide
the system into a “central” part that is of primary interest and, thus is treated accurately, and a “surroundings” part that is treated on a simpler level. The focus is not on strongly correlated systems so an NEGF treatment with the common selfenergies that were discussed above is sufficient.
Such embedding schemes have been developed and successfully applied in many fields, including quantum chemistry~\cite{WARSHEL1976227}, the statistical theory of open systems~\cite{ness_pre_14}, and plasma-surface interaction~\cite{Bonitz_fcse_19,bronold_jap_22}. In NEGF simulations, the coupling of the central and surroundings part can be described by a two-time embedding selfenergy, which, similarly as the hybridization function in DMFT, is added to the correlation selfenergy of the central subsystem, and allows for an efficient treatment of nonequilibrium problems and ultrafast electron dynamics, 
limiting the computational effort to the solution of the KBE of the central part. Applications of this embedding approach include
quantum transport in nanoscale junctions coupled to macroscopic leads~\cite{khosvari_prb_12,rabani_jcp_13}, the excitation dynamics of excitonic insulators~\cite{tuovinen_prb_20}, the photoionization of atoms in strong laser fields~\cite{perfetto_pra_15}, or the Auger decay in molecules~\cite{covito_pra_18}, for a text book overview, see Ref.~\cite{stefanucci_nonequilibrium_2013}.
A third example is the resonant charge transfer between an ion impacting a solid during which the ion deposits energy (stopping power) and is being (partially) neutralized. Using a simplified description of the ion, the resonant charge transfer could be modeled efficiently \cite{Bonitz_fcse_19,bonitz_pss_18, balzer_cpp_21}. This approach yielded very good agreement with experiments on highly charged ions impacting 2D quantum materials \cite{niggas_prl_22}. 
In the following we briefly recall the embedding concept, starting with its two-time formulation, and then present results for the time-local embedding approach within the G1-G2 scheme which allows one to fully take advantage of the improved CPU time scaling \cite{balzer_prb_23}. Finally, we discuss possible future extensions of the embedding scheme.

\subsection{Embedding selfenergy for two-time NEGF}\label{ss:embedding-2time}
Consider the electron Hamiltonian of the many-body system and subdivide it into a system~(s) part and its ``environment''~(e) [we denote $\Omega=\{\tn{e},\tn{s}\}$ and do not write the spin index explicitly],
\begin{align}
\label{eq.ham}
\hat{H}_{\textup{total}}(t) = &\sum_{\alpha\beta\in\Omega}\sum_{ij}h^{\alpha\beta}_{ij}(t)\hat{c}^{\alpha\dagger}_i\hat{c}^\beta_j
+\frac{1}{2}\sum_{\alpha\beta\gamma\delta\in\Omega}\sum_{ijkl}w^{\alpha\beta\gamma\delta}_{ijkl}\hat{c}^{\alpha\dagger}_i\hat{c}^{\beta\dagger}_j\hat{c}^{\gamma}_k\hat{c}^{\delta}_l\,,
\end{align}
where, the operator $\hat{c}^{\alpha\dagger}_i$ ($\hat{c}^{\alpha}_i$) creates (annihilates) an electron in the state $|i\rangle$ of part $\alpha$. The one-particle Hamiltonian, $h(t)=T+V(t)$, is the sum of kinetic and the (in general, time-dependent) potential energy. Further, $w$ accounts for the elec\-tron-electron  interactions within and between the two parts. 
At the same time this part of the system can be very complex and heterogeneous, consisting of many sub-parts, so the index ``e'' can be a multi-index describing many baths~\cite{Bonitz_fcse_19, balzer_cpp_21}, see also Sec.~\ref{sss:multilayer}. 
Embedding approaches have been applied to a large variety of problems. This includes the analysis of long-time effects, such as thermalization and emergence of irreversibility which were considered in the NEGF or density operator formalisms, see e.g. Refs.~\cite{haug_2008_quantum, galperin_epjst_21,bonitz_cpp18}.

Here, in contrast, we study short-time phenomena. 
Our central quantities ares the one-particle nonequilibrium Green function, $G^{\alpha\beta}_{ij}(t,t')$, as introduced above, and its time-diagonal elements -- the density matrix. These quantities now carry an additional $2\times2$ matrix structure ($\alpha, \beta=\Omega$),
\begin{align}
\label{eq.negf}
 G^{\alpha\beta}_{ij}(z,z') &= \frac{1}{\i\hbar} \langle T_C \hat{c}^\alpha_{i}(z)\hat{c}_{j}^{\beta\dagger}(z')\rangle\,,\\
  \rho_{ij}^{\alpha\beta}(t) &= \frac{\hbar}{\i} G^{\beta\alpha}_{ji}(t,t^+)\,.
\label{eq:g-dm}  
\end{align}
In addition to the density matrices of the system parts, $\rho_{ij}^{\tn{ss}}$ and $\rho_{ij}^{\tn{ee}}$, the off-diagonal matrix components  $G_{ij}^{\tn{es}}$ and $\rho_{ij}^{\tn{es}}$ are related to charge and energy transfer processes between  system and environment and will be of special interest in the following. 

The equations of motion for the NEGF are the Keldysh-Kadanoff-Baym equations, Eqs.~(\ref{eq:kbe-sigma-form1}) and (\ref{eq:kbe-sigma-form2}), extended
to the total system,
\begin{align}
 &\i\hbar\partial_tG^{\alpha\beta}_{ij}(t,t') - \sum_{\delta=\tn{e},\tn{s}} h^{\rm HF, \alpha\delta}_{ik}(t)G^{\delta\beta}_{kj}(t,t')
 \label{eq.kbe_interface}
 =\delta^{\alpha\beta}_{ij}\delta_C(t,t')+\sum_{\delta=\tn{e},\tn{s}} \int_C\!\!\!\d\bar{t}\,\Sigma^{\alpha\delta}_{ik}(t,\bar{t})G^{\delta\beta}_{kj}(\bar{t},t')\,,
\end{align}
where here and below we simplify the notation by  implying summation over repeating orbital indices $k$.

We will now rewrite this equation in components assuming that, in the environment, correlations can be neglected, $\Sigma^{\tn{ee}}=0$, but we still retain interaction effects on the mean-field level. Furthermore, it is assumed, that for the coupling between system and environment, correlations are negligible as well, i.e. $\Sigma^{\tn{se}}=\Sigma^{\tn{es}}=0$, whereas Hartree-Fock terms are retained, cf. r.h.s. of Eqs.~(\ref{eq:gss-equation}) and~(\ref{eq:ges-equation}). Note that, in many applications where the environment represents e.g. a large ``bath'', a gas phase or macroscopic leads, interaction effects in the environment play a minor role, and these approximations are appropriate, e.g.~\cite{stefanucci_nonequilibrium_2013}.

Further, we use simpler notations for the diagonal terms: $G^{\tn{ss}} \to G^{\tn{s}}$, $G^{\tn{ee}} \to G^{\tn{e}}$ 
 $h^{\rm HF,\alpha \alpha} \to h^{\rm HF,\alpha}$ and $\Sigma^{\alpha \alpha} \to \Sigma^\alpha$,
 and use underlined indices for the orbitals in the environment, for a better distinction:
\begin{align}
\left\{\i\hbar\partial_t\delta_{ik}-h^{{\rm HF},\tn{s}}_{ik}(t)\right\}G^{\tn{s}}_{kj}(t,t')&
\label{eq:gss-equation}
=h^{{\rm HF},\tn{se}}_{i\,\underline k}(t)G^{\tn{es}}_{\underline k \,j}(t,t')
+\delta_{ij}\delta_C(t,t') 
  + \int_C\!\!\!\d\bar{t}\,\Sigma^{\tn{s}}_{ik}(t,\bar{t})
  G^{\tn{s}}_{kj}(\bar{t},t')\,,\\
  \left\{\i\hbar\partial_t\delta_{\underline i \,\underline k}-h^{{\rm HF},\tn{e}}_{\underline i \,\underline k}(t)\right\}G^{\tn{es}}_{\underline k \,j}(t,t')&=h^{{\rm HF},\tn{es}}_{\underline i\, k}(t)G^{\tn{s}}_{kj}(t,t')\,,
  \quad\label{eq:ges-equation}\\
  \left\{\i\hbar\partial_t\delta_{\underline i \,\underline k}-h^{{\rm HF},\tn{e}}_{\underline i \,\underline k}(t)\right\}g^{\tn{e}}_{\underline k\,\underline j}(t,t')&=\delta_{\underline i\, \underline j}\delta_C(t,t')\,,
  \label{eq:gee-equation}
\\
    \left\{\i\hbar\partial_t\delta_{\underline i\, \underline k}-h^{{\rm HF},\tn{e}}_{\underline i\,\underline k}(t)\right\}G^{\tn{e}}_{\underline k\,\underline j}(t,t')
  &=
  h^{\tn{HF},\tn{es}}_{\underline i\,k}(t)G^{\tn{se}}_{k\,\underline j}(t,t')
  +\delta_{\underline i\, \underline j}\delta_C(t,t')\,.
  \label{eq:gee-equation-new}   
\end{align}
Here we also introduced a second environment function, $g^\tn{e}$, that obeys an equation that is decoupled from the system. This function plays a central role in the two-time embedding selfenergy approach.
Notice that the equations for $g^{\tn{e}}$ and $G^{\tn{es}}$ contain the same term on the left hand side (parentheses) which is the inverse Green function, $g^{\tn{e}\,-1}_{\underline i\, \underline k}$. Multiplication of \eref{eq:ges-equation} by $g^{\tn{e}}_{\underline l\, \underline i}(t,t')$ and integration over the time contour, yields:
\begin{align}
    G^{\tn{es}}_{\underline l\,j}(t,t') = \int_C\!\!\!\d\bar{t}\, g^{\tn{e}}_{\underline l\,\underline i}(t,\bar t)\, h^{{\rm HF},\tn{es}}_{\underline i\,k}(\bar t)\,G^{\tn{s}}_{kj}(\bar t,t')\,.\label{eq:ges-solution}
\end{align}
Equation~\eqref{eq:ges-solution} allows us to eliminate $G^{\tn{es}}$ from the equation for $G^{\tn{s}}$ and to rewrite this term in the form of an additional ``embedding'' selfenergy, $\Sigma^{\rm emb}$:
\begin{align}
\label{eq.kbe.embedding}
&\left\{\i\hbar\partial_t\delta_{ik}-h^{{\rm HF},\tn{s}}_{ik}(t)\right\}G^{\tn{s}}_{kj}(t,t')
=\delta_{ij}\delta_C(t,t')
 +\int_C\!\!\!\d\bar{t}\,\left\{\Sigma^{\rm emb}_{ik}(t,\bar{t})+\Sigma^{\tn{s}}_{ik}(t,\bar{t})\right\}G^{\tn{s}}_{kj}(\bar{t},t')\,,
\end{align}
which is given by
\begin{align}
\label{eq.sigma.ct}
  \Sigma^{\rm emb}_{ij}(t,t') &= 
  h^{{\rm HF},\tn{se}}_{i\,\underline k}(t)g_{\underline k \, \underline l}^{\tn{e}}(t,t') h^{{\rm HF},\tn{es}}_{\underline l \,j}(t')\,,
\\
h^{\rm HF, \tn{se}}_{i\,\underline j}(t) &= \int\!\!\d^3r\,\phi^{\tn{s}*}_i(\mathbf{r})[\hat{T}+\hat{V}^{\rm HF}(t)]\chi_{\underline j}^{\tn{e}}(\mathbf{r};t)\,,
\label{eq:hsp}
\end{align}
and involves the system-environment coupling Hamiltonian $h^{\rm HF,\tn{se}}$, which is renormalized by the Hartree-Fock mean field. 
The purpose of the embedding concept is to retain a closed KBE~\eqref{eq.kbe.embedding} for the system part without the need to solve equations for the system-environment coupling and for the environment whereas the effect of the latter is fully accounted for by the additional selfenergy, $\Sigma^{\textup{emb}}(t,t')$.

A few comments are in order: 
\begin{enumerate}
    \item Equations (\ref{eq.kbe.embedding}) and (\ref{eq.sigma.ct}) are the two-time versions of the embedding NEGF approach. Correspondingly, these equations retain the cubic scaling with $N_\tn{t}$.
    \item Interaction effects in the coupling of the system parts (e.g. charge transfer) are captured only on the mean-field level. This restriction can be omitted, as we will show in Sec.~\ref{ss:embedding-beyond}. 
    \item The embedding selfenergy (\ref{eq.sigma.ct}) involves the isolated Green function of the environment, $g^{\tn{e}}$, which is decoupled from the system, cf. Eq.~(\ref{eq:gee-equation}). This equation ignores particle number loss (or gain) of the environment due to transfer processes to the central system. However, for the embedding selfenergy (\ref{eq.sigma.ct}) this is irrelevant. 
    \item A modified equation for the environment NEGF that takes into account charge transfer processes is also presented above, cf. Eq.~(\ref{eq:gee-equation-new}), and its solution is denoted by $G^{\tn{e}}$. As we will see in Sec.~\ref{ss:g1-g2-embedding}, for the time-local version of the embedding approach an equation of motion for $G^{\tn{e}}$ has to be used instead of an equation for $g^{\tn{e}}$. It is easy to see that both environment functions are related by 
\begin{align}
 \,\,G^{\tn{e}}_{\underline l\,\underline j}(t,t')
&=g^{\tn{e}}_{\underline l\,\underline j}(t,t') + \int_C\!\!\!\d\bar{t}\,g^{\tn{e}}_{\underline l\,\underline i}(t,\bar t)  h^{\tn{HF},\tn{es}}_{\underline i\, k}(\bar t)G^{\tn{se}}_{k\,\underline j}(\bar t,t')\,.
\quad
\label{eq:connection-ge-Ge}
\end{align}    
\end{enumerate}

\subsection{Time-linear G1--G2 embedding approach}\label{ss:g1-g2-embedding}
As discussed above, the embedding scheme for the two-time NEGF, Eqs.~(\ref{eq.kbe.embedding}, \ref{eq.sigma.ct}), suffers from the unfavorable (at least) cubic scaling with $N_\tn{t}$. If correlation selfenergies are taken into account for the environment, the scaling is even worse, cf. Sec.~\ref{sss:correlated-embedding}.
Therefore, it is tempting to apply the HF-GKBA and transform these equations into the time-local  G1--G2 scheme, in order to retain the time-linear scaling. This was achieved  in Ref.~\cite{balzer_prb_23}, and here we summarize the key results.

\subsubsection{Extended G1--G2 embedding equations}
The equation of motion for the single-particle Green function $G^<(t)=G^{\tn{s},<}(t,t)$ on the time diagonal (first equation of the G1--G2 scheme) becomes 
\begin{align}
    \i \hbar\partial_t G^{<}_{ij}(t) &-  
     \left[ 
    h^{\rm HF},G^{<}\right]^{\tn{s}}_{ij,t} = 
    \left(I(t)+ I^\dagger(t)\right)_{ij}\,,
    \label{eq:time-diagonal-gless-equation}
\qquad
I_{ij}(t) = I^{\rm cor}_{ij}(t)+I^{\rm emb}_{ij}(t)\,,\\    
I^{\rm cor}_{ij}(t) &= \int_{t_0}^t \d\bar t 
    \left\{\Sigma^{>}_{ik}(t,\bar{t})G^{<}_{kj}(\bar{t},t) - \Sigma^{<}_{ik}(t,\bar{t})G^{>}_{kj}(\bar{t},t)\right\}\,,
\label{eq:collision-integral}
\\
    I^{\rm emb}_{ij}(t) &= \int_{t_0}^t \d\bar t 
    \bigg\{\Sigma^{\rm emb, >}_{ik}(t,\bar{t})G^{<}_{kj}(\bar{t},t)  
-\Sigma^{\rm emb, <}_{ik}(t,\bar{t})G^{>}_{kj}(\bar{t},t)\bigg\}\,.
\label{eq:collision-integral-emb}
\end{align}
The first equation, Eq.~(\ref{eq:time-diagonal-gless-equation}), now contains two collision integrals that replace the two selfenergies of the two-time KBE scheme. The first, as before, accounts for correlation effects whereas the integral $I^{\rm emb}$ is derived from the embedding selfenergy, Eq.~(\ref{eq.sigma.ct}).
Further, we denoted
\begin{align}
    \left[ 
     A,B\right]^\alpha_{ij,t} &= (AB)^\alpha_{ij,t} - (BA)^\alpha_{ij,t}\,,\quad \alpha=\tn{s}, \tn{e}\,,
     \label{eq:def-commutator}\\
     (AB)^\tn{s}_{ij,t} &= \sum_{k\in \tn{s}} A_{ik}(t)B_{kj}(t)\,,
     \label{eq:def-product}
\qquad     (AB)^\tn{e}_{ij,t} = \sum_{\underline k\in \tn{e}} A_{i\,\underline k}(t)B_{\underline k\, j}(t)\,,
\end{align}
where the superscript stands for the sub-space over which the internal summation is performed.

Equation~\eqref{eq:time-diagonal-gless-equation} is not closed for the time-diagonal Green function, $G^<(t)$, but still involves two-time functions under the integral. As before, this problem is solved, by applying the GKBA, cf. Sec.~\ref{s:GKBA}. However, it turns out that, using the previous version of the GKBA, Eq.~(\ref{eq:gkba}), and including only the system (s) contributions in the GKBA, leads to wrong results, as it violates particle number conservation \cite{balzer_prb_23}. Instead, one has to adopt the GKBA to the present multi-component character of the problem 
\begin{align}
&G^{\tn{s},\gtrless}_{ij}(t,t') = 
\i \hbar\left[G_{ik}^{\tn{s},{\rm R}}(t,t') G_{kj}^{\tn{s},\gtrless}(t')-G_{ik}^{\tn{s},\gtrless}(t) G_{kj}^{\tn{s},{\rm A}}(t,t')\right]
+\i \hbar\left[G_{i \,\underline k}^{\tn{se},{\rm R}}(t,t') G_{\underline k \,j}^{\tn{es},\gtrless}(t')-G_{i \,\underline k}^{\tn{se},\gtrless}(t) G_{\underline k\, j}^{\tn{es},{\rm A}}(t,t')\right]
\,,\quad 
\label{eq:GKBA-ss}
\end{align}
and include also contributions from the retarded and advanced functions that couple the two different system parts, $G^{\tn{se},{\rm R/A}}$. This gives rise to the following \textit{extended time-local embedding scheme} \cite{balzer_prb_23}
\begin{align}
    \i\hbar \frac{\d }{\d t} G^{<}_{ij}(t) &-  
     \left[ 
    h^{\rm HF},G^{<}\right]^s_{ij,t} = 
    \left(I(t)+ I^\dagger(t)\right)_{ij}\,,
\quad
I_{ij}(t) = \pm \i\hbar w_{iklp}(t) \mathcal{G}^s_{lpjk}(t) +  h^{\tn{HF,se}}_{i\,\underline k}(t) G^{\tn{es},<}_{\underline k\, j}(t)\,,
    \label{eq:time-diagonal-gless-equation-gkba}
\\
    \i\hbar \frac{\d }{\d t} \mathcal{G}^s_{ijkl}(t) &- \Big[ h^{(2),\tn{HF}}(t),\mathcal{G}^s(t) \Big]_{ijkl} = \hat\Psi^{s\pm}_{ijkl}(t)
 \, ,\quad  \label{eq:G1-G2_soa_em}
 \\
\i\hbar \frac{\d }{\d t}G^{\tn{es},<}_{\underline i\,j}(t)
&=  \left( h^{\tn{HF,es}} G^{\tn{s},<}\right)^\tn{s}_{\underline i\,j,t}
 - \left(G^{\tn{e},<}h^{\tn{HF},\tn{es}}\right)^\tn{e}_{\underline i\,j,t}
 \label{eq:ges-new}
+ \left( h^{\tn{HF},\tn{e}}G^{\tn{es},<}\right)^\tn{e}_{\underline i\,j,t}
 - \left(G^{\tn{es},<}h^{\tn{HF},\tn{s}}\right)^\tn{s}_{\underline i\,j,t}\,,
\\
\i\hbar \frac{\tn {d}}{\tn{d}t} G^{\tn{e},<}_{\underline i\, \underline j}(t) &= \left[ h^{\rm HF,\tn{e}},G^{\tn{e},<}\right]^\tn{e}_{\underline i \,\underline j ,t} 
    \label{eq:gee-new}
+ \left(h^{\tn{HF},\tn{es}}G^{\tn{se},<}\right)^\tn{s}_{\underline i\, \underline j,t}- \left(G^{\tn{es},<}h^{\tn{HF},\tn{se}}\right)^\tn{s}_{\underline i\, \underline j,t}\,.
\end{align}
Note that equation for $G^{\tn{s},<}$, Eq.~(\ref{eq:time-diagonal-gless-equation-gkba}), now contains two collision integrals arising from the correlation and embedding selfenergies: the first is due to correlations inside the system and couples to the pair correlation function, $\mathcal{G}^{\tn{s}}$. The second couples to the charge transfer function, $G^{\tn{es},<}$.
The equation for the pair correlation function has been written for the second order Born approximation, Eq.~(\ref{eq:g1g2_2B}), and can be improved by using any of the more accurate approximations for the selfenergies that were discussed above. In addition to the standard G1--G2 scheme, we now have to solve two additional equations: one for the charge transfer function, $G^{\tn{es},<}$, and one for the Green function of the environment.
Note that it is crucial that the time evolution of the environment is propagated selfconsistently via the Green function $G^{\tn{e}}$, which includes the coupling to the system part, but not the isolated function $g^{\tn{e}}$ that enters the embedding selfenergy.

We point out that Eq.~(\ref{eq:G1-G2_soa_em}) is an approximation because it neglects the coupling to the environment. The proper generalization is presented in Ref.~\cite{bonitz_prb_24}.
Finally, compared to the two-time version of the embedding equations, the present system is significantly more complex. While the two-time version contains just a single closed equation for $G^{\tn{s},<}(t,t')$, here we have to solve four time-local equations. However, this allows for a dramatic acceleration arising from the time-linear scaling vs. cubic scaling.


\subsubsection{Example: Ultrafast charge transfer from a highly charged ion to quantum materials}
The embedding scheme is very efficient to treat ultrafast and very intense charge transfer processes. Such situations occur,  e.g. during the stopping of highly charge ions in quantum materials such as graphene and TMDCs which was studied in Ref.~\cite{niggas_prl_22}. In that reference a two-time NEGF embedding simulation was reported which allowed to explain the experimental observations. Nevertheless, the high computational cost of these simulations stimulated the development of a time local embedding scheme, as discussed above.
 The behavior of the time-local extended embedding scheme, Eqs.~(\ref{eq:time-diagonal-gless-equation-gkba})--(\ref{eq:gee-new}), has been analyzed in Ref.~\cite{balzer_prb_23}, for an instructive model of charge transfer between a ``system'' and an ``environment''.
There, the system was a linear Hubbard chain with a finite number of sites, $1 \dots L$, that couples via site $1$ to an external energy level (an attached site or the impacting ``ion''). The goal was to compute and analyze the time-dependent charge transfer between the system and the additional site as it occurs, e.g., upon impact of a highly charged ion on a correlated material, e.g., Ref.~\cite{niggas_prl_22}. In this case, the transfer of electrons into energetically high Rydberg levels of the ion, forming a (partially) neutral hollow atom, is a rapid process with an amplitude that is maximal shortly before the ion hits the surface and which can be modeled to a good approximation with a Gaussian time dependence \cite{balzer_cpp_21}, cf. also Eq.~(\ref{eq:gamma}) below.
In all cases studied in Ref.~\cite{balzer_prb_23}, the results of the extended embedding scheme, Eqs.~(\ref{eq:time-diagonal-gless-equation-gkba})--(\ref{eq:gee-new}), were indistinguishable from full two-time KBE simulations, where either the full Green function of the joint system (system plus environment) had been computed directly or the full two-time embedding selfenergy had been used.
In contrast, an alternative version of the time-local scheme which uses the GKBA only for the system components, i.e., does not explicitly take into account the system-environment coupling in the GKBA [second term in Eq.~(\ref{eq:GKBA-ss})] and the environment function $g^\textup{e}$, instead of $G^\textup{e}$, fails for strong charge transfer amplitudes $\gamma_0$. 
Here, the electron density on the additional site was found to exceed unity whereas the charge on the neighboring site 1 on the chain became negative. Thus, this approximation is  un-physical and has to be discarded.

In the following, we present novel results only for the \textit{extended time-local embedding scheme} which involves the complete GKBA, Eq.~(\ref{eq:GKBA-ss}). We apply it to the resonant charge transfer between a 50-site Hubbard ring away from half filling  and an environment in form of a single orbital (mimicking an impacting ion) of energy $\epsilon^\textup{e}$ that is attached to a single site of the ring. To be precise on the system components, we refer to the Hubbard ring as the system $\textup{s}$ (with electron creation and annihilation operators $\hat{c}^\dagger_{i\sigma}$ and $\hat{c}_{i\sigma}$) and label parameters of the environment with superscript $\textup{e}$ (operators $\hat{b}^\dagger_\sigma$ and $\hat{b}_\sigma$), where $\sigma$ denotes the spin index. Considering electron-electron interactions on the Hartree-Fock level, the Hamiltonian is given by an extension of the Hubbard model, Eq.~(\ref{eq:h-hubbard}),
\begin{align}
\hat{H}(t)&=\sum_{ij\sigma}h^\textup{s}_{ij\sigma}(t)\hat{c}^\dagger_{i\sigma} \hat{c}_{j\sigma}+\varepsilon^\textup{e}\, \hat{b}_\sigma^\dagger\hat{b}_\sigma+\gamma^{\textup{es}}(t)\sum_{\sigma}\left(\hat{b}_\sigma^\dagger\hat{c}_{1\sigma} + \textup{h.c.}\right)\,,&
h^\textup{s}_{ij\sigma}(t)=-J\,\delta_{\langle i,j\rangle}+\delta_{ij}\,U\left(\langle \hat{n}_{i\sigma}\rangle(t)-\frac{1}{2}\right)\,,
\label{eq:emb-example-ham}
\end{align}
 \begin{figure}[h]
\centering
\includegraphics[width=0.55\textwidth]{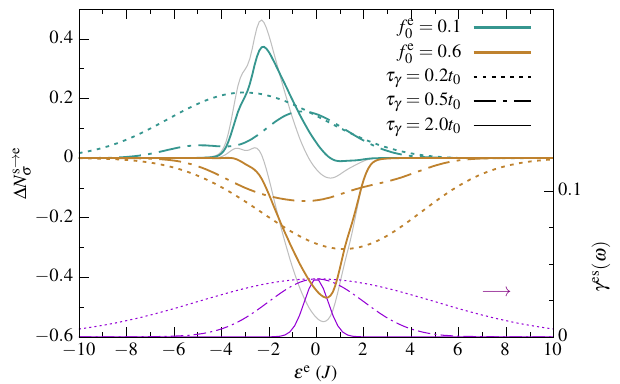}
\caption{Total charge transfer $\Delta N^{\textup{s}\rightarrow\textup{e}}_\sigma$ between a system $\textup{s}$ (Hubbard ring at finite temperature, $\beta^{-1}=1J$,  with initial filling $f_0^\textup{s}=0.3$) and an environment $\textup{e}$ with a single orbital with energy $\epsilon^\textup{e}$. Positive values denote charge transfer to the ion. The green and brown lines are horizontal cuts through panels of Fig.~\ref{fig:emb-example-2}. The violet lines show the coupling $\gamma^\textup{es}$ in frequency space, cf.~Eq.~(\ref{eq:gamma}), right axis. The gray solid lines indicate the change of charge transfer for the green, respectively, brown solid lines when the temperature is reduced to $\beta^{-1}=0.4J$ ($\beta=2.5J^{-1}$), compare with Fig.~\ref{fig:emb-example-2}d.}
\label{fig:emb-example-1}
\end{figure}
%
where $J$ denotes the nearest-neighbor hopping, $U=4J$ is the on-site Hubbard interaction and $\langle\hat{n}_{i\sigma}\rangle(t)= -\i\hbar G^{\textup{s},<}_{ii\sigma}(t,t)$ is the time-dependent spin density on site $i$. We generally measure energies in units of $J$, times in units of $t_0=\hbar J^{-1}$ and prepare the Hubbard ring at a finite inverse temperature $\beta$ and filling $f_0^\textup{s}=\langle\hat{n}_{i\sigma}\rangle(-\infty)=0.3$. The attached orbital is initially occupied to a variable degree $f_0^\textup{e}\in [0,1]$, whereas the coupling $\gamma^{\textup{es}}(t)$ is time-dependent and chosen as \cite{balzer_cpp_21}
\begin{align}
\gamma^{\textup{es}}(t)&=\gamma_0 \,e^{-\frac{t^2}{2\tau_\gamma^2}}\,,& \textup{Fourier transform:}\quad\,\,\gamma^{\textup{es}}(\omega)= \frac{1}{2\pi\gamma_0} \, e^{-\tau_\gamma^2\omega^2/2} \,.
\label{eq:gamma}
\end{align}
Thus, around time $t=0$, for a temporal duration $\tau_\gamma$, both systems, $\textup{s}$ and $\textup{e}$, undergo charge exchange. In the course of this, the net transfer of electrons onto the orbital,
\begin{align}
\Delta N^{\textup{s}\rightarrow\textup{e}}_\sigma=\sum_{i}\langle\hat{n}_{i\sigma}\rangle(-\infty)-\sum_{i}\langle\hat{n}_{i\sigma}\rangle(+\infty)\,,
\end{align}
is expected to depend primarily on the energy $\varepsilon^\textup{e}$ and the initial occupation $f_0^\textup{e}$.

\begin{figure}[p]
\centering
\includegraphics[width=0.425\textwidth]{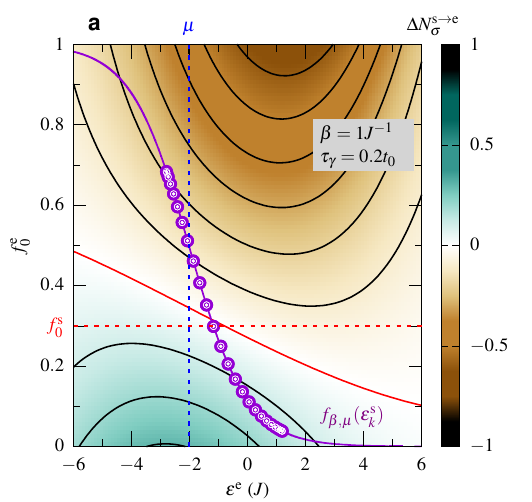}
\includegraphics[width=0.425\textwidth]{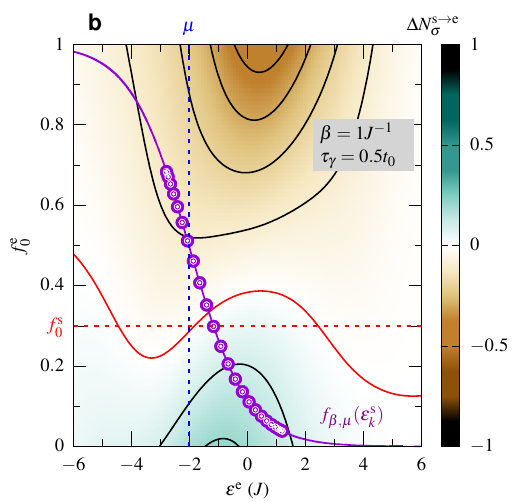}\\
\includegraphics[width=0.425\textwidth]{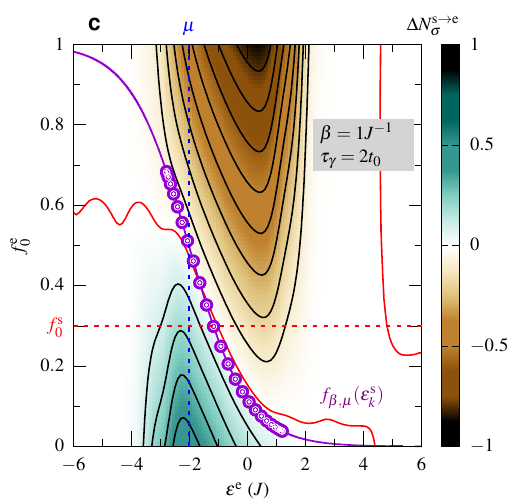}
\includegraphics[width=0.425\textwidth]{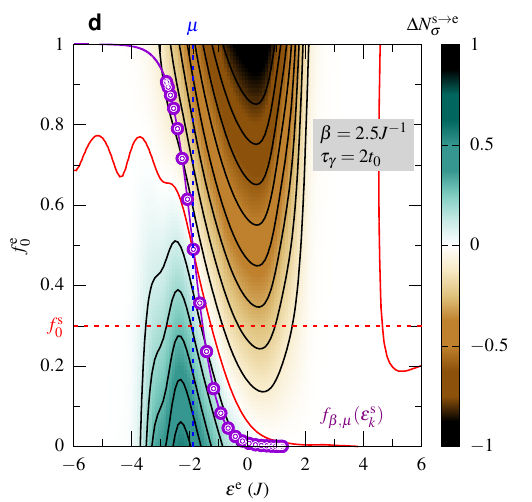}
\caption{Charge transfer $\Delta N^{\textup{s}\rightarrow\textup{e}}_\sigma$ between a $50$-site Hubbard ring (system index $\textup{s}$) at filling $f_0^\textup{s}=0.3$ (horizontal red dashed line) and a single site (environment index $\textup{e}$) as a function of energy $\varepsilon^\textup{e}$ and occupation $f_0^\textup{e}$ of the external site (environment). Initially, the Hubbard ring is prepared in a finite-temperature equilibrium state with inverse temperature $\beta=1J^{-1}$, panels a)-c), and $\beta=2.5J^{-1}$, panel d), with Hubbard interaction $U=4J$ (treated on the Hartree-Fock level). The 
ring connects at one site via a time-dependent coupling $\gamma(t)$, Eq.~(\ref{eq:gamma}), to the attached external site. As charge transfer amplitude we chose $\gamma_0=4J$, and the interaction time $\tau_\gamma$ varies between a) $0.2t_0$, b) $0.5t_0$, and c) and d) $2t_0$. The violet open circles indicate the HF self-consistent field eigenstates that have energy $\varepsilon_k^\textup{s}$ and follow a Fermi-Dirac distribution, $f_{\beta,\mu}(\varepsilon_k^{\textup{s}})$, cf. the violet solid line, whereas the vertical blue dashed line marks the chemical potential $\mu$. The solid black lines are contours of constant charge transfer and are spaced by $0.1$. The red contour line marks a value of $\Delta N^{\textup{s}\rightarrow\textup{e}}_\sigma=0$.}
\label{fig:emb-example-2}
\end{figure}
We solve the extended time-local embedding equations for the model~(\ref{eq:emb-example-ham}) and compute the charge transfer, first, as function of $\varepsilon^\textup{e}$ only, i.e. for fixed occupations $f_0^\textup{e}$, and present the results in Fig.~\ref{fig:emb-example-1}. Throughout, we consider the non-perturbative regime, where $\gamma_0=4J$, which is large compared to the system bandwidth. As it is to be expected, the result for $\Delta N^{\textup{s}\rightarrow\textup{e}}_\sigma$ is generally negative, for $f_0^\textup{e}>f_0^\textup{s}$, where electrons move into the system (brown curves). In contrast, the total charge transfer is positive, for $f_0^\textup{e}<f_0^\textup{s}$, where the external orbital becomes more strongly populated (green curves). As the exchange of electrons between system and environment, however, requires single-particle states of both parts to be in resonance, the yield of transferred electrons sensitively depends on the orbital energy $\varepsilon^{\textup{e}}$. Moreover, for an interaction time $\tau_\gamma$ much larger than the characteristic timescale of a charge transfer, $\tau_\gamma \gg t_\textup{ct}=\hbar\gamma^{-1}_0=1/4t_0$, the yield's energy profile is rather narrow, cf. the full brown and green lines. In contrast, the yield becomes strongly Gaussian broadened, for the case of small interaction times, according to the Fourier transform $\gamma^{\textup{es}}(\omega)$ of the coupling (\ref{eq:gamma}), cf. dashed lines. A coupling which is sufficiently narrow in frequency space will, hence, allow us to directly probe the single-particle energies in the Hubbard ring as well as their occupation. 

The overall dependence of the charge transfer upon the orbital energy $\varepsilon^e$ and the initial energy-resolved filling (distribution function) of the system is analyzed in more detail in Fig.~\ref{fig:emb-example-2}.    
First, from all panels of Fig.~\ref{fig:emb-example-2}, we can extract information about the filling, $f_0^\textup{s}$, of the Hubbard ring by identifying lines of vanishing charge transfer, $\Delta N^{\textup{s}\rightarrow\textup{e}}=0$ (red solid lines). These lines fluctuate as function of orbital energy around the value of system filling, $f_0^\textup{s} = 0.3$ (red dashed lines).
Furthermore, the sign of the charge transfer serves as a probe of the occupied (positive sign, green area) and unoccupied (negative sign, brown area) eigenstates of the Hubbard ring. Of course, this direct relation holds only in case of an energetically narrow ``probe pulse'' $\gamma(\omega)$, i.e., for the case of a slow charge exchange (long interaction time $\tau_\gamma$), cf. panel c). In contrast, for a rapid charge exchange (short interaction time), e.g., panel a), areas of positive and negative charge exchange are broadened. 
Finally, it is also interesting to explore the dependence of the charge transfer on the temperature of the system. This can be seen in panels c) and d) where, for a fixed interaction time $\tau_\gamma=2t_0$ the temperature is reduced by a factor $2.5$, from panel c) to panel d). 
Clearly, reducing the temperature causes an energetic narrowing (broadening) of the occupied (unoccupied) HF eigenstates in the Hubbard ring, i.e., less energy levels $\epsilon_k^\textup{s}$ are dominantly present in the system $\textup{s}$. As a consequence, the energy range for which we observe positive (negative) charge transfer $\Delta N^{\textup{s}\rightarrow\textup{e}}$ decreases (increases). At the same time, the focus on energy levels with an occupation closer to 1 respectively zero in Fig.~\ref{fig:emb-example-2}d leads to an enhanced charge transfer at these energies, compare with the gray lines in Fig.~\ref{fig:emb-example-1}. 

\subsection{Extension of the embedding selfenergy approach}\label{ss:embedding-beyond}
Here, we present two extensions of the embedding scheme for the two-time nonequilibrium Green functions. The first is that we allow for correlation selfenergies in the environment and in the system-environment coupling. The second is that we allow for a system which is more complex, consisting of three different parts. Both ideas are well suitable to address the memory problem of the G1--G2 schem making larger systems feasible. On the other hand, these concepts are suitable to systematically include longer range correlations, similar to the multi-impurity-site (cluster) extensions developed in DMFT, cf. the discussion at the beginning of Sec.~\ref{s:embedding}.

\subsubsection{Inclusion of correlation selfenergies $\Sigma^{\tn{e}}, \Sigma^{\tn{es}}$}\label{sss:correlated-embedding}
We now reconsider the Keldysh-Kadanoff-Baym equations and allow for correlation selfenergies in the environment ($\Sigma^{\tn{e}}\ne 0$) and in the system-environment coupling ($\Sigma^{\tn{es}}\ne 0$),
%
\begin{align}
\left\{\i\hbar\partial_t\delta_{ik}-h^{{\rm HF},\tn{s}}_{ik}(t)\right\}G^{\tn{s}}_{kj}(t,t')&
\label{eq:gss-equation-i}
=h^{{\rm HF},\tn{se}}_{i\,\underline k}(t)G^{\tn{es}}_{\underline k \,j}(t,t')
  +\delta_{ij}\delta_C(t,t') 
  + \int_C\!\!\!\d\bar{t}\,\Sigma^{\tn{s}}_{ik}(t,\bar{t})G^{\tn{s}}_{kj}(\bar{t},t')
  + \int_C\!\!\!\d\bar{t}\,\Sigma^{\tn{se}}_{i\underline k}(t,\bar{t})G^{\tn{es}}_{\underline k j}(\bar{t},t')\,,\quad\\
    \left\{\i\hbar\partial_t\delta_{\underline i\, \underline k}-h^{{\rm HF},\tn{e}}_{\underline i\,\underline k}(t)\right\}G^{\tn{e}}_{\underline k\,\underline j}(t,t') 
    &-\int_C\!\!\!\d\bar{t}\,\Sigma^{\tn{e}}_{\underline i\underline k}(t,\bar{t})G^{\tn{e}}_{\underline k\underline j}(\bar{t},t')\,
  =
  \delta_{\underline i\, \underline j}\delta_C(t,t')\,
  + h^{\tn{HF},\tn{es}}_{\underline i k}(t)G^{\tn{se}}_{k\,\underline j}(t,t')
  + \int_C\!\!\!\d\bar{t}\,\Sigma^{\tn{es}}_{\underline i k}(t,\bar{t})G^{\tn{se}}_{k\underline j}(\bar{t},t')\,,\quad
  \label{eq:gee-equation-new-i}   
\\
  \left\{\i\hbar\partial_t\delta_{\underline i \,\underline k}-h^{{\rm HF},\tn{e}}_{\underline i \,\underline k}(t)\right\}g^{\tn{e}}_{\underline k\,\underline j}(t,t')
  &-\int_C\!\!\!\d\bar{t}\,\Sigma^{\tn{e}}_{\underline i\underline k}(t,\bar{t})g^{\tn{e}}_{\underline k\underline j}(\bar{t},t')\,
  =\delta_{\underline i\, \underline j}\delta_C(t,t')\,,
  \label{eq:gee-equation-i}
\\
  \left\{\i\hbar\partial_t\delta_{\underline i \,\underline k}-h^{{\rm HF},\tn{e}}_{\underline i \,\underline k}(t)\right\}G^{\tn{es}}_{\underline k \,j}(t,t')
  &- \int_C\!\!\!\d\bar{t}\,\Sigma^{\tn{e}}_{\underline i\underline k}(t,\bar{t})G^{\tn{es}}_{\underline k j}(\bar{t},t')\,
  =h^{{\rm HF},\tn{es}}_{\underline i\, k}(t)G^{\tn{s}}_{kj}(t,t') + \int_C\!\!\!\d\bar{t}\,\Sigma^{\tn{es}}_{\underline i k}(t,\bar{t})G^{\tn{s}}_{kj}(\bar{t},t')\,\,,
  \quad\label{eq:ges-equation-i}  
\end{align}
where the equations for $g^e$ and $G^e$ are identical, except for the two final terms on the r.h.s. of Eq.~\eqref{eq:gee-equation-new-i} that contain $G^{\tn{se}}$.
We again recognize that the equations for $g^{\tn{e}}$ and $G^{\tn{es}}$ contain an identical term on the left-hand-side which now contains, in addition, $\Sigma^\tn{e}$ and which is nothing but the inverse of the (now correlated) Green function, $g^{\tn{e}\,-1}_{\underline i\, \underline k}$, 
\begin{align}
  g^{\tn{e}\,-1}_{\underline i\, \underline k}(t,\bar t) &= 
  \left[\i\hbar\partial_t\delta_{\underline i \,\underline k}-h^{{\rm HF},\tn{e}}_{\underline i \,\underline k}(t)\right]\delta_C(t,\bar t)-\Sigma^{\tn{e}}_{\underline i\,\underline k}(t,\bar{t})\,.
\end{align}
Thus, multiplying \eref{eq:ges-equation-i} by $g^{\tn{e}}_{\underline l\, \underline i}(t,t')$, summing over $\underline{i}$ and integrating over the time contour, we obtain an improved explicit solution for $G^{\tn{es}}$:
\begin{align}
    G^{\tn{es}}_{\underline l\,j}(t,t') = \int_C\!\!\!\d\bar{t}\, g^{\tn{e}}_{\underline l\,\underline i}(t,\bar t)\, h^{{\rm HF},\tn{es}}_{\underline i\,k}(\bar t)\,G^{\tn{s}}_{kj}(\bar t,t') 
    + \int_C\!\!\!\d\bar{t}\,\int_C\!\!\!\d{\tilde t}\, g^{\tn{e}}_{\underline l\,\underline i}(t,\bar t)\, \Sigma^{\tn{es}}_{\underline i\,k}(\bar t,\tilde t)\,G^{\tn{s}}_{kj}(\tilde t,t') \,,\label{eq:ges-solution-correlated}
\end{align}
where the first term recovers our previous result, Eq.~(\ref{eq:ges-solution}) and the second is the correlation correction.
With this result we can eliminate the two appearances of $G^{\tn{es}}$ from the system equation, Eq.~\eqref{eq:gss-equation-i}:
\begin{align}
  \int_C\!\!\!\d\bar{t}\,\left\{ h^{{\rm HF},\tn{se}}_{i\,\underline k}(\bar t)\,\delta_C(t,\bar t) + \Sigma^{\tn{se}}_{i\,\underline k}(t,\bar t)
  \right\}G^{\tn{es}}_{\underline k\,j}(\bar t,t') =& 
  \nonumber\\
    \int_C\!\!\!\d\bar{t}\,\left\{ h^{{\rm HF},\tn{se}}_{i\,\underline k}(\bar t)\,\delta_C(t,\bar t) + \Sigma^{\tn{se}}_{i\,\underline k}(t,\overline{t})
  \right\}\int_C\!\!\!\d\overline{\overline t}\,\int_C\!\!\!\d{\tilde t}\, g^{\tn{e}}_{\underline k\,\underline l}(\bar t,\overline{\overline t})\,\left\{ h^{{\rm HF},\tn{es}}_{\underline l\,m}(\overline{\overline t})\,\delta_C(\overline{\overline t},\tilde t) + \Sigma^{\tn{es}}_{\underline l\,m}(\overline{\overline t},\tilde t)
  \right\}G^{\tn{s}}_{mj}(\tilde t,t') \equiv & \int_C\!\!\!\d\tilde{t}\,\Sigma^{\rm emb}_{ik}(t,\tilde t)G^s_{kj}(\tilde t,t')\,,
\end{align}
where we have identified the improved result for the embedding selfenergy
\begin{align}
  \Sigma^{\rm emb}_{ij}(t,\tilde t) = \int_C\!\!\!\d{\bar t}\,\int_C\!\!\!\d\overline{\overline t}\,  \left\{ h^{{\rm HF},\tn{se}}_{i\,\underline k}(\bar t)\,\delta_C(t,\bar t) + \Sigma^{\tn{se}}_{i\,\underline k}(t,\overline{t})
  \right\} g^{\tn{e}}_{\underline k\,\underline l}(\bar t,\overline{\overline t})\,\left\{ h^{{\rm HF},\tn{es}}_{\underline l\,j}(\overline{\overline t})\,\delta_C(\overline{\overline t},\tilde t) + \Sigma^{\tn{es}}_{\underline l\,j}(\overline{\overline t},\tilde t)
  \right\}\,,
\label{eq:sigma-embedding-i}
\end{align}
%
which differs from the previous result by the additional correlation selfenergies $\Sigma^\tn{se}$. 

With this, we have again obtained a single closed equation for the system Green function that contains, in addition to the selfenergy $\Sigma^\tn{s}$, the embedding selfenergy. Interestingly, this results is exact, since all previous approximations for the embedding Green function and for $G^\tn{se}$ have been avoided. Obviously, this scheme will be advantageous (compared to a full simulation without subdivision into system parts), if the selfenergies $\Sigma^\tn{se}$ and $\Sigma^\tn{e}$ can be chosen significantly simpler than the system selfenergy $\Sigma^\tn{s}$.

Again we observe that, even though the environmental Green function $G^\tn{e}$ couples to $G^\tn{se}$, the embedding selfenergy involves only the isolated environment Green function $g^\tn{e}$. Note that the embedding selfenergy again contains the previously studied pure Hartree-Fock situation, as a special case, in that case the time integrals vanish. If, however, correlation effects in the system-environment coupling are included, there appear two additional time integrals. If, furthermore, $g^e$ is computed with a correlation selfenergy, the effort for $g^e$ alone scales as $N^3_\tn{t}$ and that for (any set of time arguments of) $\Sigma^{\rm emb}$ as $N^5_\tn{t}$. 

\subsubsection{Extension to a multi-layer embedding scheme}\label{sss:multilayer}
Let us now demonstrate how the present scheme can be extended to a system that consists of more than two parts, e.g. $\{s, e, f\}$. Then the idea is to construct a hierarchical embedding scheme, treating $e$ as a ``new system'' which couples to its ``own environment'' $f$. Again the advantage would be the possibility to use different selfenergy approximations for $e$ and $f$. Such a multi-layer approach could be very efficient if the final layer is large and can be treated as ideal or on the HF level. 

To simplify the total system, we will assume that there is no direct coupling between $s$ and $f$. Let us write down the coupled KBEs for the three systems ``s, e, f'' on the Keldysh contour. As before in Sec.~\ref{sss:correlated-embedding}, we allow for correlation selfenergies to exist in all system parts. The key difference is now that there is one system (``e'') that couples to both other system part. To distinguish the basis sets, we will denote the orbitals of system ``e'' by underlined indices and the orbitals of ``f'' by greek letters.
%
%
\begin{align}
\left\{\i\hbar\partial_t\delta_{ik}-h^{{\rm HF},\tn{s}}_{ik}(t)\right\}G^{\tn{s}}_{kj}(t,t')&
\label{eq:gss-equation-ml}
=h^{{\rm HF},\tn{se}}_{i\,\underline k}(t)G^{\tn{es}}_{\underline k \,j}(t,t')
  +\delta_{ij}\delta_C(t,t') 
  + \int_C\!\!\!\d\bar{t}\,\Sigma^{\tn{s}}_{ik}(t,\bar{t})G^{\tn{s}}_{kj}(\bar{t},t')
  + \int_C\!\!\!\d\bar{t}\,\Sigma^{\tn{se}}_{i\underline k}(t,\bar{t})G^{\tn{es}}_{\underline k j}(\bar{t},t')\,,\quad\\
    \left\{\i\hbar\partial_t\delta_{\underline i\, \underline k}-h^{{\rm HF},\tn{e}}_{\underline i\,\underline k}(t)\right\}G^{\tn{e}}_{\underline k\,\underline j}(t,t') 
    &-\int_C\!\!\!\d\bar{t}\,\Sigma^{\tn{e}}_{\underline i\underline k}(t,\bar{t})G^{\tn{e}}_{\underline k\underline j}(\bar{t},t')\,
  =
  \delta_{\underline i\, \underline j}\delta_C(t,t')\,
  + h^{\tn{HF},\tn{es}}_{\underline i k}(t)G^{\tn{se}}_{k\,\underline j}(t,t')
  + \int_C\!\!\!\d\bar{t}\,\Sigma^{\tn{es}}_{\underline i k}(t,\bar{t})G^{\tn{se}}_{k\underline j}(\bar{t},t')\,\quad
  \nonumber\\
   & \qquad\qquad\qquad \qquad \qquad\qquad\qquad\qquad\quad
  + h^{\tn{HF},\tn{ef}}_{\underline i \alpha}(t)G^{\tn{fe}}_{\alpha\,\underline j}(t,t')
  + \int_C\!\!\!\d\bar{t}\,\Sigma^{\tn{ef}}_{\underline i \alpha}(t,\bar{t})G^{\tn{fe}}_{\alpha\underline j}(\bar{t},t')\,,\quad
  \label{eq:gee-equation-new-ml}   
\\
  \left\{\i\hbar\partial_t\delta_{\underline i \,\underline k}-h^{{\rm HF},\tn{e}}_{\underline i \,\underline k}(t)\right\}g^{\tn{e}}_{\underline k\,\underline j}(t,t')
  &-\int_C\!\!\!\d\bar{t}\,\tilde\Sigma^{\tn{e}}_{\underline i\underline k}(t,\bar{t})g^{\tn{e}}_{\underline k\underline j}(\bar{t},t')\,
  =\delta_{\underline i\, \underline j}\delta_C(t,t')\,,
  \label{eq:gee-equation-ml}
\\
  \left\{\i\hbar\partial_t\delta_{\underline i \,\underline k}-h^{{\rm HF},\tn{e}}_{\underline i \,\underline k}(t)\right\}G^{\tn{es}}_{\underline k \,j}(t,t')
  &- \int_C\!\!\!\d\bar{t}\,\tilde\Sigma^{\tn{e}}_{\underline i\underline k}(t,\bar{t})G^{\tn{es}}_{\underline k j}(\bar{t},t')\,
  =h^{{\rm HF},\tn{es}}_{\underline i\, k}(t)G^{\tn{s}}_{kj}(t,t') + \int_C\!\!\!\d\bar{t}\,\Sigma^{\tn{es}}_{\underline i k}(t,\bar{t})G^{\tn{s}}_{kj}(\bar{t},t')\,,
  \quad\label{eq:ges-equation-ml}  \\
    \left\{\i\hbar\partial_t\delta_{\alpha\, \gamma}-h^{{\rm HF},\tn{f}}_{\alpha\,\gamma}(t)\right\}G^{\tn{f}}_{\gamma\,\beta}(t,t') 
    &-\int_C\!\!\!\d\bar{t}\,\Sigma^{\tn{f}}_{\alpha \gamma}(t,\bar{t})G^{\tn{f}}_{\gamma\beta}(\bar{t},t')\,
  =
  \delta_{\alpha\, \beta}\delta_C(t,t')\,
  + h^{\tn{HF},\tn{fe}}_{\alpha \underline k}(t)G^{\tn{ef}}_{\underline k\,\beta}(t,t')
  + \int_C\!\!\!\d\bar{t}\,\Sigma^{\tn{fe}}_{\alpha \underline k}(t,\bar{t})G^{\tn{ef}}_{\underline k\beta}(\bar{t},t')\,,\quad
  \label{eq:gff-equation-new-ml}   
\\
  \left\{\i\hbar\partial_t\delta_{\alpha \,\underline k}-h^{{\rm HF},\tn{f}}_{\alpha\,\gamma}(t)\right\}g^{\tn{f}}_{\gamma\,\beta}(t,t')
  &-\int_C\!\!\!\d\bar{t}\,\Sigma^{\tn{f}}_{\alpha\gamma}(t,\bar{t})g^{\tn{f}}_{\gamma \beta}(\bar{t},t')\,
  =\delta_{\alpha \beta}\delta_C(t,t')\,,
  \label{eq:gff-equation-ml}
\\
  \left\{\i\hbar\partial_t\delta_{\alpha \beta}-h^{{\rm HF},\tn{f}}_{\alpha \beta}(t)\right\}G^{\tn{fe}}_{\beta \underline j}(t,t')
  &- \int_C\!\!\!\d\bar{t}\,\Sigma^{\tn{f}}_{\alpha\beta}(t,\bar{t})G^{\tn{fe}}_{\beta\underline j}(\bar{t},t')\,
  =h^{{\rm HF},\tn{fe}}_{\alpha\, \underline k}(t)G^{\tn{e}}_{\underline k\underline j}(t,t') + \int_C\!\!\!\d\bar{t}\,\Sigma^{\tn{fe}}_{\alpha \underline k}(t,\bar{t})G^{\tn{e}}_{\underline k\underline j}(\bar{t},t')\,\,,
  \quad\label{eq:gfe-equation-ml}  
\end{align}
where the equations for $g^{\rm f}$ and $G^{\rm f}$ are identical, except for the two terms on the r.h.s. of Eq.~\eqref{eq:gff-equation-new-ml} that contain $G^{\rm ef}$. While the equation for $G^\tn{e}$ contains the complete couplings to ``s'' and ``f'', the precise form of the equations for $g^\tn{e}$ and $G^{\tn es}$ has still to be established. To account for their proper coupling to ``f'', we introduced the modified selfenergy $\tilde \Sigma^\tn{e}$ which will be specified below.

Let us start from the lowest layer -- the equation for the ``f''-system.
We again recognize that the equations for $g^{\tn{f}}$ and $G^{\tn{fe}}$ contain identical terms on the left-hand-side which are nothing but the inverse of the (correlated) Green function, $g^{\tn{f}\,-1}_{\alpha \beta}$,
\begin{align}
  g^{\tn{f}\,-1}_{\alpha \beta}(t,\bar t) &= 
  \left[\i\hbar\partial_t\delta_{\alpha\beta}-h^{{\rm HF},\tn{f}}_{\alpha \beta}(t)\right]\delta_C(t,\bar t)-\Sigma^{\tn{f}}_{\alpha\beta}(t,\bar{t})\,,
\end{align}
Thus, multiplying \eref{eq:gfe-equation-ml} by $g^{\tn{f}}_{\delta \alpha}(t,t')$, summing over $\alpha$ and integrating over the time contour, we obtain an explicit solution for $G^{\tn{fe}}$:
\begin{align}    G^{\tn{fe}}_{\delta \underline j}(t,t') = \int_C\!\!\!\d\bar{t}\, g^{\tn{f}}_{\delta\gamma}(t,\bar t)\, h^{{\rm HF},\tn{fe}}_{\gamma \underline k}(\bar t)\,G^{\tn{e}}_{\underline k\underline j}(\bar t,t') 
    + \int_C\!\!\!\d\bar{t}\,\int_C\!\!\!\d{\tilde t}\, g^{\tn{f}}_{\delta\,\gamma}(t,\bar t)\, \Sigma^{\tn{fe}}_{\gamma \underline k}(\bar t,\tilde t)\,G^{\tn{e}}_{\underline k\underline j}(\tilde t,t') \,.\label{eq:gfe-solution}
\end{align}
With this result we can eliminate the two appearances of $G^{\tn{fe}}$ from  Eq.~\eqref{eq:gee-equation-new-ml}:
\begin{align}
  \int_C\!\!\!\d\bar{t}\,\left\{ h^{{\rm HF},\tn{ef}}_{\underline i \gamma}(\bar t)\,\delta_C(t,\bar t) + \Sigma^{\tn{ef}}_{\underline i\gamma}(t,\bar t)
  \right\}G^{\tn{fe}}_{\gamma \underline j}(\bar t,t') =& 
  \nonumber\\
    \int_C\!\!\!\d\bar{t}\,\left\{ h^{{\rm HF},\tn{ef}}_{\underline i \gamma}(\bar t)\,\delta_C(t,\bar t) + \Sigma^{\tn{ef}}_{\underline i \gamma}(t,\overline{t})
  \right\}\int_C\!\!\!\d\overline{\overline t}\,\int_C\!\!\!\d{\tilde t}\, g^{\tn{f}}_{\gamma\delta}(\bar t,\overline{\overline t})\,\left\{ h^{{\rm HF},\tn{fe}}_{\delta \underline k}(\overline{\overline t})\,\delta_C(\overline{\overline t},\tilde t) + \Sigma^{\tn{fe}}_{\delta \underline k}(\overline{\overline t},\tilde t)
  \right\}G^{\tn{e}}_{\underline k\underline j}(\tilde t,t') \equiv & \int_C\!\!\!\d\tilde{t}\,\Sigma^{\rm emb, e}_{\underline i\underline k}(t,\tilde t)G^e_{\underline k\underline j}(\tilde t,t')\,,
\end{align}
where we have introduced the embedding selfenergy for the system ``e'' which is given below in Eq.~\eqref{eq:sigma-embedding-e}.
This result allows us to rewrite the equations for $g^\tn{e}$,  $G^\tn{e}$ and $G^{\tn{es}}$ with a generalized selfenergy, $\tilde\Sigma^\tn{e}:=\Sigma^\tn{e}+\Sigma^\tn{emb,e}$, which takes over the role of $\Sigma^\tn{e}$ in the previous equations. With this, the functions $g^\tn{f}$ and $G^\tn{f}$ have been completely eliminated.

The final set of equations is, therefore,
\begin{align}
\left\{\i\hbar\partial_t\delta_{ik}-h^{{\rm HF},\tn{s}}_{ik}(t)\right\}G^{\tn{s}}_{kj}(t,t')&
\label{eq:gss-equation-ml-final}
=
  \delta_{ij}\delta_C(t,t') 
  + \int_C\!\!\!\d\bar{t}\,\tilde\Sigma^{\tn{s}}_{ik}(t,\bar{t})G^{\tn{s}}_{kj}(\bar{t},t')\,,
\qquad \tilde\Sigma^{\tn{s}} := \Sigma^{\tn{s}} + \Sigma^{\tn{emb,s}}
\\
  \Sigma^{\rm emb, s}_{ik}(t,\tilde t) &= \int_C\!\!\!\d{\bar t}\,\int_C\!\!\!\d\overline{\overline t}\,  \left\{ h^{{\rm HF},\tn{se}}_{i\,\underline l}(\bar t)\,\delta_C(t,\bar t) + \Sigma^{\tn{se}}_{i\,\underline l}(t,\overline{t})
  \right\} g^{\tn{e}}_{\underline l\,\underline m}(\bar t,\overline{\overline t})\,\left\{ h^{{\rm HF},\tn{es}}_{\underline m\,k}(\overline{\overline t})\,\delta_C(\overline{\overline t},\tilde t) + \Sigma^{\tn{es}}_{\underline m\,k}(\overline{\overline t},\tilde t)
  \right\}\,,
\label{eq:sigma-embedding-s}
\\
  \left\{\i\hbar\partial_t\delta_{\underline i \,\underline k}-h^{{\rm HF},\tn{e}}_{\underline i \,\underline k}(t)\right\}g^{\tn{e}}_{\underline k\,\underline j}(t,t')
  &=\delta_{\underline i\, \underline j}\delta_C(t,t')+ 
  \int_C\!\!\!\d\bar{t}\,\tilde\Sigma^{\tn{e}}_{\underline i\underline k}(t,\bar{t})g^{\tn{e}}_{\underline k\underline j}(\bar{t},t')\,,
  \quad
\qquad \tilde\Sigma^{\tn{e}} := \Sigma^{\tn{e}} + \Sigma^{\tn{emb,e}}\,,
  \label{eq:gee-equation-ml-final}
\\
  \Sigma^{\rm emb, e}_{\underline i\underline j}(t,\tilde t) &= \int_C\!\!\!\d{\bar t}\,\int_C\!\!\!\d\overline{\overline t}\,  \left\{ h^{{\rm HF},\tn{ef}}_{\underline i\,\gamma}(\bar t)\,\delta_C(t,\bar t) + \Sigma^{\tn{ef}}_{\underline i\,\gamma}(t,\overline{t})
  \right\} g^{\tn{f}}_{\gamma\delta}(\bar t,\overline{\overline t})\,\left\{ h^{{\rm HF},\tn{fe}}_{\delta\underline j}(\overline{\overline t})\,\delta_C(\overline{\overline t},\tilde t) + \Sigma^{\tn{fe}}_{\delta \underline j}(\overline{\overline t},\tilde t)
  \right\},\quad
\label{eq:sigma-embedding-e}
\\
  \left\{\i\hbar\partial_t\delta_{\alpha\gamma}-h^{{\rm HF},\tn{f}}_{\alpha\gamma}(t)\right\}g^{\tn{f}}_{\gamma\beta}(t,t')
  &-\int_C\!\!\!\d\bar{t}\,\Sigma^{\tn{f}}_{\alpha\gamma}(t,\bar{t})g^{\tn{f}}_{\gamma\beta}(\bar{t},t')\,
  =\delta_{\alpha \beta}\delta_C(t,t')\,.
  \label{eq:gff-equation-ml-final}
\end{align}
These are the two-time embedding equations for a hierarchy of three layers: $\tn{f} \to \tn{e} \to \tn{s}$. The structure is very clear and indicates that this scheme can be extended to more layers. The main advantage of this approach comes about in situations when the systems s, e, and f can be described by increasingly simple selfenergies, and if f (and e) is large. Then the embedding scheme will provide significant speedup, compared to a description of the entire system by the most complicated selfenergy $\Sigma^{\tn{s}}$.

Note that we presented the above two extensions of the common embedding scheme for the two-time Keldysh-Kadanoff-Baym equations. While the structure within the two-time framework is very clear and compact, the scaling of this system with the number of time steps is highly unfavorable, at least when correlation selfenergies for the environment and/or the system-environment coupling are taken into account. The situation is completely different within the time local scheme which always retains the time-linear scaling. On the other hand, the embedding scheme will allow one to avoid the bottleneck of the G1--G2 scheme -- the large computer memory for storage of $\mathcal{G}$ -- only if with the embedding the dimension of the $\mathcal{G}$-matrix can be significantly reduced. This will be the case if, again, the environment is treated uncorrelated and the correlated part (the ``system'') is a small part of the whole system. The transformation of the above two-time embedding equations into  a time local form  will be presented in forthcoming work.

\section{NEGF-based Quantum fluctuations approach}\label{s:quantum-fluctuations}
This is an approach to the many-particle problem that is qualitatively different from, both, standard NEGF theory or the G1--G2 scheme. It allows one to avoid the memory-expansive computation of the two-particle correlation function $\mathcal{G}$. The idea is to replace, in the G1--G2 scheme, the computation of $\mathcal{G}$ -- a rank 4 tensor -- by the computation of products of single-particle quantities, i.e. rank 2 tensors. This situation is familiar from the mean field approximation but its accuracy is often no sufficient. The extension of such an approach to correlated classical systems goes back to Klimontovich \cite{klimontovich_75} and similar ideas for quantum systems  were introduced by Ayik, Lacroix and co-workers, e.g. \cite{ayik_plb_08,lacroix_epj_14,lacroix_prb14}.
Recently, we have developed a more general quantum fluctuations concept that we discuss in the following. It was presented in Ref.~\cite{schroedter_cmp_22} and was extended to two-time quantities in Ref.~\cite{schroedter_23}. The relation of the approach to the Bethe--Salpeter approach was investigated in Ref.~\cite{schroedter_pssb23}.
Here we briefly summarize the idea.

\subsection{Idea of the quantum fluctuations approach}
The main idea of the present quantum fluctuations approach \cite{schroedter_cmp_22,schroedter_23,schroedter_pssb23} is to consider fluctuations of the single-particle NEGF on the time diagonal that are defined as
\begin{align}
    \delta\hat{G}_{ij}(t)&\coloneqq \hat{G}_{ij}^<(t)-G^<_{ij}(t) 
    \equiv \hat{G}^>_{ij}(t)-G^>_{ij}(t)\,,
\end{align}
where $\hat{G}^\gtrless$ denotes the operator corresponding to the lesser/greater component of the single-particle NEGF,
\begin{align*}
    \hat G^<_{ij}(t,t') &=\pm \frac{1}{\i\hbar} \chat c_j^\dagger(t') \chat c_i(t)\,,
    \\
    \hat G^>_{ij}(t,t') &= \frac{1}{\i\hbar} \chat c_i(t) \chat c_j^\dagger(t') \,.
\end{align*}

While, within the G1--G2 scheme,  the focus is on the correlated part of the two-particle NEGF, $\mathcal{G}$, the idea of the quantum fluctuations approach is to instead consider the two-particle exchange-correlation (XC) function $L$ instead. This function comprises all exchange and  correlation contributions to the two-particle NEGF, and its definition on the Keldysh contour was given in Eq.~(\ref{eq:definition_XC_function}).
Interestingly, the following real-time components of the XC function can also be identified as correlation functions of single-particle fluctuations, i.e.,
\begin{align}\label{eq:l-2time}
    L_{ijkl}(t,t') &\coloneqq \left\langle \delta\hat{G}_{ik}(t)\delta\hat{G}_{jl}(t')\right\rangle\,,\\
    L_{ijkl}(t)&\coloneqq L_{ijkl}(t,t)\,.
\end{align}
Considering a reformulation of the theory in terms of two-particle fluctuations rather than the two-particle correlation functions, $\mathcal{G}$, holds a number of advantages. On the one hand, all results within the single-time G1--G2 scheme also hold within the framework of the quantum fluctuations approach. On the other hand, this new approach is closely related to the classical theory of fluctuations developed by Klimontovich that was mentioned above \cite{klimontovich_75}. This allows us to use approximations that have already been explored for classical systems and extend them to the quantum case. 
Furthermore, it is possible to combine the fluctuations approach with a stochastic procedures that allows for a significant reduction of the  computational effort as compared to, for example, the G1--G2 scheme. In particular, this stochastic approach allows one to completely eliminate the costly computation of the rank-four tensor of $\mathcal{G}$ which is currently the bottleneck in the use of the G1--G2 scheme, as was discussed in Sec.~\ref{ss:g1-g2-problems}. Aside from the mentioned advantages, the present fluctuations scheme has also limitations: 
the application of stochastic methods is presently restricted to the weak coupling regime.

\subsection{Dynamics of fluctuations}\label{ss:flucutations-dynamics}
Expressing the EOM for the lesser NEGF in terms of two-particle fluctuations eliminates the exchange contribution from the single-particle Hartree--Fock Hamiltonian, $h^\mathrm{HF}$, and transfers it to the collision term. Thus, the EOM takes the following form
\begin{align}
    \mathrm{i}\hbar\frac{\mathrm{d}}{\mathrm{d}t}G^<_{ij}(t)+\Big[h^\mathrm{H},G^<\Big]_{ij}(t)+\Big[\mathcal{I}+\mathcal{I}^\dagger\Big]_{ij}(t)\,, \label{eq:EOM_G<_fluc}
\end{align}
where $h^\mathrm{H}$ denotes the single-particle Hartree Hamiltonian that follows from Eq.~\eqref{eq:h_HF} by the replacement $w^\pm\rightarrow w$, whereas $\mathcal{I}$ denotes the fluctuations collision term defined as
\begin{equation}
    \mathcal{I}_{ij}(t)\coloneqq \pm\mathrm{i}\hbar \sum_{klp}w_{ikjl}(t) L_{plkj}(t)\,. \label{eq:definition_fluc_collision_term}
\end{equation}
As was shown in Ref.~\cite{schroedter_cmp_22}, the dynamics of quantum many-body systems can be described using the EOM for the lesser NEGF and the EOM for single-particle fluctuations since the latter provides the basis for the EOMs for all $n$-particle fluctuations. Single-particle fluctuations obey the following EOM
\begin{align}
    \mathrm{i}\hbar\frac{\mathrm{d}}{\mathrm{d}t}\delta\hat{G}_{ij}(t)=& \Big[h^\mathrm{H},\delta\hat{G}\Big]_{ij}(t)+\Big[\delta\hat{\Sigma}^\mathrm{H},G^<\Big]_{ij}(t)
    +\Big[\delta\hat{\mathcal{I}}+\delta\hat{\mathcal{I}}^\dagger\Big]_{ij}(t)\,, \label{eq:EOM_1pFluctuations}
\end{align}
where we introduced the fluctuations Hartree selfenergy defined as
\begin{equation}
    \delta\hat{\Sigma}^\mathrm{H}_{ij}(t)\coloneqq \pm\mathrm{i}\hbar\sum_{kl} w_{ikjl}(t) \delta\hat{G}_{lk}(t)\,,\label{eq:def_fluctuations_Hartree_selfenergy}
\end{equation}
and the second-order fluctuations collision term given by 
\begin{equation}
    \delta\hat{\mathcal{I}}_{ij}(t)\coloneqq \pm\mathrm{i}\hbar\sum_{klp} w_{iklp}(t)\delta\hat{L}_{plkj}(t)\,. \label{eq:def_2nd-order_collision_term}
\end{equation}
where we have second-order fluctuations, i.e., fluctuations of fluctuations defined as 
\begin{align}
    \delta\hat{L}_{ijkl}(t)\coloneqq \delta\hat{G}_{ik}(t)\delta\hat{G}_{jk}(t)-L_{ijkl}(t)\,.\label{eq:def_2nd-order_fluctuations}
\end{align}
The appearance of second order fluctuations gives rise to a hierarchy of equations that is similar to the BBGKY hierarchy of reduced density operators \cite{bonitz_qkt}. Within the framework of fluctuations, the EOM for two-particle fluctuations couples to three-particle fluctuations and so on \cite{schroedter_cmp_22}.

\subsection{Approximations}\label{ss:fluctuations-approximations}
Given the direct connection between the fluctuations approach and the general theory of NEGF, it is possible to translate the known approximations of the latter [cf. Sec.~\ref{s:sigmas}] into the framework of the former. Additional approximations can be obtained by using results of the classical fluctuations theory by Klimontovich \cite{klimontovich_75} and constructing their quantum counterparts. However, the main advantage of the quantum fluctuations approach lies in the existence of certain approximations that can be expressed entirely in terms of  single-particle fluctuations. The reason is, that for these approximations semiclassical methods can then be applied which allow eliminate any explicit dependence on any rank-four tensor such as $\mathcal{G}$ and $L$.

\subsubsection{Approximations of moments}
The simplest approximations, within the quantum fluctuations approach, are the approximations of moments. In the approximation of first moments, all contributions to the fluctuations hierarchy except for the first moment are neglected, i.e., $L\approx 0$. At the level of single-particle fluctuations this corresponds to $\delta\hat{G}\approx 0$ and, consequently, leads to the well-known Hartree approximation.\\
Another approximation is given by the approximation of second moments and corresponds to neglecting all contributions due to three-particle fluctuations. On the level of single-particle fluctuations this corresponds to $\delta\hat{L}\approx 0$ and thus leads to the following EOM
\begin{align}
    \mathrm{i}\hbar\frac{\mathrm{d}}{\mathrm{d}t}\delta\hat{G}^\mathrm{2M}_{ij}(t)=& \Big[h^\mathrm{H},\delta\hat{G}^\mathrm{2M}\Big]_{ij}(t)+\Big[\delta\hat{\Sigma}^\mathrm{H},G^<\Big]_{ij}(t)\,.
\end{align}
Unlike the approximation of first moments, this approximation to the fluctuations hierarchy does not directly correspond to any of the well-known approximations within NEGF theory and its validity range is unclear. We, therefore, will concentrate on the quantum polarization approximation which has been explored in some detail before.
\subsubsection{Quantum polarization approximation}\label{sss:qpa}
The quantum polarization approximation (QPA) is a weak coupling approximation within the quantum fluctuations approach. Therefore, it is closely related to the $GW$ approximation of NEGF theory \cite{schroedter_cmp_22,schroedter_pssb23}, cf. Secs.~\ref{ss:negf-gw} and \ref{s:g1g2-approx}. Within the QPA it is assumed that two-particle fluctuations are much larger than two-particle correlations, i.e., $|\mathcal{G}|\ll |L|$, (weak coupling). At the level of single-particle fluctuations, this approximation is equivalent to
\begin{align}
    \delta\hat{L}_{ijkl}(t)\approx \pm\big\{G^>_{il}(t)\delta\hat{G}_{jk}(t)+\delta\hat{G}_{il}(t)G^<_{jk}(t)\big\}\,,
\end{align}
leading to the following EOM for the single-particle fluctuations
\begin{align}
    \mathrm{i}\hbar\frac{\mathrm{d}}{\mathrm{d}t}\delta\hat{G}^\mathrm{P}_{ij}(t)=& \Big[h^\mathrm{HF},\delta\hat{G}^\mathrm{P}\Big]_{ij}(t)+\Big[\delta\hat{\Sigma}^\mathrm{HF},G^<\Big]_{ij}(t)\,. \label{eq:EOM_1pFluc_PA}
\end{align}
One immediately recognizes the appearance of the Hartree-Fock hamiltonian on the l.h.s., meaning that exchange contributions that are missing for the approximation of second moments are restored. Using this equation and relation (\ref{eq:l-2time}) allows one to derive an equation of motion for $L$ which was given in Ref.~\cite{schroedter_23}. There it was shown that this equation is, in the weak coupling limit, very close to the $GW$ approximation within the G1--G2 scheme but, in addition, includes further exchange contributions. In particular, compared to the GW approximation, cf. Eq.~(\ref{eq:g1g2_GW_exp}), the QPA replaces $w\rightarrow w^\pm$, both, in the SOA and polarization contributions. In this sense, the QPA is an approximation that is intermediate between the $GW$ approximation (which does not include exchange contributions) and the ASP which includes exchange contributions but also particle-hole T-matrix diagrams, for details, see table~\ref{tab:g1g2approx}. 

The EOMs for the single-particle fluctuations, Eq.~(\ref{eq:EOM_1pFluc_PA}), are not directly solvable as they are operator equations. A solution to this problem is given in the form of the stochastic mean field approximation (SMF) \cite{ayik_plb_08,lacroix_epj_14,lacroix_prb14} where quantum mechanical operators are replaced by random variables, i.e., $\delta\hat{G}\rightarrow \Delta G^\lambda$, where $\lambda$ denotes a sufficiently large set of random realizations. Furthermore, the quantum-mechanical expectation value is replaced by a standard stochastic expectation value over the realizations, i.e., $\langle\cdot\rangle\rightarrow \overline{(\cdot)}$. Ideally, the random variables should be constructed such that all symmetric moments of the single-particle fluctuations for the initial state are exactly reproduced\footnote{The symmetrized moments have to be considered as the random variables commute whereas this does not hold for the operators.} which, however, is known to be not possible \cite{Lacroix2019}. Within standard SMF theory this gives rise to deviations resulting from contributions of higher moments. However, introducing approximations that eliminate nonlinear terms in the EOM for the single-particle fluctuations makes these contributions negligible \cite{schroedter_cmp_22}. Application of the SMF approach to the QPA, has been termed stochastic polarization approximation (SPA), the result of which are in very good agreement with the $GW$ approximation within the G1--G2 scheme \cite{schroedter_cmp_22}. \\

While the previous results of the SPA were restricted to observables that depend on one time argument, this excluded more complex quantities such as correlation functions.
Only recently, this problem could be solved. The modified method was called the multiple ensembles (ME) approach and presented in Ref.~\cite{schroedter_23}. This method allows to evaluate expectation values of non-commuting operators which is not possible in a straightforward scheme that samples semiclassical quantities which always commute. The idea is to introduce not one ensemble, but two ensembles of random variables. Then operators are replaced by random variables in a way that depends on their ordering, i.e., 
\begin{align}
    \delta\hat{G}_{ij}&\rightarrow \left(\Delta G^{(1),\lambda}_{ij},\Delta G^{(2),\lambda}_{ij}\right)\,\\
    \delta\hat{G}_{ij}\delta\hat{G}_{kl}&\rightarrow \Delta G^{(1),\lambda}_{ij}\Delta G^{(2),\lambda}_{kl}\,,
\end{align}
where the ensembles satisfy the following two conditions on their lowest moments\footnote{Notice that an ideal initial state is assumed as the correlated initial state can be produced by means of the adiabatic switching method.}
\begin{align}
    \overline{\Delta G^{(1),\lambda}_{ij}(t_0)}&= \overline{\Delta G^{(2),\lambda}_{ij}(t_0)}=0\,\\
    \overline{\Delta G^{(1),\lambda}_{ij}(t_0)\Delta G^{(2),\lambda}_{kl}(t_0)}&= -\frac{1}{\hbar^2}\delta_{il}\delta_{jk}n_j(t_0)[1\pm n_i(t_0)]\,.
\end{align}
The equations of motion, within the SPA-ME, follow from Eqs.~\eqref{eq:EOM_G<_fluc} and \eqref{eq:EOM_1pFluc_PA} by the aforementioned replacements of operator expressions, i.e., 
\begin{align}
    \mathrm{i}\hbar \frac{\mathrm{d}}{\mathrm{d}t} G^<_{ij}(t)=&\Big[h^\mathrm{H},G^<\Big]_{ij}(t)+\Big[\mathcal{S}+\mathcal{S}^\dagger\Big]_{ij}(t)\nonumber 
    +\Big[ \mathcal{I}^\mathrm{ME}+\mathcal{I}^{\mathrm{ME}\dagger}\Big]_{ij}(t)\,,
    \\
    \mathrm{i}\hbar \frac{\mathrm{d}}{\mathrm{d}t} \Delta G^{(m),\lambda}_{ij}(t)=& \Big[ h^\mathrm{HF},\Delta G^{(m),\lambda} \Big]_{ij}(t)\nonumber
    +\Big[\Delta \Sigma^{\mathrm{HF},(m),\lambda}, G^<\Big]_{ij}(t)\,,
\end{align}
where $m=1,2$, and we further introduced the abbreviations
\begin{align}
    \mathcal{S}_{ij}(t)&\coloneqq \frac{1}{2}\sum_{kl}w_{kljk}(t) G^<_{il}(t)\,,\\
    \mathcal{I}^\mathrm{ME}_{ij}(t)&\coloneqq \pm\frac{\mathrm{i}\hbar}{2}\sum_{klp}w_{iklp}(t)\big\{ L^\mathrm{ME}_{plkj}(t)+L^\mathrm{ME}_{lpjk}(t)\big\}\,,\\
    L^\mathrm{ME}_{ijkl}(t)&\coloneqq \overline{\Delta G^{(1),\lambda}_{ik}(t)\Delta G^{(2),\lambda}_{jl}(t)}\,.
\end{align}

\subsubsection{Scaling of the Quantum Fluctuations approach. Numerical example}\label{sss:qpa}
The advantage of the SPA-ME is twofold; the computational complexity of the equations is significantly lower compared to the equivalent equations within the G1--G2 scheme: for the SPA-ME the memory consumption scales as $\mathcal{O}(N_\lambda N_\mathrm{b}^2)$, where $N_\lambda$ is the number of sampled realizations, compared to $\mathcal{O}(N_\mathrm{b}^4)$ for the $GW$--G1--G2 approximation. This reduction by a factor of $N_\lambda / N_\mathrm{b}^2$ also applies for the computational complexity. For the special case $N_\lambda \sim N_\mathrm{b}^2$ the statistical error of the SPA-ME approach vanishes and the scaling of both methods is identical. However, in practice it was observed that a constant number of samples of $N_\lambda\approx 10^4$, independent of the system size, is sufficient to achieve good accuracy~\cite{schroedter_cmp_22}. This makes this approach especially efficient for large systems where $N_\tn{b}^2 > 10^4$. The other advantage of this approach is the direct access to observables that depend on products of fluctuations for (in general) independent time arguments such as the density response function or the dynamic structure factor~\cite{schroedter_23}. Most importantly, this approach can be applied to describe systems not only in but also far from equilibrium. Thus, it becomes possible to calculate, for example, the dynamic structure factor for systems that undergo a rapid dynamics, entirely from single-particle single-time calculations.\\
\begin{figure}[h]
\centering
\includegraphics[width=1.00\columnwidth]{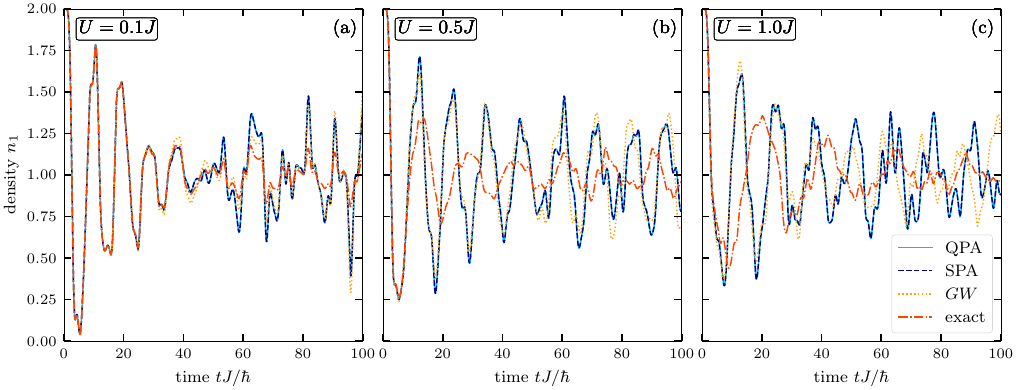}
\caption{Density dynamics on the first site of a half-filled eight-site Hubbard chain for three coupling strengths [$U=0.1J$ (a), $U=0.5J$ (b) and $U=1.0J$ (c)]. In the initial state, the leftmost four sites are fully occupied whereas the remaining four ones are empty. Moreover, the initial state is uncorrelated. The dynamics are calculated using QPA and SPA and are compared to calculations using the $GW$ approximation and exact diagonalization.}
\label{fig:comparison_fluctuations}
\end{figure}

Finally, we illustrate the SPA approach with a numerical example. We consider a half-filled Hubbard chain with eight lattice sites. We choose an initial state such that the leftmost four lattice sites are fully occupied, whereas the remaining four are empty.
As before, we examine the density dynamics on the first site. We compare results from the QPA and the SPA for varying on-site interactions. 
In addition we include results from $GW$--G1--G2 as well as from exact diagonalization  calculations. The results are depicted in Fig. \ref{fig:comparison_fluctuations}. The first observation is that, for all coupling strengths, there is excellent agreement between QPA and SPA. Let us now compare SPA (and QPA) to the $GW$ approximation and exact diagonalization, starting with the weak coupling case,  $U=0.1J$ in panel (a). All calculations exhibit very good agreement, up to a time of $t\approx 40 \hbar/J$. Subsequently, all three approximations exhibit small deviations from the exact result in the amplitude of the density oscillations, whereas the frequency of the oscillations is well captured. Another difference is that the exact dynamics are damped with increasing time, whereas the approximations show a revival. This is attributed to the use of HF propagators in the GKBA. Furthermore, it is demonstrated that the polarization approximation shows good agreement with the $GW$ approximation. Next, consider panel (b) where the coupling is increased to $U=0.5J$. Here, differences between the exact calculation and the approximations become apparent already at $t\approx 10\hbar/J$ . Nevertheless, the two polarization approximations and the $GW$ approximation exhibit very good agreement, up to $t\approx 50\hbar/J$, after which minor deviations become visible. Finally, for $U=1.0J$, in panel (c), we observe that the approximations follow the exact dynamics only up to a time of $t\approx 5\hbar/J$. Again, all three approximations agree very well with each other for times $t\lesssim 45\hbar/J$. Subsequently, the approximations continue to have similar amplitudes of the density oscillations,
but the frequencies differ significantly. This behavior is not surprising, as each of the three approximations is based on the assumption of weak coupling which is not fulfilled in panel c). On the other hand, the equivalence of the polarization approximation and the $GW$ approximation holds only in the limiting case of weak coupling, i.e. for panels a) and b). Numerical results illustrating the multiple ensembles approach have been reported in Ref.~\cite{schroedter_23} and confirm the excellent behavior of the polarization approximations, in the weak coupling limit.

\section{Discussion and outlook}\label{s:discussion}
Nonequilibrium Green functions theory has become one of the main methods to simulate correlated quantum many-body systems in and out of equilibrium in a wide range of fields, including nuclear matter, optically excited semiconductors, quantum plasmas, correlated materials and ultracold atoms in traps and optical lattices. The main advantage of NEGF is the systematic approach to improved many-body approximations via Feynman diagram techniques and the straightforward way to construct approximations that satisfy the relevant conservation laws. At the same time, in some cases better accuracy is achieved if some symmetries of the exact solution are given up (partially), which is known as ``L\"owdin's symmetry dilemma'' and was demonstrated for ground state NEGF simulations \cite{joost_cpp_21}. Similar behavior can be expected for nonequilibrium situations.
%
NEGF not only allow one to incorporate important many-body effects qualitatively in a consistent way, they have also been shown to provide quantitative agreement with benchmarks. For example, very good agreement with cold atom experiments in optical lattices, e.g. \cite{schluenzen_prb16}, and available exact theoretical results for Hubbard systems \cite{schluenzen_prb17} could be demonstrated,  provided the proper selfenergy approximations were used.

However, full two-time NEGF simulations are computationally expansive, as they suffer from a cubic scaling of the computation time with the simulation duration. This is a result of the use of two-time Green functions, on one hand, and the formal closure of the Martin-Schwinger hierarchy with help of the one-particle selfenergy, on the other hand, which comes at the cost of a memory time integral, cf. Eqs.~(\ref{eq:kbe-sigma-form1}) and (\ref{eq:kbe-sigma-form2}). This scaling could be drastically reduced to a time-linear one by introducing the G1--G2 scheme -- an exact reformulation of the generalized Kadanoff-Baym ansatz with Hartree-Fock propagators \cite{schluenzen_prl_20}. Remarkably, this scaling is achieved quickly, cf. Fig.~\ref{fig:scaling}, and also for high-level selfenergies, including the nonequilibrium $GW$ and $T$-matrix approximations and the  dynamically screened ladder approximation \cite{joost_prb_20,joost_prb_22}. This makes long NEGF simulations with thousands of time steps possible that were previously out of reach -- speedup factors can easily reach $10^3\dots 10^5$ -- and gives also access to important approximations, such as the DSL, that were unfeasible before. These advantages of the G1--G2 scheme have led, within a short period, to a remarkable number of applications which include finite Hubbard clusters, strongly correlated solids, 2D quantum materials, electron-boson systems, and dense plasmas.

Of course, the dramatic speedup of NEGF simulations that is achieved by the G1--G2 scheme does not come for free: the elimination of the selfenergy by the correlated part of the two-particle Green function $\mathcal{G}$ requires to store, in addition to standard NEGF simulations, a rank-four tensor, $\mathcal{G}_{ijkl}$. This shifts the computational load to a computer memory problem which requires different computational strategies. Aside from the (obvious) transition to massively parallel computing there exist additional options to reduce the size of the rank-four tensor, two of which we discussed in this paper. The first is the use of NEGF embedding schemes, cf. Sec.~\ref{s:embedding}, which subdivides the system into an ``important'' one (the system of interest) and a ``less important'' (the ``environment'') one where only the former is treated taking correlations fully into account whereas the effect of the environment is treated approximately, e.g. on the mean field level. This allows one to reduce the dimensionality of the two-particle Green function to the dimensionality of the ``system''. We demonstrated that the NEGF embedding concept can be transformed into a time local form and integrated into the G1--G2 scheme, fully benefiting from time linear scaling \cite{balzer_prb_23}, and part of these results were reproduced in Ref.~\cite{tuovinen_prl_23}. The strength of this scheme was demonstrated for charge transfer between a highly charged ion and a correlated quantum material \cite{niggas_prl_22}. We also pointed out that the time local embedding scheme requires special care in order to guarantee that the conservation laws are fulfilled: it is necessary to include into the dynamics of the ``environment'' Green function $G^\tn{e}$ the coupling to the system [which is in contrast to the two-time NEGF embedding formalism where only the isolated environment Green function $g^\tn{e}$ appears] and to use the extended version of the HF-GKBA, Eq.~(\ref{eq:GKBA-ss}), which fully  includes the coupling between the system parts. We also presented two extensions of the NEGF embedding approach: the first allows one to take into account correlation effects in the environment or/and in the system-environment coupling, cf. Sec.~\ref{ss:embedding-beyond}. The second extension allow one to incorporate another ``environment layer'' into the description which might be relevant complex transport problems, hierarchical models and also to describe secondary electron emission occuring during ion impact, e.g. \cite{Bonitz_fcse_19}. A question that will be solved in a forthcoming paper, Ref.~\cite{bonitz_prb_24}, is the incorporation of the system-environment coupling into the dynamics of the correlation function $\mathcal{G}$. We also discussed similarities of our embedding scheme with the embedding approach used in dynamical mean field theory \cite{freericks_2006,aoki_2014} to efficiently treat strongly correlated systems \cite{vollhardt_2019}. Possible synergies of both methods are subject of ongoing work.

In this paper, in Sec.~\ref{sss:g1g2-uniform}, we also presented new results for spatially uniform systems that illustrated chances and open challenges of the G1--G2 scheme. 
For homogeneous systems, the dependence of the two-particle Green function $\mathcal{G}$ on three momentum vectors make simulations a technical challenge. In particular, aliasing effects easily occur in long simulations in the HF-GKBA, as a result of insufficient $k-$points. A phenomenological ansatz that uses a damping constant in the GKBA is able to cure the signs of aliasing, but this comes at the cost of a violation of energy conservation. Therefore, it is important to develop refined or alternative concepts in order to benefit from long time simulations. This is in particular important in 2D and 3D systems where the memory scaling currently only allows for rather coarse $k$-point grids, and aliasing occurs early, although it is more concealed than in 1D.

We also presented an application of the G1--G2 scheme to graphene where we studied the excitation of electrons by a short laser pulse with linear or circular polarization and a photon energy that is resonant with the energy gap at the M-points. The polarization of the electric field plays a central role in the selection of the excited $\mathbf{k}$-states. In more extreme examples, one can even achieve valley depolarisation, where one Dirac point is strongly excited while the other one is not. This happens readily in TMDCs, cf. Refs.~\cite{Caruso_ChiralValleyExcitons_2022,Caruso_NanoLett2023}, but can also be achieved in gap-less semiconductors as graphene by using very strong linearly polarized fields, e.g. Refs.~\cite{rost_fd_22,rost_prr_22}, or superpositions of different circular polarized lasers, cf. Refs.~\cite{mrudul_jpb_21, mrudul_optica_21}. Our model incorporates the full Brillouin zone and thus is suited to also investigate these processes. In order to include excitonic effects, our implementation can be supplemented to also include long-range interactions as in the PPP model, and higher selfenergies such as $GW$ and DSL. The former improvement does, in fact, not even increase the cost of the simulations.
We approximated graphene as a two-band system with energies given by the tight-binding model. A more realistic description that considers additional bands, especially those in the $\Gamma$-point region can be easily integrated in the G1--G2 scheme and will be important for the dynamics on longer time scales or for increased laser intensities.


In Sec.~\ref{s:quantum-fluctuations} we discussed a second approach that is capable to reduce the storage problem for 
$\mathcal{G}$: the NEGF-based quantum fluctuations approach. This approach allows one to eliminate all rank-four tensors from the model and to compute two-particle (four-point) functions entirely from one-particle (two-point) functions. This would be exact within the Hartree-Fock approximation. However, the quantum fluctuations approach allows one to extend such a description to correlated systems via a semiclassical average over an ensemble of random dynamical trajectories. This approach is based on the fluctuations approach of Klimontovich, e.g.~\cite{schroedter_cpp_24,bonitz_cpp_24}, and the stochastic mean field concept of Ayik and co-workers \cite{ayik_plb_08}, which was extended to NEGF theory by Schroedter et al. in Refs.~\cite{schroedter_cmp_22,schroedter_23}. We introduced the stochastic polarization approximation (SPA) which was shown to be equivalent to the GW approximations, in the weak coupling limit. Our numerical results within SPA agreed very well with GW simulations, however, the numerical cost is drastically reduced. This opens the door for more efficient G1--G2 simulations for systems were the basis dimension is presently too large to handle, such as spatially uniform systems, including jellium and plasmas, as we discussed in Sec.~\ref{sss:g1g2-uniform}. An issue of high interest is to extend the class of approximations, that are representable in terms of products of single-particle fluctuations, towards strong coupling.

Finally, further improvements of the successful G1--G2 scheme would be possible by going beyond Hartree-Fock propagators (beyond the HF-GKBA). The use of correlated propagators was shown to be essential in jellium simulations, cf. Sec.~\ref{sss:g1g2-uniform}, where it allows to avoid aliasing effects. Ideas how to systematically include correlated propagators without violating energy conservation have been discussed e.g. in Refs. \cite{bonitz-etal.99epjb,bonitz_qkt}. Further  improvements of the GKBA are possible by inclusion of vertex corrections \cite{kalvova_epl_18, kalvova_epl_23}.

\medskip
\textbf{Acknowledgements}
We acknowledge stimulating discussions with Iva Brezinova, Niclas Schlünzen, Hannes Ohldag, Fabio Caruso, and Claudio Verdozzi.
This work is supported by the German Science Foundation (DFG) via project BO1366-16.

\medskip


\providecommand{\WileyBibTextsc}{}
\let\textsc\WileyBibTextsc
\providecommand{\othercit}{}
\providecommand{\jr}[1]{#1}
\providecommand{\etal}{~et~al.}

\end{document}